\documentclass[prd,preprint,superscriptaddress,nofootinbib,preprintnumbers]{revtex4}
\usepackage{graphicx,epsfig}
\usepackage{ifpdf}
\usepackage{amsmath}
\usepackage{bm}
\usepackage{cancel}
\usepackage[english]{babel}
\usepackage{graphicx}%
\usepackage{amsfonts}%
\usepackage{amssymb}
\usepackage{braket}
\usepackage[normalem]{ulem}
\usepackage{enumerate}
\usepackage[usenames,dvipsnames]{color}
     \definecolor{hgreen}{rgb}{0,.3,0}
     \definecolor{hred}{rgb}{.3,0,0}
     \definecolor{hblue}{rgb}{0,0,.3}
     \definecolor{LightGray}{gray}{0.95}
\usepackage[backref=page,
            colorlinks=true,
            linkcolor=hblue,
            citecolor=hgreen,
            filecolor=hblue,
            urlcolor=hred]{hyperref}

\renewcommand*{\backref}[1]{}

\definecolor{nicered}{rgb}{0.7,0.1,0.1}
\definecolor{nicegreen}{rgb}{0.1,0.5,0.1}
\hypersetup{colorlinks,citecolor= nicegreen,linkcolor= nicered}

\newcommand{\beq}{\begin{equation}}
\newcommand{\eeq}{\end{equation}}
\newcommand{\bea}{\begin{eqnarray}}
\newcommand{\eea}{\end{eqnarray}}

\definecolor{Red}{rgb}{1.,0.,0.}

\def\SlashD{\,\slash\negthickspace \negmedspace\negmedspace D}

\def\mysection#1{{{\bf #1}.~}}
\def\OMIT#1{}

\preprint{FERMILAB-PUB-15-134-T}
\preprint{MITP/15-020}

\begin{document}

\def\Cincy{Department of Physics, University of Cincinnati, Cincinnati, Ohio 45221,USA}
\def\Bangkok{Department of Physics, Srinakharinwirot University,Wattana, Bangkok 10110 Thailand}
\def\Mainz{PRISMA Cluster of Excellence \& Mainz Institute for Theoretical Physics, Johannes Gutenberg University, 55099 Mainz, Germany}
\def\Fermilab{Theoretical Physics Department, Fermilab, P.O. Box 500, Batavia, IL 60510}

\title{Nonstandard Yukawa Couplings and Higgs Portal Dark Matter}

\author{Fady Bishara} 
\email[Electronic address:]{fadybishara@gmail.com} 
\affiliation{\Cincy}
\affiliation{\Fermilab}

\author{Joachim Brod} 
\email[Electronic address:]{joachim.brod@uni-mainz.de} 
\affiliation{\Mainz}

\author{Patipan Uttarayat} 
\email[Electronic address:]{patipan@g.swu.ac.th} 
\affiliation{\Bangkok}

\author{Jure Zupan} 
\email[Electronic address:]{zupanje@ucmail.uc.edu} 
\affiliation{\Cincy}

\begin{abstract}
We study the implications of non-standard Higgs Yukawa couplings to
light quarks on Higgs-portal dark matter phenomenology.  Saturating
the present experimental bounds on up-quark, down-quark, or
strange-quark Yukawa couplings, the predicted direct dark matter
detection scattering rate can increase by up to four orders of
magnitude. The effect on the dark matter annihilation cross section,
on the other hand, is subleading unless the dark matter is very light
-- a scenario that is already excluded by measurements of the Higgs
invisible decay width. We investigate the expected size of corrections
in multi-Higgs-doublet models with natural flavor conservation, the
type-II two-Higgs-doublet model, the Giudice-Lebedev model of light
quark masses, minimal flavor violation new physics models,
Randall-Sundrum, and composite Higgs models. We find that an
enhancement in the dark matter scattering rate of an order of
magnitude is possible. Finally, we point out that a discovery of
Higgs-portal dark matter could lead to interesting bounds on the
light-quark Yukawa couplings.
\end{abstract}

\maketitle

\section{Introduction} \label{sec:intro}
In Higgs-portal models~\cite{Patt:2006fw, MarchRussell:2008yu,
  Andreas:2008xy, Englert:2011yb, Lebedev:2011iq, LopezHonorez:2012kv,
  Djouadi:2012zc, Greljo:2013wja, Fedderke:2014wda, Craig:2014lda} of
dark matter (DM) the Higgs is usually assumed to be completely
Standard Model (SM) like apart from its couplings to
DM. Experimentally, only the couplings of the Higgs to the heaviest
particles of the SM are currently well constrained. The couplings to
gauge bosons are found to be in agreement with the SM predictions at
the ${\mathcal O}(20\%)$ level, while the constraints on the couplings
to third-generation fermions are somewhat
weaker~\cite{ATLAS-CONF-2013-034, CMS-PAS-HIG-13-005}.
Much less is known experimentally about the couplings of the Higgs to
the first two generations of fermions. The couplings to $u,d,s$, and
$c$ quarks could be as large as the SM bottom Yukawa coupling or be
absent altogether~\cite{Kagan:2014ila, Perez:2015aoa,
  Delaunay:2013pja, Efrati:2015eaa}. 
The Higgs couplings to top and bottom quarks will be quite well known
by the end of the high-luminosity LHC run. Some progress is also
expected on the measurements of Higgs couplings to charm and strange
quarks~\cite{Kagan:2014ila, Perez:2015aoa, Delaunay:2013pja}.  

Large $u$-, $d$-, and $s$-quark Yukawa couplings, comparable in size
to the $b$-quark Yukawa, generically require fine-tuning. A large
Yukawa coupling implies a large contribution to the quark mass from
the Higgs vacuum expectation value (vev). This would then need to be
cancelled by a different contribution to the $u$-, $d$-, and $s$-quark
masses, unrelated to the Higgs vev. The opposite limit, where the
observed Higgs does not couple to the light quarks at all is easier to
entertain. It simply requires a separate source of the light-quark
masses (for an extreme example see, e.g., \cite{Porto:2007ed}).

Modified light-quark Yukawa couplings could, in principle, have
important implications for DM phenomenology. In this article we
investigate how the Higgs-portal DM predictions change if the Higgs
couplings to the light quarks differ from the SM expectations.
We first allow for an arbitrary flavor structure of the Higgs Yukawa
couplings, only requiring that they satisfy the current experimental
bounds. In Section~\ref{sec:arbitrary:flavor} we derive the
implications for direct DM detection, indirect DM detection and the
collider searches.
We show that vanishing couplings of the Higgs to light quarks only
have a relatively small impact on these observables. Saturating the
loose current bounds on the light-quark Yukawa couplings would, on the
other hand, lead to drastically enhanced scattering cross sections on
nuclei while leaving the relic density and annihilation cross sections
nearly unmodified.

Clearly, an enhancement of the light-quark Yukawas by factors of
${\mathcal O}(100)$ or more, as allowed by current data, requires
considerable fine tuning of the quark-mass terms and hence seems quite
unlikely.
In Section~\ref{sec:flavor:models} we, therefore, explore the
deviations in the Higgs Yukawa couplings for a number of beyond-the-SM
scenarios and flavor models. This leads to more realistic expectations
as to how large the deviations in the direct DM detection rates can be
due to the poorly known Higgs couplings to the light quarks. Note that
we assume the DM to be a flavor singlet and that the new flavor
structure of the interactions with the visible sector is only due to
the modification of the SM Higgs couplings. DM that is in a nontrivial
flavor multiplet has been investigated in~\cite{Kile:2011mn,
  Agrawal:2011ze, Masina:2012hg, Lopez-Honorez:2013wla,
  Batell:2013zwa, Agrawal:2014una, Agrawal:2014aoa, Hamze:2014wca,
  Kile:2014jea, Kilic:2015vka, Calibbi:2015sfa, Agrawal:2015tfa},
while our study is closer in spirit to the work
in~\cite{Kamenik:2011nb, Bishara:2014gwa, Kim:2013ivd} where the
flavor dependence of the DM signals for flavor-singlet DM has been
explored.

A somewhat surprising result of our investigation is that, if DM is
discovered and turns out to be a thermal relic predominantly
interacting through a Higgs portal, it could be used to constrain the
light-quark Yukawa couplings. This is discussed in more detail in
Section~\ref{sec:light_quark_yukawas}. 

We summarize our results in Section~\ref{sec:conclusions}.

\section{Higgs portal with non-trivial flavor structure}
\label{sec:arbitrary:flavor}
We assume that DM and the SM fields are the only light degrees of
freedom. The remaining new physics (NP) particles can be integrated
out so that one can use an Effective Field Theory (EFT) approach. The
couplings of DM to the SM are given by the Higgs-portal Lagrangian
\begin{equation}\label{eq:LDMSM}
	\mathcal{L}_\chi =
	 \left\{ 
        \begin{aligned}
	& g_\chi \chi^\dagger\chi H^\dagger H\,,\hspace{4cm}\text{scalar DM;}\\
	& g_\chi \frac{1}{\Lambda} \bar{\chi}\chi H^\dagger H + i \tilde g_\chi
          \frac{1}{\Lambda} \bar{\chi} \gamma_5 \chi H^\dagger H\,,\hspace{0.5cm}\text{fermion  DM;}\\
	&\frac{g_\chi}{2} \chi^\mu\chi_\mu H^\dagger H,\hspace{3.7cm}\text{vector DM.}
	\end{aligned}\right.
\end{equation}
Above, the fermion DM can be either a Dirac or Majorana fermion (in
either case we use four component notation).  After electroweak
symmetry breaking (EWSB) we have
\begin{equation}
H^\dagger
H=\frac12\big(v_W^2+2v_W h+h^2\big),
\end{equation}
where $v_W=246$ GeV is the vacuum expectation value (vev) of the Higgs
field. The above interactions therefore lead to annihilation of DM
into both single Higgs, $\chi\bar\chi\to h$, and double Higgs,
$\chi\bar\chi\to h h$, final states.

The Higgs-portal operator for fermionic DM has mass dimension five and
is suppressed by the new physics scale $\Lambda$. The Higgs-portal
interaction for fermionic DM can also be re-written as ${\cal
  L}_{\chi}=(g_\chi+i\tilde g_\chi) \bar \chi_L \chi_R H^\dagger
H/\Lambda +{\rm h.c.}$. For $\tilde g_\chi\ne0$ the interaction is
thus both $P$- and $CP$-violating. The interaction for vector DM is
most probably also due to a higher-dimensional operator in the full
theory. For instance, if $\chi_\mu$ arises from a spontaneously broken
gauge symmetry in the dark sector, then $g_\chi\sim
v_{D}^2/\Lambda^2$, where $v_D$ is the vev of the field that breaks
the dark sector gauge invariance, while $\Lambda$ is the mass of the
mediator between DM and the Higgs.

The relevant terms, after EWSB, in the effective Lagrangian for the
Higgs couplings to the SM particles are given by 
\begin{equation}\label{eq:Lh}
\begin{split}
	\mathcal{L}_{\rm eff} &=- \kappa_q\frac{m_q}{v_W}\bar{q}qh -
        \kappa_\ell\frac{m_\ell}{v_W}\bar{\ell}\ell h +
        \kappa_V\left(\frac{2m_W^2}{v_W}W^{+\mu}W^-_\mu+ 
        \frac{m_Z^2}{v_W}Z^\mu Z_\mu\right)h \\
  & - \kappa_\lambda \frac{m_h^2}{2v_W} h^3 + \kappa_g^{\rm
          NP}\frac{\alpha_s}{12\pi v_W} h G^a_{\mu\nu}G^{a\mu\nu}
\end{split}
\end{equation}
where the $\kappa_i$ are real. A sum over the SM quarks,
$q=u,d,s,c,b,t$, and charged leptons, $\ell=e,\mu,\tau$ is implied,
and we have assumed custodial symmetry. The $h\to\gamma\gamma$
coupling is not relevant for DM phenomenology, since its effects are
suppressed compared to the Higgs couplings to gluons.\footnote{It
  could be relevant for direct detection if the scattering on
  electrons dominates. This requires very light DM, of order the
  electron mass. Such light Higgs-portal DM is excluded by the
  constraints on the Higgs invisible branching ratio.}  The couplings
are normalized such that $\kappa_q = \kappa_\ell = \kappa_V =
\kappa_\lambda = 1$ correspond to the SM. The experimental constraints
on the couplings of the light quarks to the Higgs, obtained from a
global fit to current data, are $|\kappa_u|<0.98 m_b/m_u$,
$|\kappa_d|<0.93 m_b/m_d$, $|\kappa_s|<0.70 m_b/m_d$, where only one
of the light Yukawa couplings was left to float in the fit, while all
the other Higgs couplings are set to the SM
values~\cite{Kagan:2014ila}. Higgs couplings to the light quarks of a
size comparable to the coupling to the $b$ quark are thus still
allowed. In \eqref{eq:Lh} we do not allow for flavor violating Higgs
couplings, since these are already tightly constrained from both Higgs
decays and low-energy observables \cite{Harnik:2012pb,
  Khachatryan:2015kon, Blankenburg:2012ex, Goudelis:2011un}.

In the SM, keeping the Higgs on shell, the $hGG$ coupling arises
predominantly from the one-loop top-quark contribution. The
$\kappa_g^{\rm NP}$ in \eqref{eq:Lh} encodes only the potential NP
contributions, and vanishes in the SM. In the global fits a parameter
$\kappa_g$ is introduced that gives the total $h\to gg$ amplitude,
including the SM contributions
\cite{ATLAS-CONF-2013-034,CMS-PAS-HIG-13-005}. We have (see, e.g.,
\cite{Carmi:2012in})
\begin{equation}
\kappa_g\simeq 1.03 \kappa_t+\kappa_g^{\rm NP}.
\end{equation}

At present, significant $CP$-violating Higgs couplings to fermions and
gluons are still allowed experimentally (see,
e.g.,~\cite{Brod:2013cka}), so we also discuss their effect on the
Higgs interactions with DM:
\begin{equation}\label{eq:Lh:CPV}
	\mathcal{L}_{\rm eff, CPV}=- i \tilde \kappa_q\frac{m_q}{v_W}\bar{q}\gamma_5 qh -
       i \tilde \kappa_\ell\frac{m_\ell}{v_W}\bar{\ell}\gamma_5 \ell h +
        \tilde \kappa_g^{\rm NP} \frac{\alpha_s}{8\pi v_W} h
        G^a_{\mu\nu} \widetilde G^{a\mu\nu}\,.
\end{equation}
Here, the $\tilde \kappa_i$ are real parameters; in the SM, we have
$\tilde \kappa_i=0$. Moreover, $G_{\mu\nu}^a$ is the gluon
field-strength tensor and $\widetilde G^{a,\mu\nu} = \tfrac{1}{2}
\epsilon^{\mu\nu\alpha\beta} G_{\alpha\beta}^a$ its dual. The
normalization of the $hG\tilde G$ term is chosen such that integrating
out the top at one loop one obtains $\tilde
\kappa_g=\tilde\kappa_t+\tilde \kappa_g^{NP}$. Accordingly, we have
Br$(h\to gg)\propto \kappa_g^2+(3 \tilde \kappa_g/2)^2$. The
$CP$-violating couplings of the Higgs to $ZZ$ and $WW$ are already
well constrained, and we thus set them to zero. In the numerical
analysis below, we will also assume $\tilde \kappa_i=0$, for
simplicity.

The modified Higgs couplings change the usual Higgs-portal predictions
for DM annihilation rates, the relic abundance, and direct detection
rates. In the following, we discuss these modifications in detail.

\subsection{Annihilation cross sections}
\begin{figure}[t]
\begin{center}
\includegraphics[height=2.15cm]{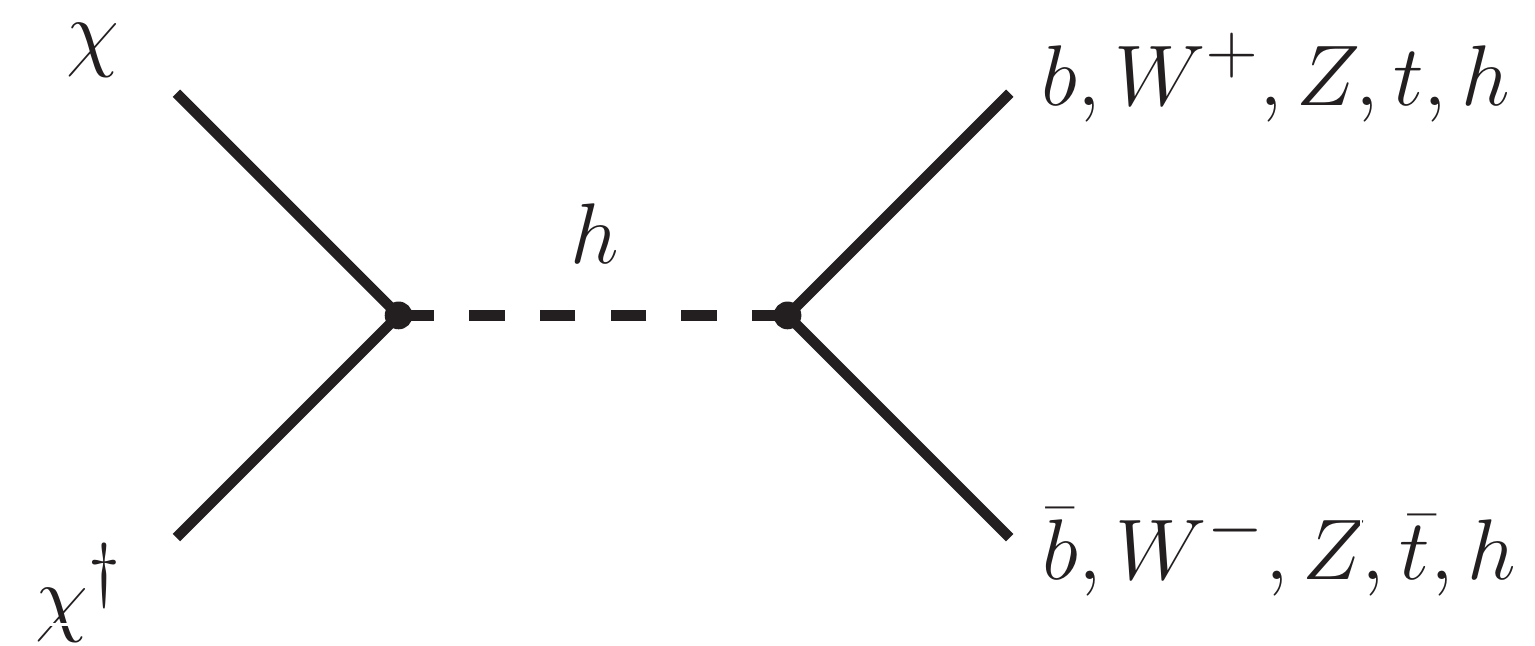}~~~
\includegraphics[height=2.15cm]{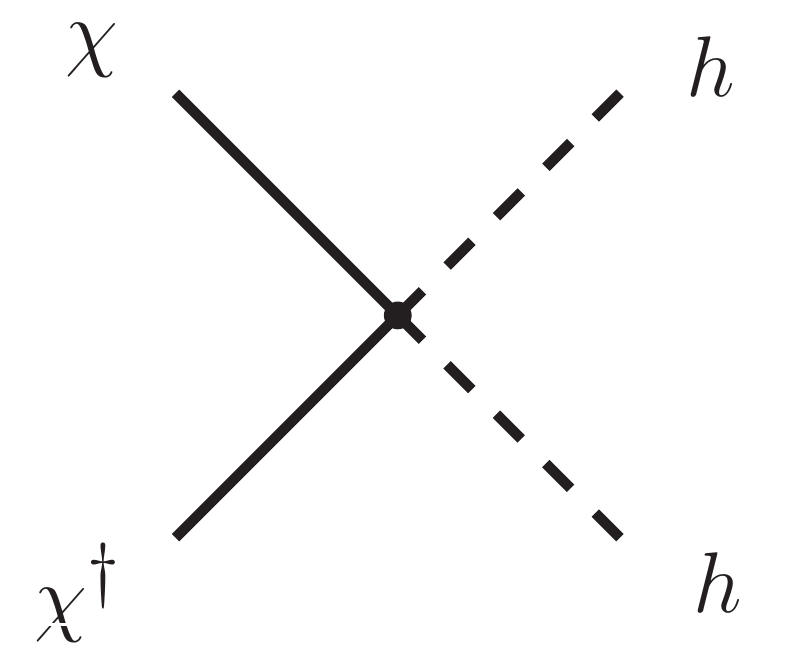}~~~
\includegraphics[height=3cm]{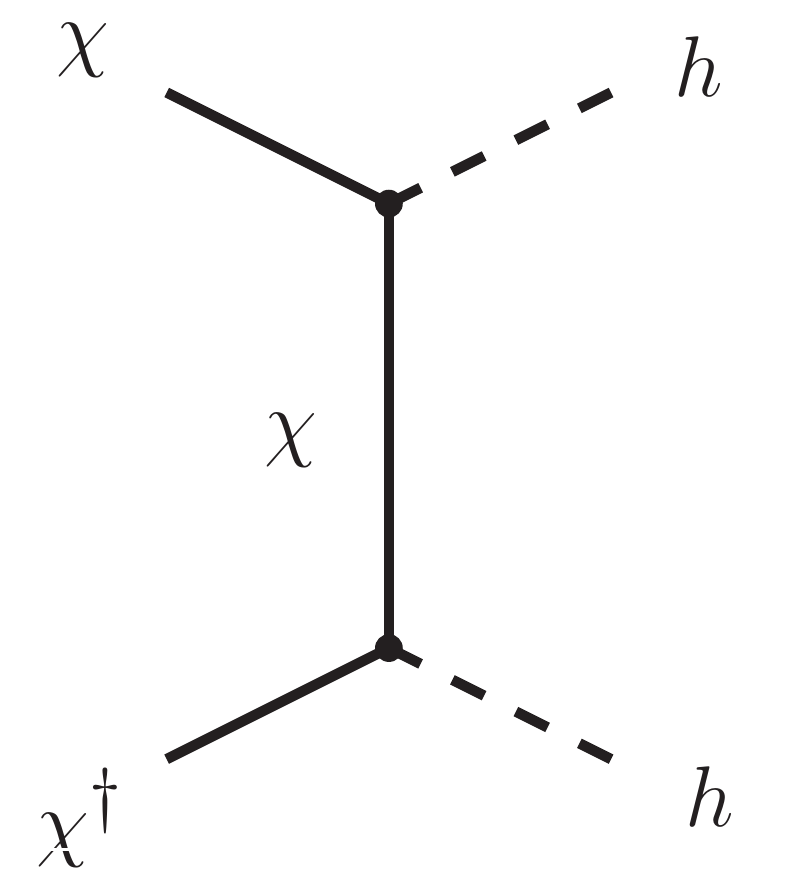}~~~
\includegraphics[height=3cm]{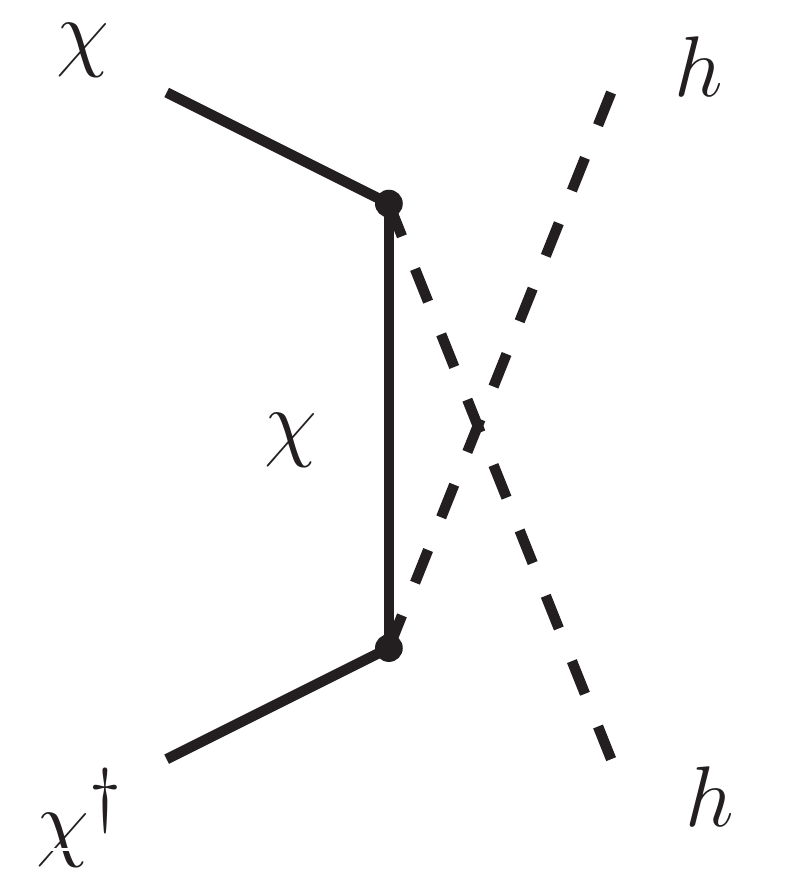}
\end{center}
\caption{DM annihilation channels in the Higgs-portal models. \label{fig:DMdecay} }
\end{figure}

The dominant DM annihilation cross sections in the Higgs-portal models
are $\chi \bar \chi\to b\bar b, W^+W^-, ZZ, t\bar t$, and $\chi \bar
\chi \to h h$. The first four proceed through the $s$-channel Higgs
exchange, while $\chi \bar \chi \to h h$ receives additional
contributions from $t$- and $u$-channel $\chi$ exchange as well as
from the four-point contact interaction
(cf. Fig.~\ref{fig:DMdecay}). The $\chi \bar \chi\to b\bar b$ channel
is only relevant if the other channels are not kinematically allowed,
i.e. for light DM masses, $m_\chi < m_W$.

The $\chi \bar \chi \to \bar b b$ annihilation cross section assuming
SM Higgs couplings is given for scalar ($S$), Dirac fermion ($DF$),
and vector ($V$) DM by
\begin{align}
	\left(\sigma_{b\bar b}^Sv_{\rm rel}\right)_{SM}&=
        \frac{N_c}{4\pi }
        \frac{g_{\chi}^2 m_b^2\beta^3_{b}}{\left(s-m_h^2\right)^2+m_h^2\Gamma_h^2}\,,
        \label{eq:sigmaSbb}
        \\ 
	\left(\sigma_{b\bar b}^{DF} v_{\rm rel}\right)_{SM} &=
        \frac{N_c}{8\pi}\,\frac{m_b^2}{\Lambda^2}\,\frac{g_\chi^2
          \left(s-4m_\chi^2\right) + \tilde g_\chi^2 \, s}{\left(s-m_h^2\right)^2+m_h^2\Gamma_h^2}\beta^3_{b}\,,
         \label{eq:sigmaFbb}
        \\ 
	\left(\sigma_{b\bar b}^V v_{\rm rel}\right)_{SM}  &=
        \frac{N_c}{9}\frac{g_\chi^2}{16\pi}\frac{m_b^2}{m_\chi^4}\beta_{b}^3
        \left(1-r_\chi+\frac34 r_\chi^2\right)\frac{s^2}{\left(s-m_h^2\right)^2+m_h^2\Gamma_h^2}\,,
         \label{eq:sigmaVbb}
\end{align}
where, here and below, $\sqrt s$ is the center-of-mass energy, $r_k =
4m_k^2/s$, $\beta_k = \sqrt{1-r_k}$ is the velocity of particle $k$,
and $v_\text{rel} = 2 \beta_\chi$ is the relative velocity of the DM
particles. If the Higgs coupling to the $b$-quarks differs from the SM
value, the annihilation cross section is rescaled as
\begin{equation}
\sigma_{b\bar b}= \big( \kappa_b^2 + \tilde \kappa_b^2 / \beta_b^2
\big) \sigma_{b\bar b}^\text{SM} \,.
\end{equation} 
The annihilation cross sections $\sigma_{f\bar f}$ to the other
fermions are obtained with the obvious replacement $b\to f$ in the
above expressions. Since the Higgs couplings to the light quarks are
poorly constrained experimentally, the DM annihilation to two light
quarks can be comparable to $\chi \bar \chi\to b\bar b$ and can be
important for light DM, $m_\chi<m_W$.

For heavy DM, $m_\chi > m_W$, the annihilation cross-sections
into a pair of $W$ or $Z$ bosons are
\begin{equation}
\sigma_{VV} =\kappa_V^2 \sigma_{VV}^\text{SM}\,,
\end{equation}
$V = W,Z$. The annihilation cross sections assuming the SM Higgs
couplings to $W$ are given by
\begin{align}
	(\sigma_{WW}^S v_{\text{rel}})_{\rm SM} &= \frac{g_{\chi}^2 }{8\pi
        }\beta_{W} \left(1-r_W+\frac34 r_W^2\right)
        \frac{s}{\left(s-m_h^2\right)^2+m_h^2\Gamma_h^2}\,,\\ 
	(\sigma_{WW}^{DF} v_{\text{rel}})_{\rm SM} &=
        \frac{1}{16\pi\Lambda^2}\beta_{W} \left(1-r_W+\frac34 r_W^2\right)
        \frac{s\left[g_{\chi}^2\left(s-4m_\chi^2\right) + \tilde
            g_{\chi}^2 \, s\right]}{\left(s-m_h^2\right)^2+m_h^2\Gamma_h^2} \,,\\
	(\sigma_{WW}^{V} v_{\text{rel}})_{\rm SM} &=
        \frac{g_{\chi}^2}{288\pi}\frac{s}{m_\chi^4}\beta_{W}
        \left(1-r_W+\frac34 r_W^2\right)\left(1-r_\chi+\frac34r_\chi^2\right)
	\frac{s^2}{\left(s-m_h^2\right)^2+m_h^2\Gamma_h^2}\,,
\end{align}
for scalar, Dirac fermion, and vector DM, respectively. The $\chi \bar
\chi \to ZZ$ annihilation cross section is obtained by replacing $W\to
Z$, and multiplying all expressions by an extra factor of 1/2 since
one has two indistinguishable particles in the final state.

The $\chi \bar \chi \to hh$ annihilation cross sections for scalar,
Dirac fermion, and vector DM, are given by
\begin{align}
	\sigma_{hh}^S v_{\text{rel}} &= \frac{\beta_{h} g_\chi^2}{64\pi m_\chi^2}
        \bigg[ 1 + \frac{3 \kappa_\lambda M_h^2}{4m_\chi^2-M_h^2} -
            \frac{2v_W^2g_\chi}{M_h^2-2m_\chi^2} \bigg]^2  \,,\\[3mm] 
	\sigma_{hh}^{DF} v_{\text{rel}} &= \frac{\beta_{h} \big(\tilde
          g_\chi^2 + g_\chi^2 \beta_\chi^2 \big) }{32\pi \Lambda^2} 
        \bigg[ 1 + \frac{3 \kappa_\lambda M_h^2}{4m_\chi^2-M_h^2} 
         + \frac{4 g_\chi m_\chi v_W^2}{\Lambda (2m_\chi^2 - M_h^2) }
         \bigg]^2 \,,\\[3mm]  
	\sigma_{hh}^{V} v_{\text{rel}} &= \frac{\beta_{h}}{576\pi m_\chi^2}
         \bigg[ 3 g_\chi^2\left( \frac{3 \kappa_\lambda
            M_h^2}{4m_\chi^2-M_h^2} + 1\right)^2 + \frac{4g_\chi^4
             v_W^4}{(2m_\chi^2-M_h^2)^2} \left( 
           6 - \frac{4M_h^2}{m_\chi^2} + \frac{M_h^4}{m_\chi^4}
           \right) \notag \\[3mm] 
           & \hspace{2cm} + \frac{16g_\chi^3 v_W^2 
             }{2m_\chi^2-M_h^2} \left( \frac{3 \kappa_\lambda
            M_h^2}{4m_\chi^2-M_h^2} + 1\right) \left( 1 -
           \frac{M_h^2}{4 m_\chi^2} \right) \bigg] \,.
\end{align}
In this result we display only the leading terms in the expansion in
powers of $\beta_\chi$. The contribution of the $s$-channel Higgs
exchange diagram is proportional to the rescaling of the trilinear
Higgs coupling, $\kappa_\lambda$. The latter is completely unknown
experimentally, at present, but can be measured to ${\mathcal
  O}(20\%)$ at the end of the LHC~\cite{Goertz:2013kp,
  deLima:2014dta}. Since we are mostly interested in the effects of
flavor modifications we will set it to the SM value,
$\kappa_\lambda=1$, in the numerics below. All cross sections for
Majorana DM can be obtained by multiplying the corresponding Dirac DM
cross sections by a factor of 4.

\begin{figure}[t]
\begin{center}
\includegraphics[width=0.5\textwidth]{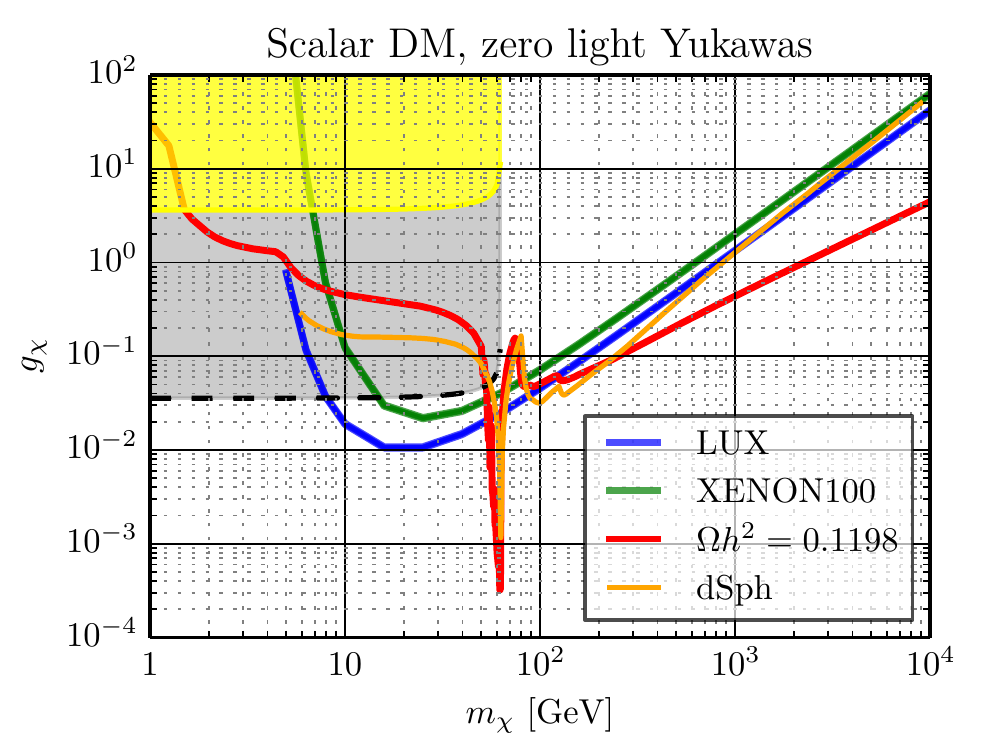}~
\includegraphics[width=0.5\textwidth]{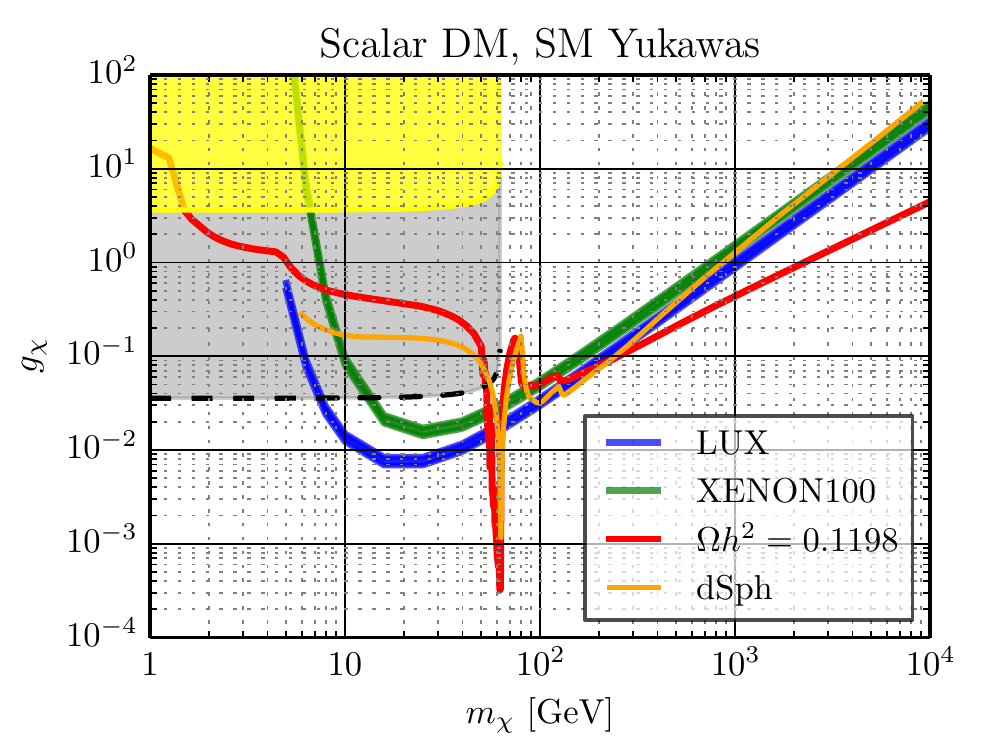}
\end{center}
\caption{Bounds from LUX (blue band), XENON100 (green band) and the
  invisible Higgs decay width (black dashed line and grey region) on
  the Higgs-portal coupling $g_\chi$ for scalar DM, assuming vanishing
  (left) and SM (right) Yukawa couplings to $u$, $d$, $s$ quarks. The
  red line denotes $g_\chi$ as a function of DM mass, $m_\chi$, for
  which the correct relic abundance is obtained, while $g_\chi$ in the
  yellow region leads to non-perturbatively large Higgs decay width,
  $\Gamma_h> m_h$, and is excluded. Constraints from Fermi-LAT
  searches for DM annihilation in dwarf spheroidal galaxies are
  denoted by the orange band. \label{fig:gchiscalar} }
\end{figure}

\begin{figure}[t]
\begin{center}
\includegraphics[width=\textwidth]{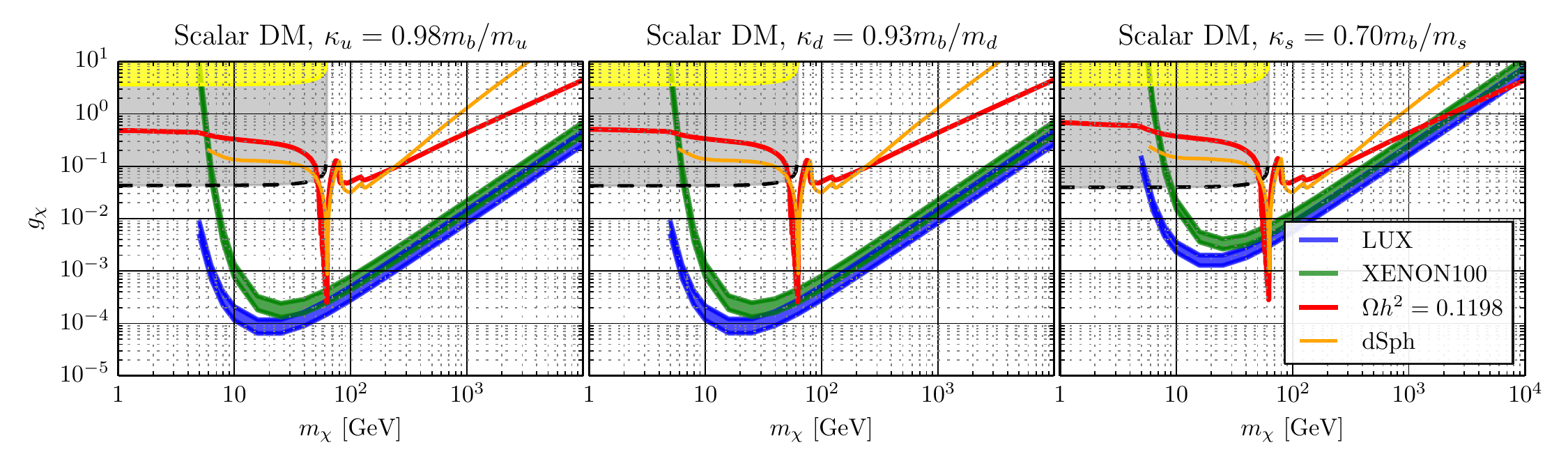}
\end{center}
\caption{Bounds on the Higgs-portal coupling $g_\chi$ for scalar DM,
  assuming maximal allowed values for the Yukawa couplings to the $u$,
  $d$, $s$ quarks (left to right), keeping all the other couplings to
  their SM values. The color coding is the same as in
  Fig. \ref{fig:gchiscalar}. \label{fig:gchiscalary}}
\end{figure}

\subsection{Relic abundance}
\begin{figure}[t]
\begin{center}
\includegraphics[width=0.5\textwidth]{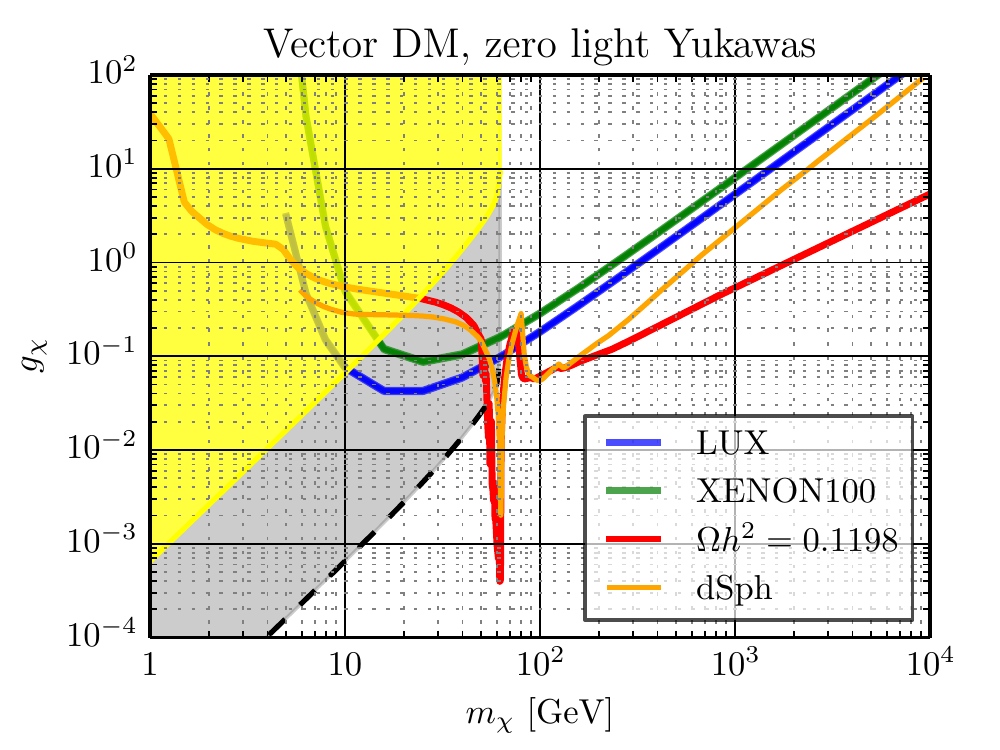}~
\includegraphics[width=0.5\textwidth]{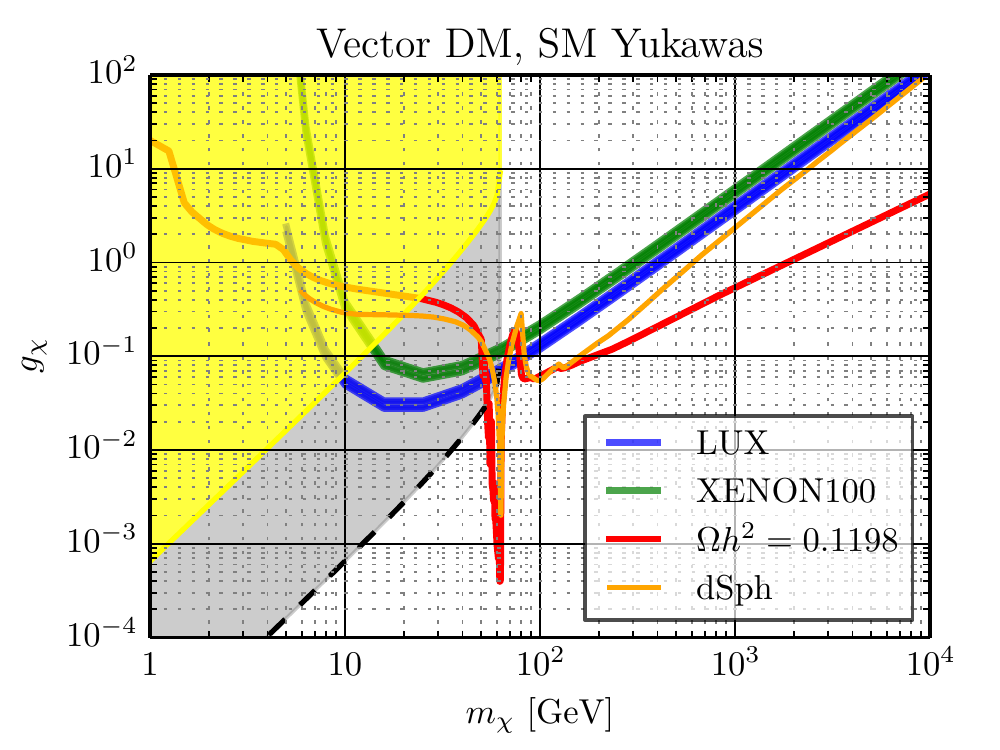}
\end{center}
\caption{Bounds on the Higgs-portal coupling for vector DM, assuming
  vanishing (left) and SM (right) Yukawa couplings to $u$, $d$, $s$
  quarks. The color coding is the same as in
  Fig. \ref{fig:gchiscalar}.\label{fig:gchivector}}
\end{figure}

\begin{figure}[t]
\begin{center}
\includegraphics[width=\textwidth]{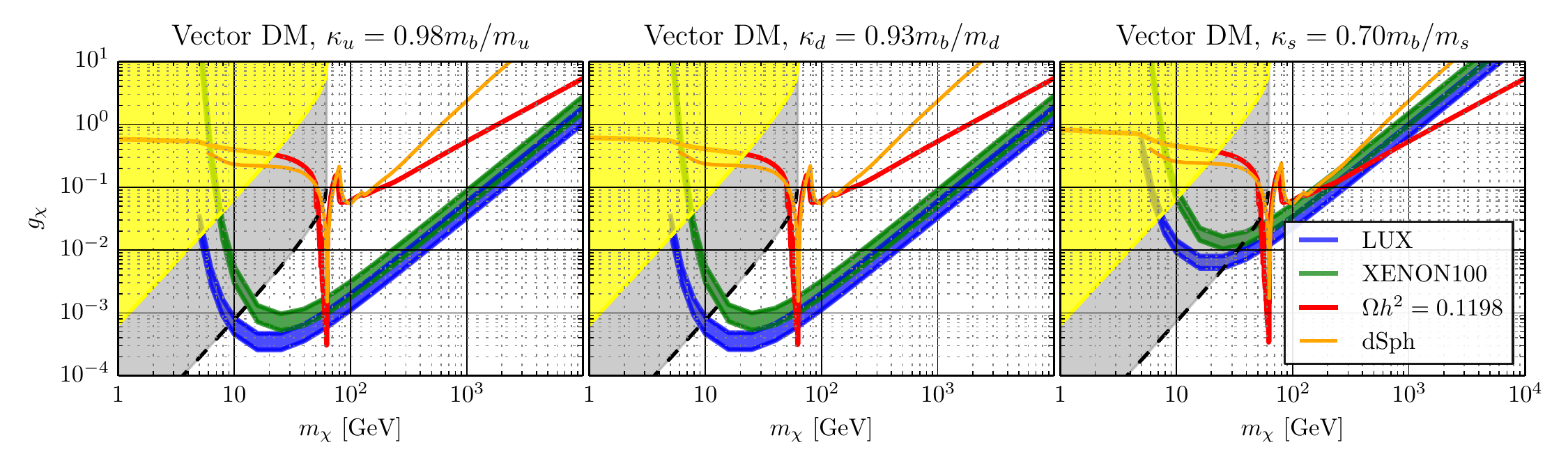}
\end{center}
\caption{Bounds on the Higgs-portal coupling for vector DM, assuming
  maximal allowed values for the Yukawa couplings to the $u$, $d$, $s$
  quarks (left to right), keeping all the other couplings to their SM
  values. The color coding is the same as in
  Fig. \ref{fig:gchiscalar}. \label{fig:gchivectory}}
\end{figure}

\begin{figure}[t]
\begin{center}
\includegraphics[width=0.5\textwidth]{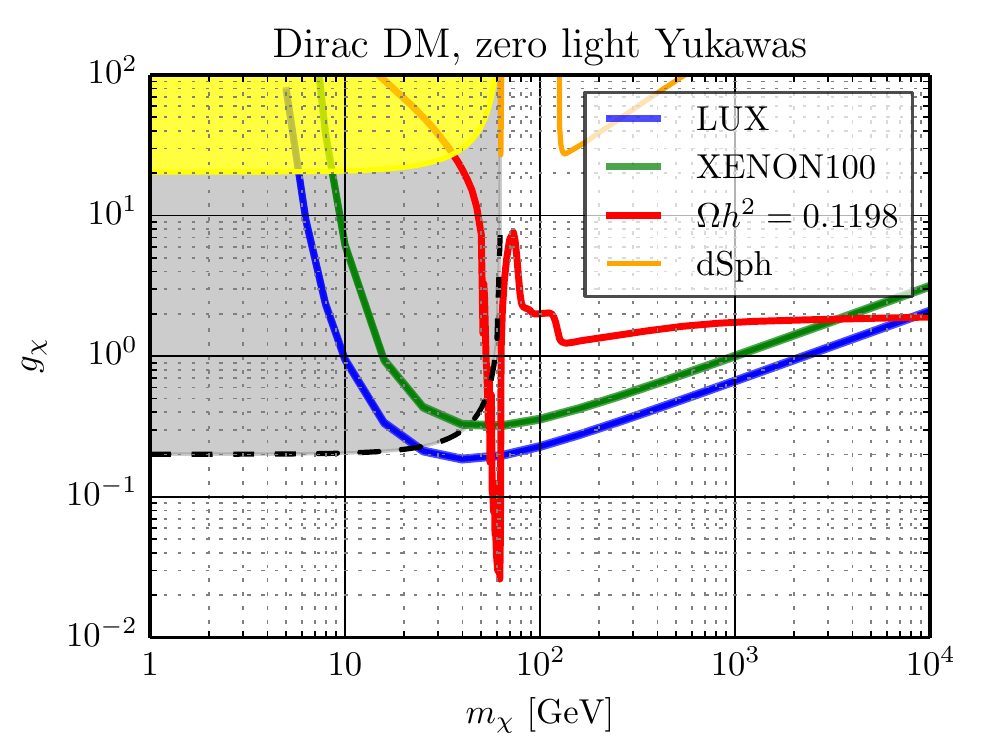}~
\includegraphics[width=0.5\textwidth]{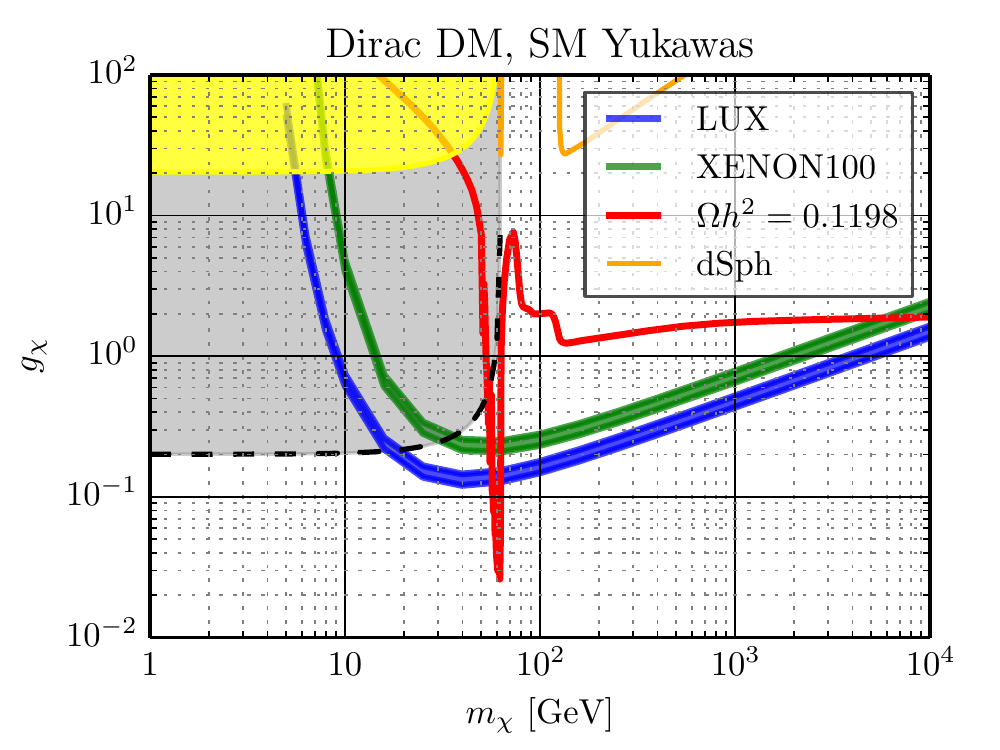}
\end{center}
\caption{Bounds on the Higgs-portal coupling for Dirac DM, assuming
  $\Lambda = 1$ TeV and vanishing (left) and SM (right) Yukawa
  couplings to $u$, $d$, $s$ quarks. The color coding is the same as
  in Fig. \ref{fig:gchiscalar}. \label{fig:gchidirac}}
\end{figure}

\begin{figure}[t]
\begin{center}
\includegraphics[width=\textwidth]{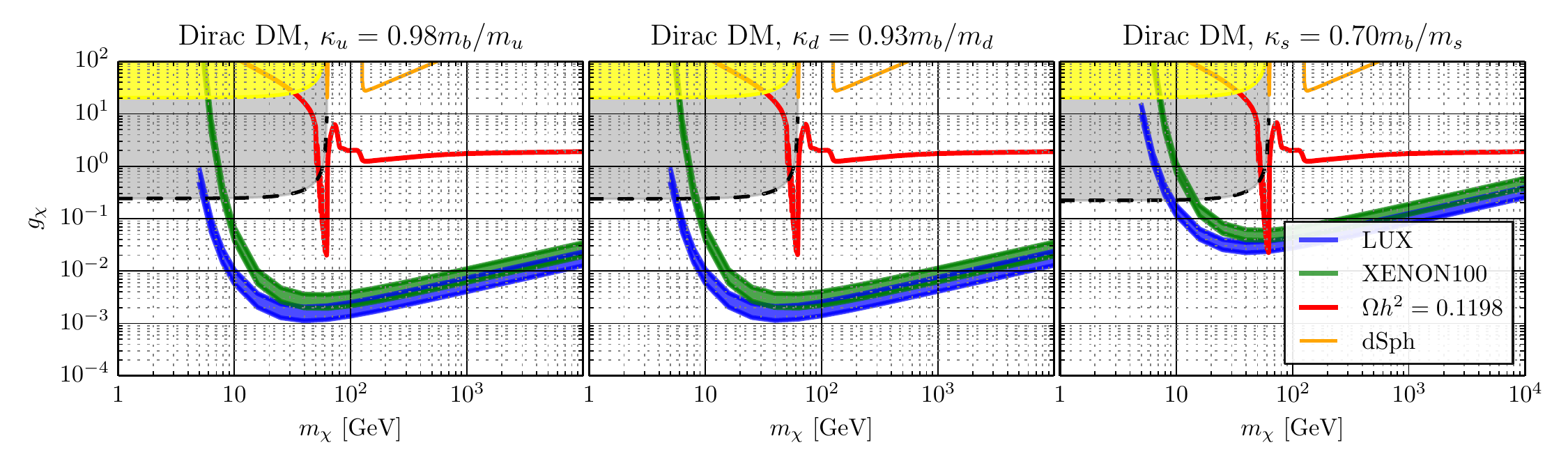}
\end{center}
\caption{Bounds on the Higgs-portal coupling for Dirac DM, assuming
  $\Lambda = 1$ TeV and maximal allowed values for the Yukawa
  couplings to the $u$, $d$, $s$ quarks (left to right), keeping all
  the other couplings to their SM values. The color coding is the same
  as in Fig. \ref{fig:gchiscalar}. \label{fig:gchidiracy} }
\end{figure}

\begin{figure}[t]
\begin{center}
\includegraphics[width=0.5\textwidth]{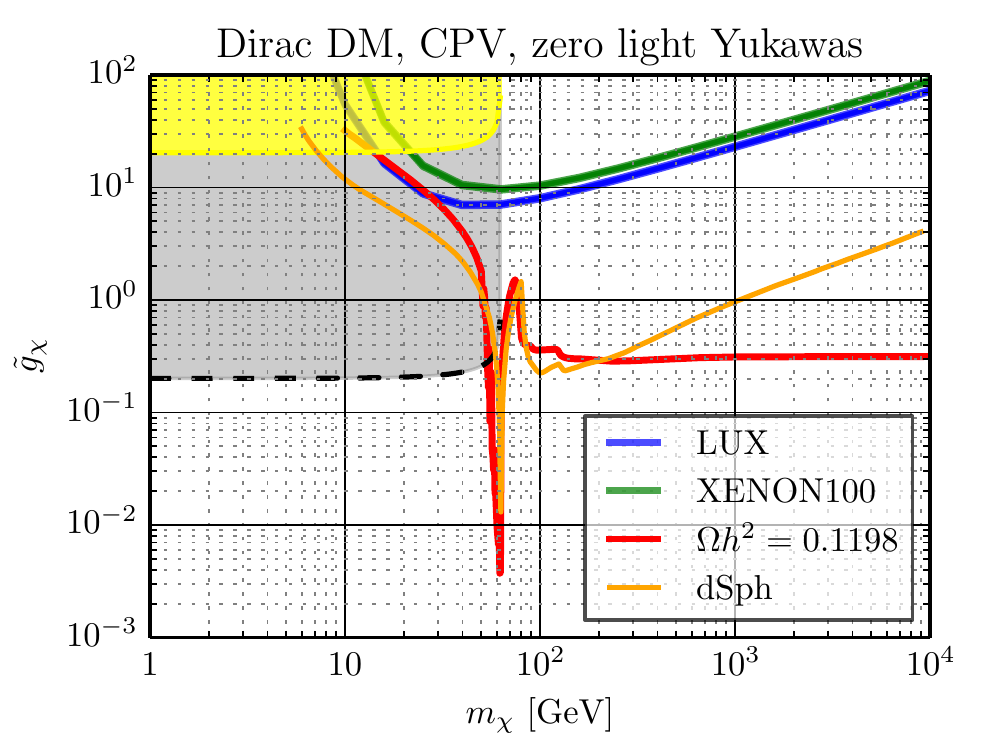}~
\includegraphics[width=0.5\textwidth]{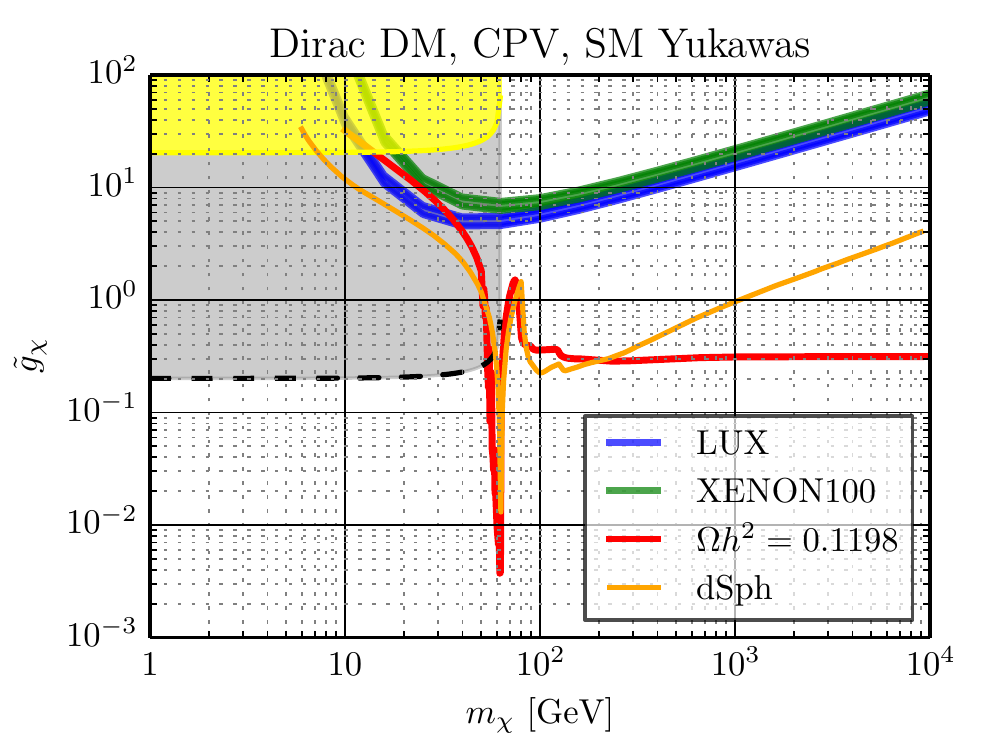}
\end{center}
\caption{Bound on the pseudoscalar Higgs-portal coupling for Dirac DM,
  assuming $\Lambda = 1$ TeV and vanishing (left) and SM (right)
  Yukawa couplings to $u$, $d$, $s$ quarks. The color coding is the
  same as in Fig. \ref{fig:gchiscalar}.  \label{fig:gchidiracp}}
\end{figure}

\begin{figure}[t]
\begin{center}
\includegraphics[width=\textwidth]{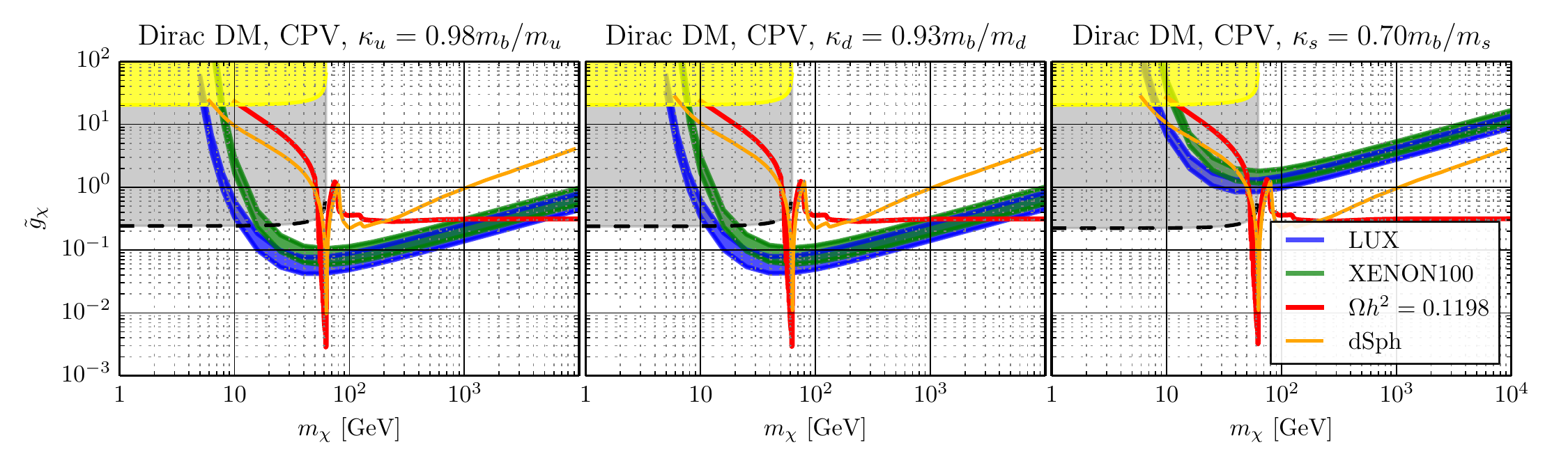}
\end{center}
\caption{Bounds on the pseudoscalar Higgs-portal coupling for Dirac
  DM, assuming $\Lambda = 1$ TeV and maximal allowed values for the
  Yukawa couplings to the $u$, $d$, $s$ quarks (left to right),
  keeping all the other couplings to their SM values. The color coding
  is the same as in
  Fig. \ref{fig:gchiscalar}. \label{fig:gchidiracpy}}
\end{figure}

The DM relic abundance $\Omega_{DM}$ is proportional to $1/{\sigma
  v_{\rm rel}}$, where $\sigma$ is the annihilation cross
section. Assuming that the DM in our scenario accounts for all of the
observed relic density, the measured value $\Omega_{DM} h^2 =
0.1198(26)$~\cite{Agashe:2014kda} fixes $g_\chi$ for a given value of
$m_\chi$. The resulting constraint in the $m_\chi$ -- $g_\chi$ plane
is denoted for the different cases by a red line in
Figs.~\ref{fig:gchiscalar} to~\ref{fig:gchidiracpy}. In
Fig.~\ref{fig:gchiscalar}, we compare two limits of the Higgs portal
for the scalar DM: the case where the Higgs does not couple to the
light quarks at all (left panel) to the case where the Higgs has SM
Yukawa couplings (right panel). The two relic abundance curves
coincide apart from very light DM, with $m_\chi$ below the charm and
tau threshold. If such light DM did not couple to the $u,d$ and $s$
quarks, this would result in noticeably reduced annihilation cross
sections and, thus, in larger relic abundance. In both cases, the
dominant annihilation process is still given by $\chi \bar \chi\to h^*
\to g g$.  For very light DM the correct relic abundance is obtained
only if the coupling of the Higgs to DM, $g_\chi$, is
nonperturbatively large. The yellow regions in
Fig.~\ref{fig:gchiscalar} denote the value of $g_\chi$ for which the
total Higgs decay width would be larger than its mass, $\Gamma_h>m_h$,
and are thus excluded.

The same comments apply to the case of vector DM, shown in
Fig.~\ref{fig:gchivector}, $CP$-conserving Dirac fermion DM, shown in
Fig.~\ref{fig:gchidirac}, $CP$-violating Dirac fermion DM, shown in
Fig.~\ref{fig:gchidiracp}, and also for Majorana fermion DM. For light
DM, $m_\chi\lesssim 30$ GeV, the correct relic density requires a
non-perturbatively large coupling $g_\chi$ so that the predictions
should be taken only as ${\mathcal O}(1)$ estimates in that region.
Note that all these non-perturbative regions are, in addition,
excluded by bounds on the decay width of the Higgs into invisible
final states (see the discussion in Section~\ref{sec:invisible}).

In
Figs.~\ref{fig:gchiscalary},~\ref{fig:gchivectory},~\ref{fig:gchidiracy},
and~\ref{fig:gchidiracpy}, we show the relic abundance curves for
$g_\chi$ as a function of $m_\chi$ for the case where the light Yukawa
couplings saturate their upper experimental bound. The left panels
show the case where $\kappa_u=0.98 m_b/m_u$ and all the other
couplings at their SM values, the middle panels the case where
$\kappa_d=0.93 m_b/m_d$, and the right panels the case where
$\kappa_s=0.70 m_b/m_s$. In all of these cases the cross section for
DM annihilation to light jets, $\sigma(\chi\bar\chi \to jj)$, coming
from DM annihilating to light quarks, is comparable to the
annihilation cross section to $b$-jets, $\sigma(\chi \bar\chi \to b
\bar b)$. For $m_\chi\lesssim m_W$ these are the two dominant
annihilation modes. Since the annihilation cross sections to $b$-jets
and light jets are comparable, the relic abundance curve show only a
small change in $g_\chi$ when the $b$-quark threshold is reached. This
should be compared with the case of the SM Yukawa couplings shown in
the right panels of
Figs.~\ref{fig:gchiscalar},~\ref{fig:gchivector},~\ref{fig:gchidirac},
and~\ref{fig:gchidiracp}. In this case the $\chi\bar\chi \to jj$
annihilation is almost exclusively due to DM annihilating to two
gluons, so that $\sigma(\chi\bar\chi\to jj)\ll \sigma(\chi\bar\chi \to
b\bar b)$, while the annihilation into two light quarks is
negligible. For $m_\chi\lesssim m_W$ and SM Yukawas, the required
$g_\chi$ is thus bigger by $30\%-40\%$ than in the case of light
Yukawas at their present experimental limits, and exhibits a
significant jump below the $b$-quark threshold.

\subsection{Invisible decay width of the Higgs}\label{sec:invisible}

The bounds on the invisible decay width of the Higgs boson provide
stringent constraints on Higgs-portal DM~\cite{Djouadi:2012zc}. The
partial $h\to \chi \bar \chi$ decay widths are given by
\begin{equation}
\begin{split}
  \Gamma_{\chi\chi}^S & = \frac{g_\chi^2}{16\pi} \frac{v_W^2}{M_h}
  \beta_{\chi} \,,\\
  \Gamma_{\chi\chi}^{DF} & = \frac{g_\chi^2}{8\pi} M_h
  \frac{v_W^2}{\Lambda^2} \beta_{\chi}^{3} + \frac{\tilde
    g_\chi^2}{8\pi} M_h \frac{v_W^2}{\Lambda^2} \beta_{\chi}^{1/2}\,,\\
  \Gamma_{\chi\chi}^V & = \frac{g_\chi^2}{128\pi} 
  \frac{M_h^3 v_W^2}{m_\chi^4} \beta_{\chi} \left( 1 - r_\chi +
  \tfrac{3}{4} r_\chi^2 \right)\,,
\end{split}
\end{equation}
where $r_\chi = 4m_\chi^2/M_h^2$ and $\beta_{\chi} = \sqrt{1 - r_\chi}$.

The current best limits on the invisible branching fraction of the SM
Higgs are obtained from $Zh$ production. The CMS collaboration gives a
95\% CL limit of Br$(h\to {\rm inv})<0.58$ for $M_h = 125$
GeV~\cite{Chatrchyan:2014tja} and ATLAS finds Br$(h\to {\rm
  inv})<0.75$ for $M_h = 125.5$ GeV~\cite{Aad:2014iia}. Note that the
increased light-quark Yukawa couplings, at their presently allowed
values, do not appreciably change the Higgs production cross section
\cite{Kagan:2014ila}. Their main effect is to increase the total decay
width of the Higgs and thus reduce the branching ratios to the other
decay modes:
\begin{equation}
\begin{split}
  \text{Br}(h\to \chi \bar \chi) = \frac{\Gamma(h\to \chi \bar
    \chi)}{\Gamma(h\to \chi \bar \chi) + \Gamma_h^\text{tot} \times
    \big[ 1 + \sum_q (\kappa_q^2-1) \text{Br}_\text{SM} (h \to q \bar
      q) \big]}\,.
\end{split}
\end{equation}
In Figs. \ref{fig:gchiscalar} to \ref{fig:gchidiracpy} we denote the
bound on $g_\chi$ corresponding to the ATLAS upper limit on Br$(h\to
{\rm inv})$ with a dashed black line and grey out the excluded region
in the $g_\chi$ vs. $m_\chi$ plane. We see that the light DM Higgs
portal, $m_\chi\lesssim m_h/2$ is excluded by the Higgs invisible
decay width.

Vector boson fusion, gluon fusion and $t\bar t H$ production, with the
off-shell Higgs going to two DM particles, can provide some limited
sensitivity to DM masses above $m_h/2$. A combination of the searches
in the three channels at a 100 TeV collider could exclude the scalar
thermal relic DM Higgs portal for DM masses in parts of the
$m_h/2\lesssim m_\chi\lesssim m_W$ interval at 95\%
C.L.~\cite{Craig:2014lda} (these result receives only a negligible
correction if light quark Yukawa couplings are enhanced). For
$m_\chi<m_h/2$ the invisible Higgs decay width is, however, always the
most constraining~\cite{Endo:2014cca}.

\subsection{Indirect detection}

\begin{figure}[t!]
\centering
\includegraphics[scale=0.7]{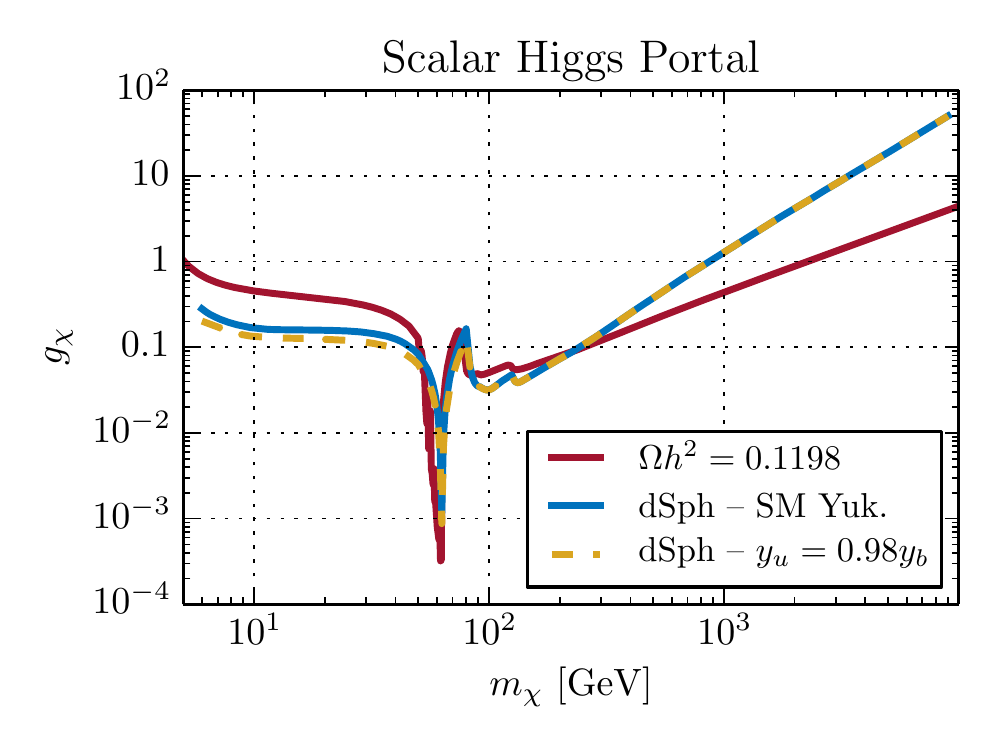}
\caption{The effect of large light-quark Yukawa couplings on the
  indirect detection bounds from Fermi-LAT observations of Milky Way
  dwarf spheroidal satellite galaxies~\cite{Ackermann:2015zua}.}
\label{fig:ID-yuk-comp}
\end{figure}

In indirect signals of DM annihilation, the effect of changing the
light-quark Yukawa couplings within the presently experimentally
allowed ranges leads to at most $\mathcal{O}(1)$ effects. Further, the
effect is present only for DM masses below the $W$ threshold where the
dominant annihilation channel is into the $b\bar b$ final state.  For
example, Fig.~\ref{fig:ID-yuk-comp} shows the recast of the Fermi-LAT
bound from dwarf spheroidals~\cite{Ackermann:2015zua} for scalar DM,
following the procedure outlined below.

Photon flux measurements with $\gamma$-ray telescopes can put a strong
bound on the annihilation cross-section of the DM.  The strongest
bound in the DM mass range of interest has been recently released by
the Fermi-LAT collaboration~\cite{Ackermann:2015zua} based on
\verb|Pass 8| observation data of the Milky Way dwarf spheroidal
satellite galaxies (dSphs). There is also a recent analysis based on
the Dark Energy Survey (DES) dSph candidates using the Fermi-LAT
data~\cite{Drlica-Wagner:2015xua}. While this bound is competitive
with the one from the known dSphs, it is still weaker on its own. One
could also consider the bounds from the isotropic gamma ray background
(IGRB)~\cite{Ackermann:2015tah}.  In our analysis, we recast the
Fermi-LAT bound on the $b\bar b$ final state using a simple
re-weighting procedure of the photon spectra which will be discussed
below.

The observed differential photon flux from the annihilation of dark
matter is given by
\begin{equation}
\frac{d\Phi}{dE_\gamma d\Omega} = \frac{1}{4\pi}
  \frac{1}{2m_\chi^2}\,J\,
  \left[ \sum_f \langle \sigma v\rangle_f\,\frac{dN_\gamma^f}{dE} \right],
\label{eq:diff-flux}
\end{equation}
where $J$ is an astrophysical factor which depends on the distance to
the source and the dark matter density profile. The factor in the
brackets is the one most interesting for our purposes. It depends on
the velocity-averaged cross-section and photon spectrum per DM
annihilation.

The Fermi-LAT analysis gives bounds for the different final states
separately while we have an admixture of final states.  In order to
recast the bound, we rely on the observation that for heavy DM the
photon spectra from DM annihilation into quarks, gauge bosons, and the
Higgs boson all peak at approximately the same photon energy and have
approximately the same shape.  Therefore, to extract the bound on the
DM Higgs portal coupling $g_\chi$, it is sufficient to find the zeros
of the polynomial
\begin{equation}
f(g_\chi) = 
  \sum_f \langle \sigma v\rangle_f\,\left(\frac{dN_\gamma^f}{dE}\right)^\text{\sc peak}
  - \langle \sigma v\rangle_b^\text{\sc fermi}\,\left(\frac{dN_\gamma^b}{dE}\right)^\text{\sc peak},
\label{eq:indirect-bound}
\end{equation}
where in the last term, $\langle \sigma v\rangle_b^\text{\sc fermi}$
is the bound from the Fermi-LAT analysis on the velocity-averaged
cross-section. The photon spectra were obtained from the interpolation
tables provided in the \verb|PPPC4DMID| package~\cite{cirelli:2010xx}.
In all cases except for the $hh$ final state, $f(g_\chi)$ has only one
zero up to a sign ambiguity. For the $hh$ final state, however, the
zero of $f(g_\chi)$ closest to the $g_\chi$ corresponding to $\chi\bar
\chi \to b\bar b$ is the one used to rescale the Fermi-LAT bound on
$g_\chi$ as a function of $m_\chi$. The resulting bounds are shown in
Figs. \ref{fig:gchiscalar} to \ref{fig:gchidiracpy}.

\subsection{Direct detection}

We have seen so far that most DM observables exhibit only a weak
dependence on the light-quark Yukawas. This is not the case for the
direct DM detection. In fact, modifying the light-quark Yukawa
couplings can significantly change the predictions for DM -- nucleus
scattering cross sections.

The differential cross section for spin-independent DM scattering on a
nucleus is given by
\begin{equation}
\frac{d\sigma}{dE_R} = \frac{m_A}{\mu_{\chi A}^2 v_\text{rel}^2}
\frac{\overline{| {\mathcal M} |^2}}{32 \pi s} \,,
\end{equation}
where $E_R$ is the nuclear recoil energy, $m_A$ is the mass of the
nucleus, $\mu_{\chi A} \equiv m_\chi m_A / (m_\chi + m_A)$ is the
reduced mass of the DM -- nucleus system, $s = (m_\chi + m_A)^2$ is
the center-of-mass energy, $v_\text{rel}$ is the DM velocity in the
detector rest frame, and $\overline{| {\mathcal M} |^2}$ is the
spin-averaged squared matrix element.

The matrix element ${\mathcal M}$ depends on the effective Higgs
couplings to the nucleus. Since the momentum exchanges in DM
scattering on nuclei are much smaller than the Higgs mass, we can
calculate $\overline{| {\mathcal M} |^2}$ by first integrating out the
Higgs and the heavy quarks ($t$, $b$, $c$). This gives an EFT with
light quarks and gluons interacting with DM through local operators,
described by the effective Lagrangians
\begin{align}\label{eq:eff-lagr-scal}
  {\cal L}_S&=\frac{g_\chi v_W}{m_h^2}(\chi^{\dagger} \chi){\cal S}_q \,,
\\
\label{eq:eff-lagr-dir}
  {\cal L}_{F} &= \frac{1}{\Lambda} \frac{g_\chi v_W}{m_h^2} (\bar \chi
  \chi){\cal S}_q + \frac{1}{\Lambda} \frac{\tilde g_\chi v_W}{m_h^2} (\bar \chi
  i \gamma_5 \chi){\cal S}_q\,.
\\  \label{eq:eff-lagr-vec}
  {\cal L}_V&=\frac{g_\chi v_W}{2m_h^2}(\chi_\mu \chi^\mu){\cal S}_q \,,
\end{align}
for scalar, fermion, and vector DM, respectively. The scalar current
is the same in all three cases:
\begin{equation}\label{eq:Sq}
{\cal S}_q = \sum_q \kappa_q\frac{m_q}{v_W} \,\bar q q -{\cal C}_g
\frac{\alpha_s}{12\pi v_W} G_{\mu\nu}^a G^{a\mu\nu}+\sum_qi\tilde
\kappa_q \frac{m_q}{v_W}\bar q\gamma_5 q - \tilde {\cal C}_g
\frac{\alpha_s}{8\pi v_W} G_{\mu\nu}^a \widetilde G^{a\mu\nu}\,.
\end{equation}
Here, the last two terms arise from $CP$-violating Higgs couplings. The
sums are over the light quarks $q=u,d,s$. The heavy quarks are
integrated out and contribute only via the gluonic terms in the
current. For the two corresponding dimensionless Wilson coefficients
we have
\begin{equation}
{\cal C}_g=\kappa_g^{\rm NP}+\kappa_t+\kappa_b+\kappa_c\,, \qquad \tilde
{\cal C}_g=\tilde \kappa_g^{\rm NP}+\tilde \kappa_t+\tilde
\kappa_b+\tilde \kappa_c\,, 
\end{equation}
where the first contribution is from tree-level matching, and the
remaining from one-loop matching, working in the limit of heavy
quarks. This is well justified for top and bottom quarks. For
scattering on heavy nuclei, e.g., on Xe or W, the maximal momentum
exchanges for DM with mass above approximately 1 TeV may, however,
start to become comparable to the charm-quark mass. We neglect these
effects, while they may need to be included in the future if such
heavy DM is discovered.

$CP$-violating Higgs couplings to light quarks lead to spin-dependent
interactions of DM with the target nuclei. The corresponding
scattering rates are suppressed relative to the spin-independent
interaction rates from $CP$-conserving Higgs couplings. We will
therefore neglect the $CP$-violating interactions in our numerical
analysis of direct detection scattering rates; i.e., we will set
$\tilde \kappa_q=0$, $\tilde{\cal C}_q=0$ from now on.

The nucleon matrix elements of the remaining terms in the scalar
current ${\cal S}_q$ are conventionally parametrized by (see,
e.g.,~\cite{Jungman:1995df}),
\begin{align}
  \langle N | m_q \bar q q | N \rangle &= m_N f_{Tq}^{(N)} \,,\\
    \langle N |
  \frac{\alpha_s}{12\pi} G_{\mu\nu} G^{\mu\nu} | N \rangle &=-
  \frac{2}{27} m_N f_{TG}^{(N)} \,.
\end{align}
In the heavy-quark limit for $t,b,c$ the trace anomaly equation leads
to the relation~\cite{Shifman:1978zn, Jungman:1995df}
\begin{equation}
f_{TG}^{(N)} = 1 - \sum_{q=u,d,s} f_{Tq}^{(N)}.
\end{equation}
We can also define the effective Higgs coupling to nucleon as the
expectation value of the scalar current,
\begin{equation}
f_{\cal S}^{(N)}\equiv \langle N| {\cal S}_q|N\rangle =
\frac{m_N}{v_W} \Big[\frac{2}{27}{\cal C}_g+\sum_q\Big(\kappa_q-\frac{2}{27}{\cal
  C}_g\Big) f_{Tq}^{(N)}\Big]\,.
\end{equation}

The exclusion curves from LUX~\cite{Akerib:2013tjd} and
Xenon100~\cite{Aprile:2012nq}, assuming a local DM density of
0.3~GeV/cm$^3$, are shown in Figs.~\ref{fig:gchiscalar}
to~\ref{fig:gchidiracpy} as blue and red bands, respectively. The
width of the exclusion curves represents the uncertainties in the
hadronic matrix elements of the light-quark scalar currents. For the
$s$ quark we use the lattice determination
$f_{Ts}^{(N)}=0.043\pm0.011$~\cite{Junnarkar:2013ac}. The matrix
elements for $u$ and $d$ quarks can be related to the $\sigma_{\pi N}$
term. A Baryon Chiral Perturbation Theory (B$\chi$PT) analysis of the
$\pi N$ scattering data gives $\sigma_{\pi N} = 59(7)$ MeV
\cite{Alarcon:2011zs}. This is in agreement with B$\chi$PT fit to
world lattice $N_f=2+1$ QCD data, which gives $\sigma_{\pi
  N}=52(3)(8)$ MeV \cite{Alvarez-Ruso:2014sma}. Including both $\Delta
(1232)$ and finite spacing parametrization in the fit shifts the
central value to $\sigma_{\pi N}=44$MeV. To be conservative we use
$\sigma_{\pi N}=(50\pm15)$MeV, which gives
$f_{Tu}^{(p)}=(1.8\pm0.5)\times 10^{-2}$,
$f_{Td}^{(p)}=(3.4\pm1.1)\times 10^{-2}$,
$f_{Tu}^{(n)}=(1.6\pm0.5)\times 10^{-2}$,
$f_{Td}^{(n)}=(3.8\pm1.1)\times 10^{-2}$, using the expressions
in~\cite{Crivellin:2013ipa,Crivellin:2014qxa}. For the effective Higgs
coupling to nucleons this gives
\begin{align}
\begin{split}
f_{\cal S}^{(p)}=\frac{m_W}{v_W}\Big[ &(1.8\pm0.5)\kappa_u+(3.4\pm1.1)\kappa_d+(4.3\pm1.1)\kappa_s\\
&+(6.70\pm0.12)\big(\kappa_c+\kappa_b+\kappa_t+\kappa_g^{\rm NP}\big)\Big]\times 10^{-2}\,,
\end{split}
\\
\begin{split}
f_{\cal S}^{(n)}=\frac{m_W}{v_W}\Big[& (1.6\pm0.5)\kappa_u+(3.8\pm1.1)\kappa_d+(4.3\pm1.1)\kappa_s\\
&+(6.69\pm0.12)\big(\kappa_c+\kappa_b+\kappa_t+\kappa_g^{\rm NP}\big)\Big]\times 10^{-2}\,.
\end{split}
\end{align}
We use the results in~\cite{Fitzpatrick:2012ix} to relate the nuclear
matrix elements to actual scattering rates on nuclei via nuclear form
factors.

We show the direct detection exclusion limits for SM
($\kappa_{u,d,s}=1$) or vanishing ($\kappa_{u,d,s}=0$) light-quark
Yukawa couplings in the right and left panels in
Figs. \ref{fig:gchiscalar},~\ref{fig:gchivector},~\ref{fig:gchidirac},
and~\ref{fig:gchidiracp}, respectively. The exclusion limits are
approximately two times weaker in the latter case; the constraint does
not vanish because, for small values of the light-quark Yukawas, the
scattering cross section is dominated by the gluon part of the scalar
current, Eq.~\eqref{eq:Sq}. When the light-quark Yukawas are taken to
be at the upper limit of their experimentally allowed range,
i.e. comparable to the SM bottom Yukawa, the direct detection bounds
on $g_\chi$ become significantly stronger, by a factor of about
$m_b/m_q$
(Figs.~\ref{fig:gchiscalary},~\ref{fig:gchivectory},~\ref{fig:gchidiracy},
and~\ref{fig:gchidiracpy}).

It is interesting to note that, because of the dominance of the gluon
contribution, for small light-quark Yukawas the theory uncertainty in
the exclusion bands is significantly smaller than if the light Yukawa
couplings are allowed to saturate the present experimental
bounds. (The nuclear matrix element of the effective gluon term has
smaller relative uncertainties than the corresponding matrix elements
of $m_q \bar q q$ since $f_{TG}^{(N)} \gg f_{Tq}^{(N)}$.)

\begin{figure}[t]
\begin{center}
\includegraphics[width=0.49\textwidth]{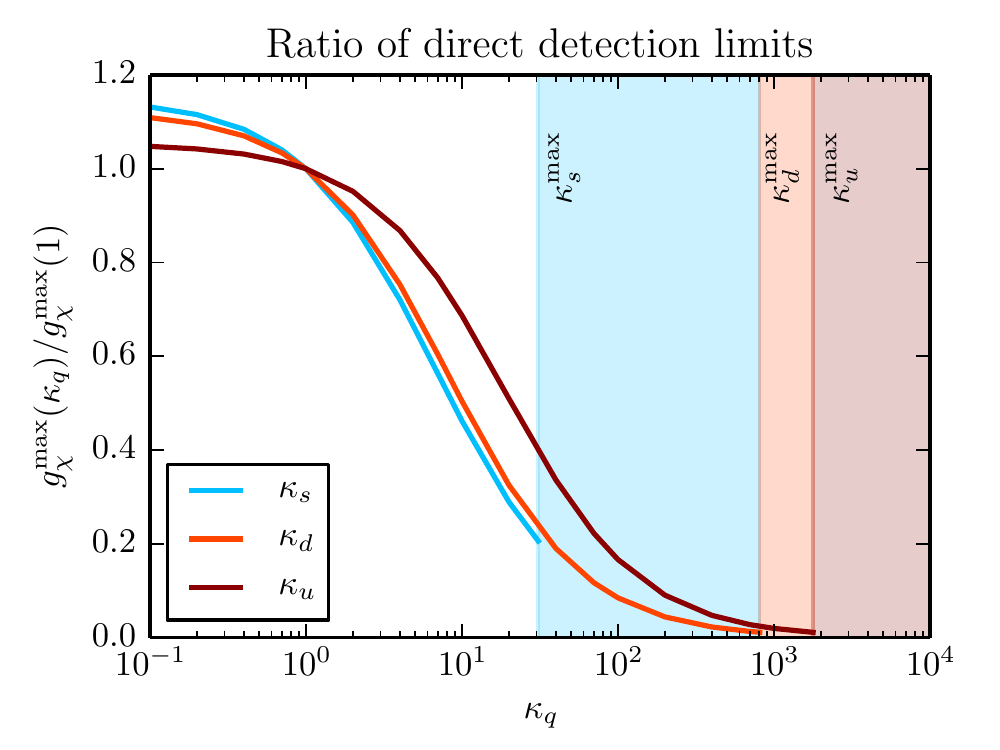}
\includegraphics[width=0.49\textwidth]{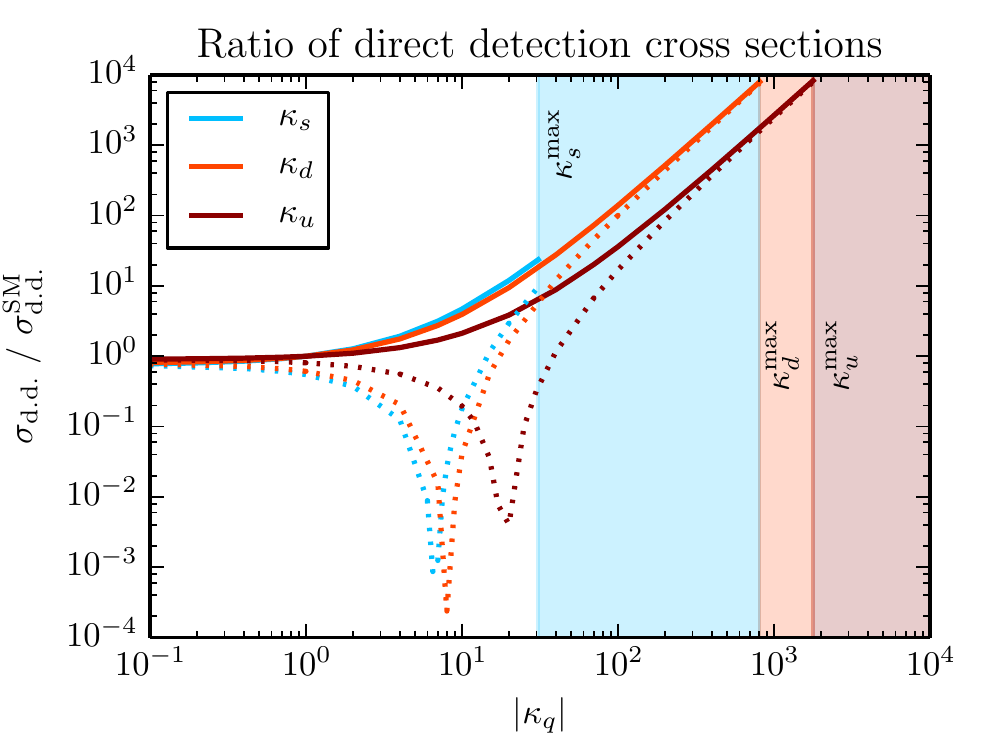}
\end{center}
\caption{Left: The ratio of direct detection bounds on $g_\chi$ from
  Xenon target varying $\kappa_u$ (dark red), $\kappa_d$ (light red),
  or $\kappa_s$ (blue), and the bound on $g_\chi$ assuming SM Higgs
  Yukawa couplings. The LHC upper bounds on $\kappa_i$ are denoted by
  vertical dashed lines with shaded regions excluded. Right: the ratio
  of predicted scattering cross sections. The dotted lines correspond
  to negative values of $\kappa_q$.
  \label{fig:ratio}
  }
\end{figure}

For $m_\chi$ smaller than a few TeV, the DM direct detection bounds
are compatible with the thermal relic Higgs-portal DM only if light
quark Yukawas are well below the present experimental bounds (the
exception is a pseudoscalar fermion DM with enhanced strange Yukawa,
where the bound is $m_\chi\gtrsim m_h/2$, see
Fig.~\ref{fig:gchidiracpy}). This means that if thermal relic DM is
discovered, it would immediately place an upper bound on
$\kappa_u,\kappa_d, \kappa_s$, assuming Higgs-portal mediation (unless
in the case of fermion DM that has purely pseudoscalar couplings). We
comment in more detail on that observation in
Section~\ref{sec:light_quark_yukawas}.

Since the DM -- nucleus scattering cross section is the only DM
observable that exhibits a rather pronounced dependence on the values
of the light-quark Yukawas, we study this dependence in more detail.

In Fig.~\ref{fig:ratio} (left) we show how the direct detection bounds
on $g_\chi$ are affected by changes in the values of the light-quark
Yukawas. We plot the ratio
\begin{equation}
\xi_{g_\chi}=\frac{g_\chi^{\rm max}(\kappa_q)}{g_\chi^{\rm max}(1)},
\end{equation}
where $g_\chi^{\rm max}(\kappa_q)$ is the upper bound on $g_\chi$
obtained from direct detection experiments for a given value of
$\kappa_q$, with $q=u,d,s$. Hence, $g_\chi^{\rm max}(1)$ is the bound
obtained assuming SM Yukawa couplings. Its value depends on $m_\chi$,
on whether DM is a scalar, fermion, or vector, and on the experiment
that measured the bounds.
Similarly, also $g_\chi^{\rm max}(\kappa_q)$ depends on $m_\chi$, the
spin of DM, and the experiment; however, all these dependences cancel
in the ratio $\xi_{g_\chi}$.  The ratio $\xi_{g_\chi}$ thus only
depends on $\kappa_q$ and on which target material was used to derive
the direct detection bounds. In Fig.~\ref{fig:ratio} we show
$\xi_{g_\chi}$ for a Xenon target, varying in turn $\kappa_u$ (dark
red line), $\kappa_d$ (light red) and $\kappa_s$ (blue), while keeping
all other parameters fixed to their SM values. We set the hadronic
matrix element $f_{Tq}^{(N)}$ to their present central values,
anticipating that in the future their uncertainties will be further
reduced.
In Fig.~\ref{fig:ratio} (right) we show a closely related quantity --
the ratio of the scattering cross sections with varied
$\kappa_{u,d,s}$ and the scattering cross section with SM Yukawa
couplings, $\sigma_{\rm d.d.}/\sigma_{\rm d.d.}^{\rm SM}$.

Fig.~\ref{fig:ratio} illustrates clearly that the difference between
the bounds where one of the light quark Yukawa couplings is taken to
be small or vanish completely, and the bounds where all the couplings
are SM-like, is very small, ${\mathcal O}(10\%)$. Saturating the
present experimental bounds on $\kappa_u$ or $\kappa_d$, the allowed
value of $g_\chi$ could lie two orders of magnitude below what one
obtains for the case of SM Yukawa couplings. Such large values for the
light-quark Yukawas are not very likely to be realized in a concrete
model, as we will discuss in the next section. However, it is very
interesting to observe that even a moderate increase of the values of
the light-quark Yukawa couplings to only a few times their SM values
can have a significant effect on the direct detection bounds,
enhancing the scattering cross sections by up to a factor of ten.

Finally, in the right panel of Fig.~\ref{fig:ratio} we show the cross
section ratios for negative values of the $\kappa_q$ (dotted
lines). We see that the interference of the light-quark contributions
with the effective gluon interaction can, in principle, lead to a
strong reduction of the scattering cross section.

\section{Changes to Yukawa couplings in new physics models}
\label{sec:flavor:models}

So far we allowed the Yukawa couplings of the Higgs to quarks to have
arbitrary values, only restricting them to lie within the bounds
obtained from global fits to LHC data.
For simplicity, we also neglected flavor violation and $CP$ violation
when discussing their impact on the DM interactions. 

Of course, changes of the Yukawa couplings by several orders of
magnitude, as allowed by current experimental constraints on the
light-quark Yukawas, are not very likely to be realized in a complete
model, and might require significant fine tuning of the corresponding
quark masses.
In this section we investigate how large the deviations from the SM
Yukawa interactions can be in popular models of NP with viable flavor
structures.

Tables~\ref{tab:upyukawa} and~\ref{tab:downyukawa} summarize the
predictions for the effective Yukawa couplings in the Standard Model
(SM), in multi-Higgs-doublet models (MHDM) with natural flavor
conservation (NFC)~\cite{Glashow:1976nt, Paschos:1976ay}, in the MSSM
at tree level, the Giudice-Lebedev model of quark masses
(GL)~\cite{Giudice:2008uua}, in NP models with minimal flavor
violation (MFV)~\cite{D'Ambrosio:2002ex}, in Randall-Sundrum models
(RS)~\cite{Randall:1999ee}, and in models with a composite Higgs,
realized as a pseudo-Nambu-Goldstone boson (pNGB)~\cite{Dugan:1984hq,
  Georgi:1984ef, Kaplan:1983sm, Kaplan:1983fs}. For completeness, we
include both the flavor-conserving and flavor-violating Yukawa
interactions, and allow for $CP$ violation. The Higgs couplings to
quarks are thus described by
\begin{equation}\label{eq:Lh:CPV2}
\begin{split}
	\mathcal{L}_{\rm eff, q} &=- \kappa_q\frac{m_q}{v_W}\bar{q}qh
        - i\tilde \kappa_q\frac{m_q}{v_W}\bar{q}\gamma_5 q h  
	- \Big[\big(\kappa_{qq'}+i \tilde \kappa_{qq'}\big) \bar{q}_L
          q_R'h +{\rm h.c.}\Big], 
\end{split}
\end{equation}
where a sum over the SM quark fields is understood. The first two
terms are flavor diagonal, with the first term $CP$ conserving and the
second term $CP$ violating, and coincide with the definitions in
eqs.~\eqref{eq:Lh} and \eqref{eq:Lh:CPV}, respectively. The terms in
square brackets are flavor violating, with the real (imaginary) part
of the coefficient $CP$ conserving (violating). In the SM we have
$\kappa_q=1$, while $\tilde \kappa_q=\kappa_{qq'}=\tilde
\kappa_{qq'}=0$. The flavor-violating couplings in the above set of NP
models are collected in Tables \ref{tab:upFVyukawa} and
\ref{tab:downFVyukawa}. These tables complement the analyses in
\cite{Dery:2014kxa,Dery:2013aba,Dery:2013rta} (see
also~\cite{Hedri:2013wea}, where implications of a negative top-quark
Yukawa were explored, and~\cite{Goertz:2014qia} for an indirect bound
on the down-quark Yukawa in alignment models).

\begin{table}
\begin{center}
\begin{tabular}{l  c  c  c c  }
\hline\hline
Model	& $\kappa_t$ & $\kappa_{c (u)}/\kappa_t$  & $\tilde \kappa_t/\kappa_t$ & $\tilde \kappa_{c (u)}/\kappa_t$ \\ \hline
SM	& 1	& 1 & 0 &0 \\
NFC & $V_{hu}\,v_W/v_u$	& 1 &  0 &0 \\
MSSM	& $\cos\alpha/\sin\beta$	&1 &0  &0\\
GL 	& $1 +{\mathcal O}(\epsilon^2)$& $\simeq 3 (7)$ & ${\mathcal O}(\epsilon^2)$ ~~& ~~${\mathcal O}(\kappa_{c(u)})$\\
GL2 	& $\cos\alpha/\sin\beta$& $\simeq 3 (7)$ & ${\mathcal O}(\epsilon^2)$ ~~& ~~${\mathcal O}(\kappa_{c(u)})$\\
MFV &$1+\frac{\text{Re}(a_uv_W^2+2b_u m_t^2)}{\Lambda^2}$
~~&~~$1-\frac{2\text{Re}(b_u)m_t^2}{\Lambda^2}$~~
&~~$\frac{\Im(a_uv_W^2+2b_u m_t^2)}{\Lambda^2}$~~ & ~~$\frac{\Im(a_u
  v_W^2)}{\Lambda^2} $ \\
RS &$1-{\mathcal O}\Big(\frac{ v_W^2}{m_{KK}^2}\bar Y^2\Big)$~~&~~$1+{\mathcal O}\Big(\frac{ v_W^2}{m_{KK}^2}\bar Y^2\Big)$~~ &~~$1+{\mathcal O}\Big(\frac{ v_W^2}{m_{KK}^2}\bar Y^2\Big)$~~ &~~$1+{\mathcal O}\Big(\frac{ v_W^2}{m_{KK}^2}\bar Y^2\Big)$ \\
pNGB & $1+{\mathcal O}\Big(\frac{ v_W^2}{f^2}\Big)+{\mathcal O}\Big(y_*^2 \lambda^2 \frac{ v_W^2}{M_*^2}\Big)$ & $1+{\mathcal O}\Big(y_*^2 \lambda^2 \frac{ v_W^2}{M_*^2}\Big)$ & ${\mathcal O}\Big(y_*^2 \lambda^2 \frac{ v_W^2}{M_*^2}\Big)$ & ${\mathcal O}\Big(y_*^2 \lambda^2 \frac{ v_W^2}{M_*^2}\Big)$ \\
\hline\hline
\end{tabular}
\caption{Predictions for the flavor diagonal up-type Yukawa couplings
  in a number of new physics models (see text for details). }
\label{tab:upyukawa}
\end{center}
\end{table}

\begin{table}
\begin{center}
\begin{tabular}{ l   c  c c c}
\hline\hline
Model	& $\kappa_b$ & $\kappa_{s(d)}/\kappa_b$ & $\tilde \kappa_b/\kappa_b$ & $\tilde \kappa_{s(d)}/\kappa_b$ \\ \hline
SM	& 1 & 1 &0 &0\\
NFC & $V_{hd}\,v_W/v_d$	& 1 &0 &0\\
MSSM	 & $-\sin\alpha/\cos\beta$	&1 &0 &0\\
GL	& $\simeq 3$	& $\simeq 5/3(7/3)$ & ${\mathcal O}(1)$ & ${\mathcal O}(\kappa_{s(d)}/\kappa_b)$ \\
GL2	& $-\sin\alpha/\cos\beta$	& $\simeq 3(5)$ & ${\mathcal O}(\epsilon^2)$ & ${\mathcal O}(\kappa_{s(d)}/\kappa_b)$ \\
MFV & $1+\frac{\text{Re}(a_d v_W^2 +2 c_d m_t^2)}{\Lambda^2}$~~&~~$1-\frac{2\text{Re}(c_d)m_t^2}{\Lambda^2}$~~&~~$ \frac{\Im(a_d v_W^2+2 c_d m_t^2)}{\Lambda^2}$~~&~~$ \frac{\Im(a_d v_W^2+2 c_d |V_{ts(td)}|^2 m_t^2)}{\Lambda^2}$ \\
RS &$1-{\mathcal O}\Big(\frac{ v_W^2}{m_{KK}^2}\bar Y^2\Big)$~~&~~$1+{\mathcal O}\Big(\frac{ v_W^2}{m_{KK}^2}\bar Y^2\Big)$~~ &~~$1+{\mathcal O}\Big(\frac{ v_W^2}{m_{KK}^2}\bar Y^2\Big)$~~ &~~$1+{\mathcal O}\Big(\frac{ v_W^2}{m_{KK}^2}\bar Y^2\Big)$ \\
pNGB & $1+{\mathcal O}\Big(\frac{ v_W^2}{f^2}\Big)+{\mathcal O}\Big(y_*^2 \lambda^2 \frac{ v_W^2}{M_*^2}\Big)$ & $1+{\mathcal O}\Big(y_*^2 \lambda^2 \frac{ v_W^2}{M_*^2}\Big)$ & ${\mathcal O}\Big(y_*^2 \lambda^2 \frac{ v_W^2}{M_*^2}\Big)$ & ${\mathcal O}\Big(y_*^2 \lambda^2 \frac{ v_W^2}{M_*^2}\Big)$\\
\hline\hline
\end{tabular}
\caption{Predictions for the flavor diagonal down-type Yukawa
  couplings in a number of new physics models (see text for details).}
\label{tab:downyukawa}
\end{center}
\end{table}

\begin{table}[tc]
\begin{center}
\begin{tabular}{l  c  c  c }
\hline\hline
Model	& $\kappa_{ct (tc)}/\kappa_t$ & $\kappa_{ut (tu)}/\kappa_t$  & $\kappa_{uc (cu)}/\kappa_t$ \\ \hline
GL \& GL2	& $\epsilon (\epsilon^2)$ & $\epsilon (\epsilon^2)$ & $\epsilon^3$ \\
MFV &$ \frac{\text{Re}\big( c_u m_b^2 V_{cb}^{(*)}\big)}{\Lambda^2}\frac{\sqrt2 m_{t(c)}}{v_W} $~&~ $ \frac{\text{Re}\big( c_u m_b^2 V_{ub}^{(*)}\big)}{\Lambda^2} \frac{\sqrt2 m_{t(u)}}{v_W}$~&~ $ \frac{\text{Re}\big( c_u m_b^2 V_{ub(cb)}V_{cb(ub)}^{*}\big)}{\Lambda^2} \frac{\sqrt2 m_{c(u)}}{v_W}$\\
RS & $\sim \lambda^{(-)2} \frac{m_{t(c)}}{v_W} \bar Y^2\frac{v_W^2}{m_{KK}^2} $~~&~~$\sim \lambda^{(-)3} \frac{m_{t(u)}}{v_W} \bar Y^2\frac{v_W^2}{m_{KK}^2} $~~&~~$\sim \lambda^{(-)1} \frac{m_{c(u)}}{v_W} \bar Y^2\frac{v_W^2}{m_{KK}^2} $ \\
pNGB & ${\mathcal O}(y_*^2 \frac{m_t}{v_W}\frac{\lambda_{L (R),2} \lambda_{L(R),3}m_W^2}{M_*^2})$ & ${\mathcal O}(y_*^2 \frac{m_t}{v_W}\frac{\lambda_{L (R),1} \lambda_{L(R),3}m_W^2}{M_*^2})$  & ${\mathcal O}(y_*^2 \frac{m_c}{v_W}\frac{\lambda_{L (R),1} \lambda_{L(R),2}m_W^2}{M_*^2})$ \\
\hline\hline
\end{tabular}
\caption{Predictions for the flavor violating up-type Yukawa couplings
  in a number of new physics models (see text for details). In the SM,
  NFC and the tree-level MSSM the Higgs Yukawa couplings are flavor
  diagonal. The estimates of the $CP$-violating versions of the
  flavor-changing transitions, $\kappa_{ij}/\kappa_t$, are the same as
  the $CP$-conserving ones, apart from substituting ``Im'' for ``Re''
  in the ``MFV'' row. }
\label{tab:upFVyukawa}
\end{center}
\end{table}

\begin{table}[tc]
\begin{center}
\begin{tabular}{l  c  c  c }
\hline\hline
Model	&   $\kappa_{bs (sb)}/\kappa_b$ & $\kappa_{bd (db)}/\kappa_b$  & $\kappa_{sd (ds)}/\kappa_b$ \\ \hline
GL \& GL2	& $\epsilon^3(\epsilon^2)$ & $\epsilon^2$ & $\epsilon^3(\epsilon^4)$ \\
MFV &$\frac{\text{Re}\big(c_d m_t^2 V_{ts}^{(*)}\big)}{\Lambda^2} \frac{\sqrt2m_{s(b)}}{v_W}$~&~$\frac{\text{Re}\big(c_d m_t^2 V_{td}^{(*)}\big)}{\Lambda^2} \frac{\sqrt2 m_{d(b)}}{v_W}$~&~$\frac{\text{Re}\big(c_d m_t^2 V_{ts(td)}^*V_{td(ts)}\big)}{\Lambda^2} \frac{\sqrt2 m_{s(d)}}{v_W}$ \\
RS & $\sim \lambda^{(-)2} \frac{m_{b(s)}}{v_W} \bar Y^2\frac{v_W^2}{m_{KK}^2} $~~&~~$\sim \lambda^{(-)3} \frac{m_{b(d)}}{v_W} \bar Y^2\frac{v_W^2}{m_{KK}^2} $~~&~~$\sim \lambda^{(-)1} \frac{m_{s(d)}}{v_W} \bar Y^2\frac{v_W^2}{m_{KK}^2} $ \\
pNGB & ${\mathcal O}(y_*^2 \frac{m_b}{v_W}\frac{\lambda_{L (R),2} \lambda_{L(R),3}m_W^2}{M_*^2})$ & ${\mathcal O}(y_*^2 \frac{m_b}{v_W}\frac{\lambda_{L (R),1} \lambda_{L(R),3}m_W^2}{M_*^2})$  & ${\mathcal O}(y_*^2 \frac{m_s}{v_W}\frac{\lambda_{L (R),1} \lambda_{L(R),2}m_W^2}{M_*^2})$\\
\hline\hline
\end{tabular}
\caption{Predictions for the flavor violating down-type Yukawa
  couplings in a number of new physics models (see text for
  details). In SM, NFC and tree level MSSM the Higgs Yukawa couplings
  are flavor diagonal. The estimates of the $CP$-violating versions of
  the flavor-changing transitions, $\kappa_{ij}/\kappa_b$, are the
  same as the $CP$-conserving ones, apart from substituting ``Im'' for
  ``Re'' in the ``MFV'' row. }
\label{tab:downFVyukawa}
\end{center}
\end{table}

\subsection{Dimension-Six Operators with Minimal Flavor Violation}

We start our discussion by considering dimension-six operators arising
from integrating out NP at a high scale $\Lambda$. In addition, we
assume that the flavor breaking in the NP sector is only due to the SM
Yukawas, i.e. that NP satisfies the Minimal Flavor Violation (MFV)
hypothesis \cite{D'Ambrosio:2002ex, Chivukula:1987py,
  Gabrielli:1994ff, Ali:1999we, Buras:2000dm, Buras:2003jf,
  Kagan:2009bn}. Integrating out the new physics states gives for the
Higgs couplings to quarks
\begin{equation}
\begin{split}\label{eq:EFT:MFV}
	\mathcal{L}_{\rm EFT} &= Y_u \bar{Q}_L H^c u_R + Y_d \bar{Q}_L
        H d_R  + \frac{Y_u^\prime}{\Lambda^2}\bar{Q}_L H^c u_R
        (H^\dagger H)+ \frac{Y_d^\prime}{\Lambda^2} \bar{Q}_L H d_R
        (H^\dagger H)+\text{h.c.}\,, 
\end{split}
\end{equation}
where $\Lambda$ is the scale of new physics and $H^c =
i\sigma_2H^\ast$. We identify the NP scales in the up- and down-quark
sectors for simplicity. There are also modifications of quark kinetic
terms through dimension-six derivative operators. These can be
absorbed in~\eqref{eq:EFT:MFV} using equations of
motion~\cite{AguilarSaavedra:2009mx}.  The quark mass matrices and
Yukawa couplings after EWSB are thus
\begin{equation}
M_q=\frac{v_W}{\sqrt2}\Big(Y_q +Y_q' \frac{v_W^2}{2 \Lambda^2}\Big)\,,
\qquad y_q=Y_q +3 Y_q' \frac{v_W^2}{2 \Lambda^2}\,, \qquad \quad q=u,d\,. 
\end{equation}
Because $Y_q$ and $Y_{q'}$ appear in two different combinations in
$M_q$ and $y_q$, the two, in general, cannot be made diagonal in the
same basis.

In MFV the coefficients of the dimension-six operators can be expanded
in terms of $Y_{u,d}$,
\begin{equation}
\begin{split}
	Y_u^{\prime} &= a_uY_u +
        b^{\phantom\dagger}_uY^{\phantom\dagger}_uY_u^\dagger
        Y^{\phantom\dagger}_u + c^{\phantom\dagger}_u
        Y^{\phantom\dagger}_dY_d^\dagger
        Y^{\phantom\dagger}_u+\cdots\,,\\  
	Y_d^{\prime} &= a_dY_d +
        b^{\phantom\dagger}_dY^{\phantom\dagger}_dY_d^\dagger
        Y^{\phantom\dagger}_d + c^{\phantom\dagger}_d
        Y^{\phantom\dagger}_uY_u^\dagger
        Y^{\phantom\dagger}_d+\cdots\,. 
\end{split}
\end{equation}
with $a_q, b_q, c_q\sim {\mathcal O}(1)$. Working to first order in
dimension-six operator insertions we can thus write for the Yukawa
couplings, in the mass eigenbases for up and down quarks respectively,
\begin{equation}
\begin{split}
	y_u &= \Big[1+\frac{v_W^2}{\Lambda^2}\Big(a_u  + b_u (y_{\rm
            SM}^u)^2 + c_u V (y_{\rm SM}^d)^2
          V^\dagger+\cdots\Big)\Big]y_{\rm SM}^u\,,\\  
	y_d &= \Big[1+\frac{v_W^2}{\Lambda^2}\Big(a_d  + b_d (y_{\rm
            SM}^d)^2 + c_d V^\dagger (y_{\rm SM}^u)^2
          V+\cdots\Big)\Big]y_{\rm SM}^d\,. 
\end{split}
\end{equation}
Here $y_{\rm SM}^{u,d}$ are the diagonal matrices of the SM Yukawa
couplings, while $V$ is the Cabibbo-Kobayashi-Maskawa (CKM) matrix. In
general, the coefficients $a_q, b_q, c_q$ are complex so that
$CP$-violating Higgs couplings arise at ${\mathcal
  O}(v_W^2/\Lambda^2)$. Flavor-violating Higgs couplings arise first
from the terms proportional to $c_{u,d}$ in the Yukawa expansion and
are thus suppressed by the corresponding CKM matrix elements. In
Tables~\ref{tab:upyukawa}-\ref{tab:downFVyukawa} we collect the values
for flavor-conserving and flavor-violating Yukawa couplings in the
``MFV'' row, assuming that all the coefficients $a_q, b_q, c_q$ are
${\mathcal O}(1)$, and show only the numerically leading non-SM
contributions. In the expressions we also set $V_{tb}$ to unity.

The corrections to DM phenomenology are dominated by changes of the
third-generation Yukawa couplings. The MFV corrections to light-quark
Yukawa couplings are all either additionally CKM suppressed or involve
extra insertions of light-quark masses. Hence the theory error in
Higgs-portal DM phenomenology due to Yukawa coupling uncertainties
will be small in MFV models of NP once the Higgs couplings to top and
bottom quarks are well measured.

\subsection{Multi-Higgs-doublet model with natural flavor conservation}
In MHDMs there are no tree-level FCNCs if natural flavor conservation
is assumed~\cite{Glashow:1976nt, Paschos:1976ay}. Under this
assumption we can choose a Higgs doublet basis in which only one
doublet, $H_u$, couples to the up-type quarks, and only one Higgs
doublet, $H_d$, couples to the down-type quarks\footnote{Note that
  $H_u = H_d$ is included as a special case.}. After EWSB the two
doublets obtain the vevs $v_u$ and $v_d$, respectively. On the other
hand, the vevs of all Higgs doublets contribute to the $W$ and $Z$
masses. They satisfy the sum rule $v_W^2=\sum_i v_i^2$, where the sum
is over all Higgs doublets.

The neutral scalar components of $H_i$ are $(v_i+h_i)/\sqrt2$, where
the dynamical fields $h_{i}$ are a linear combination of the neutral
Higgs mass eigenstates (and include $h_u$ and $h_d$). We thus have
$h_i=V_{hi} h + \ldots$, where $V_{hi}$ are elements of the unitary
matrix $V$ that diagonalizes the neutral-Higgs mass terms and we only
write down the contribution of the lightest Higgs, $h$. Under the
assumptions above, the mass and Yukawa terms can be diagonalized in
the same basis, so that there is no flavor violation and no $CP$
violation in the Yukawa interactions:
\begin{equation}
\kappa_{qq'}=\tilde \kappa_{qq'}=0\,, \qquad \tilde \kappa_q=0\,.
\end{equation}
We obtain a universal shift in all up-quark Yukawa couplings, and a
different universal shift in all down-quark Yukawa couplings, given by
\begin{equation}
	\kappa_u = \kappa_c = \kappa_t = V_{hu}\frac{v_W}{v_u}\,,
        \qquad\quad \kappa_d = \kappa_s = \kappa_b =
        V_{hd}\frac{v_W}{v_d}\,. 
\end{equation}
Since the shifts are universal over generations, the precise
measurements of the Higgs couplings to top and bottom quarks will also
determine the Higgs couplings to light quarks. Both $\kappa_t$ and
$\kappa_b$ are expected to be known with ${\mathcal O}(5\%)$ precision
after the end of the high-luminosity LHC run \cite{CMS:2013xfa,
  ATLAS:2013hta}. The uncertainties in the DM direct detection rates
due to uncertainties in the Yukawa couplings will thus be negligible,
assuming NFC. Note that the Higgs portal with an additional SM singlet
mixing with the Higgs is also described by the above modifications of
fermion couplings, with a completely universal shift
$\kappa_i=\cos\theta$, where $\theta$ is the singlet--Higgs mixing
angle \cite{Falkowski:2015iwa, Martin-Lozano:2015dja}.

Our analysis of modified Higgs-portal DM phenomenology given in
Section~\ref{sec:arbitrary:flavor} applies in the somewhat special
limit where the DM only couples to the lightest mass-eigenstate
$h$. For instance, for scalar DM the general Higgs portal is
\begin{equation}
{\cal L}_{\rm NFC}=g_{\chi,ij} \chi^\dagger \chi H_i^\dagger H_j.
\end{equation}
If the hermitian matrix of couplings $g_{\chi,ij}$ is such that it has
$h$ as the only eigenstate with nonzero eigenvalue, then our analysis
in Section \ref{sec:arbitrary:flavor} applies unchanged. In general,
however, all the expressions in Section \ref{sec:arbitrary:flavor} get
corrected by terms of order $1/m_{H_i}^2$ due to exchanges of heavy
Higgs bosons with masses $m_{H_i}$. If DM is heavy, $m_\chi >m_{H_i}$,
the presence of heavy Higgs bosons would also open new annihilation
channels.

\subsection{Type-II Two-Higgs-Doublet Model}
The MSSM tree-level Higgs potential and the couplings to quarks are
the same as in the type-II two-Higgs-doublet model (2HDM), see, e.g.,
\cite{Haber:1984rc}. This is an example of a 2HDM with natural flavor
conservation in which $v_u=\sin\beta\, v_W$, $v_d=\cos\beta\,
v_W$. The mixing of $h_{u,d}$ into the Higgs mass-eigenstates $h$ and
$H$ is given by
\begin{equation}
	\begin{pmatrix}h_u \\ h_d\end{pmatrix}
          = \begin{pmatrix}\cos\alpha & \sin\alpha \\ -\sin\alpha
            &\phantom{-}\cos\alpha\end{pmatrix}\begin{pmatrix}h
              \\ H\end{pmatrix}, 
\end{equation}
where $h$ is the observed SM-like Higgs. Thus
\begin{equation}
\begin{aligned}
	\kappa_u = \kappa_c = \kappa_t &= \frac{\cos\alpha}{\sin\beta},\\
	\kappa_d = \kappa_s = \kappa_b&= -\frac{\sin\alpha}{\cos\beta},
\end{aligned}
\end{equation}
while the flavor-violating and/or $CP$-violating Yukawas are zero. In
the decoupling limit ($\beta-\alpha=\pi/2$) the heavy Higgs bosons
become infinitely heavy, while the Yukawa couplings tend toward their
SM value, $\kappa_i=1$. The global fits to Higgs data in type-II 2HDM
already constrain $\beta-\alpha\simeq \pi/2$~\cite{Carmi:2012in,
  Falkowski:2013dza, Grinstein:2013npa} so that in this case the
corrections to Higgs-portal DM phenomenology due to non-standard Higgs
Yukawa couplings are small.

As in the case of MHDM, the DM phenomenology of Section
\ref{sec:arbitrary:flavor} remains unchanged only in the limit where
the DM couples to the light Higgs $h$ but not to the heavy Higgs
$H$. In the general case, our analysis gets corrections that are
relatively suppressed by ${\mathcal O}(m_h^2/m_H^2)$. If we are not
too far away from the decoupling limit these corrections can be
neglected, while in parts of the parameter space, where cancellation
can occur, the extra contributions are numerically important
\cite{Greljo:2013wja}.

\subsection{Higgs-dependent Yukawa Couplings}
In the model of quark masses introduced by Giudice and Ledebev
(GL)~\cite{Giudice:2008uua} the Higgs-quark interactions are written
in terms of effective operators
\begin{equation}\label{eq:Higgs-dep}
{\cal L}_{q}=c_{ij}^{u} \bigg( \frac{H^\dagger H}{M^2}
  \bigg)^{n_{ij}^{u}} \, \bar Q_{L,i} u_{R,j} H^c + c_{ij}^{d} \bigg( \frac{H^\dagger H}{M^2}
  \bigg)^{n_{ij}^{d}} \, \bar Q_{L,i} d_{R,j} H +\text{h.c.}\,.
\end{equation}
They can be thought of as arising from integrating out heavy mediators
at a large mass scale $M$. In this model the light quarks couple to
the Higgs only through operators with mass dimension higher than four,
i.e., for light quarks we have $n_{ij}^{u,d}\ne0$. The values of the
integers $n_{ij}^{u,d}$, and of the coefficients $c_{ij}^{u,d}$ that
take values of order unity, are chosen such that the hierarchies of
the observed quark masses and mixing angles are explained, after EWSB,
in terms of the expansion parameter $\epsilon \equiv v_W^2/M^2 \approx
1/60$. Thus, the Yukawa couplings are of the form
\begin{equation}\label{eq:yudrelation}
  y_{ij}^{u,d} = (2n_{ij}^{u,d} + 1) (y_{ij}^{u,d})_\text{SM} \, .
\end{equation}
After mass diagonalization the SM Yukawas are diagonal in the same
basis as the quark masses, $(y_{ij}^{u,d})_\text{SM} \propto
\delta_{ij} m_{i}^{u,d}$, while the $y_{ij}^{u,d}$ are not diagonal in
the same basis\footnote{Note that the mixing of contributions from
  different effective operators that may have large relative phases
  could lead to sizeable CP-violating contributions to the Yukawa
  couplings.}. Using the ansatz $n_{ij}^{u,d}=a_i +b_j^{u,d}$ with
$a=(1,1,0)$, $b^d=(2,1,1)$, and $b^u=(2,0,0)$, this gives the
deviations in the Yukawa couplings collected in Tables
\ref{tab:upyukawa}-\ref{tab:downFVyukawa} in the row denoted by
``GL''. Since the couplings to the bottom quark is enhanced by a large
factor, $\kappa_b \simeq 3$, the simplest version of the GL model is
already excluded by the Higgs data on $h\to WW$, $h\to ZZ$ and $h\to
\gamma\gamma$ decays.

We therefore modify the initial GL proposal and assume that we have
two Higgs doublets in \eqref{eq:Higgs-dep}, $H_u$ that only gives
masses to up-type quarks and $H_d$ that only gives masses to down-type
quarks. The correct mass and CKM angle hierarchy is obtained by using
$b^d=(1,0,0)$ in the ansatz for $n_{ij}^d$, and leaving $a$ and $b^u$
unchanged. This gives satisfactory Higgs phenomenology at present as
long as $\kappa_b=\sin\alpha/\cos\beta\simeq 1$ up to ${\mathcal
  O}(20\%)$. In this limit also $\kappa_t = \cos\alpha/\sin\beta\simeq
1$.  The scaling of Yukawa couplings for this modification of the GL
model is shown in Tables \ref{tab:upyukawa}-\ref{tab:downFVyukawa} in
the row denoted by ``GL2''.

In the GL model it is natural that the Higgs is the only state that
couples to DM. The GL model is thus an example of Higgs-portal DM
where the light-quark Yukawa couplings can substantially differ from
their SM values. For instance, in GL2 $\kappa_u\simeq 7 \kappa_t$,
$\kappa_d\simeq 5\kappa_b$, $\kappa_s\simeq 3\kappa_b$,
$\kappa_c\simeq 3 \kappa_t$. The coupling of DM to
gluons~\eqref{eq:Sq} ${\cal C}_g\simeq 4\kappa_t+\kappa_b$, so that
${\cal C}_g\sim (5/3) {\cal C}_g^{\rm SM}$, and $\tilde {\cal C}_g\sim
{\mathcal O}({\cal C}_g)$. Taking $\kappa_b\simeq 1$, this means that
the effective Higgs coupling to nucleons, governing the direct DM
detections rates, gets enhanced compared to the SM Higgs Yukawa
couplings by
\begin{equation}
\frac{f_{\cal S}^{(p)}}{f_{\cal S}^{(p)}|_{\rm SM}}\simeq
1.2\kappa_t+1.3\kappa_b\simeq 2.5\,, \quad \frac{f_{\cal
    S}^{(n)}}{f_{\cal S}^{(n)}|_{\rm SM}}\simeq
1.3\kappa_t+1.3\kappa_b\simeq 2.6\,. 
\end{equation}
Here most of the enhancement over the SM comes from enhanced
$\kappa_u$ and $\kappa_d$, which is also the reason for enlarged
isospin breaking (the difference between $f_{\cal S}^{(p)}$ and
$f_{\cal S}^{(n)}$).  As a result of larger couplings to light quarks
the spin-independent DM scattering cross section can thus be enhanced
by an order of magnitude in the GL2 model of light-quark masses.

\subsection{Randall-Sundrum models}
The Randall-Sundrum (RS) warped extra-dimensional models with the SM
fields propagating in the bulk provide a solution to the hierarchy
problem and simultaneously explain the hierarchy of the SM fermion
masses without large hierarchies in the initial five-dimensional (5D)
Lagrangian~\cite{Randall:1999ee, Gherghetta:2000qt,
  Grossman:1999ra}. The fermion zero modes are either localized toward
the UV brane (for lighter fermions) or toward the IR brane (the top,
the left-handed $b$ quark and potentially the right-handed $c$
quark)~\cite{Huber:2000ie, Huber:2003tu}. The Higgs field and the
Higgs vev are both localized toward the IR brane. Integrating out the
Kaluza-Klein (KK) modes and working in the limit of a brane-localized
Higgs, the SM quark mass matrices are given, to leading order in
$v_W^2/m_{KK}^2$, by~\cite{Azatov:2009na} (see
also~\cite{Casagrande:2008hr, Bauer:2009cf, Malm:2013jia,
  Archer:2014jca, Blanke:2008zb, Blanke:2008yr, Albrecht:2009xr,
  Agashe:2006wa, Agashe:2014jca}, and Ref.~\cite{Dillon:2014zea} for a
bulk Higgs scenario)
\begin{equation}
M^{d(u)}_{ij}=\big[F_q Y_{1(2)}^{5D}F_{d(u)}\big]_{ij} v_W\,.
\end{equation}
Here, $m_{KK}$ is the KK mass scale. The $F_{q,u,d}$ are diagonal
$3\times 3$ matrices of fermion wave-functions for the left-handed
electroweak quark doublets and the right-handed electroweak up and
down quark singlets, respectively, evaluated at the IR brane. Assuming
flavor anarchy, the 5D Yukawa matrices for up and down quarks,
$Y_{1,2}^{5D}$, are general $3\times 3$ complex matrices with
${\mathcal O}(1)$ entries. For a Higgs field propagating in the bulk,
5D gauge invariance guarantees
$Y_1^{5D}=Y_2^{5D}$~\cite{Azatov:2009na}.

At leading order in $v_W^2/m_{KK}^2$ the Higgs Yukawas are aligned
with the quark masses, i.e.,
\begin{equation}
M_{u,d}=y_{u,d} \frac{v_W}{\sqrt2}+{\mathcal O}(v_W^2/m_{KK}^2)\,.
\end{equation}
The misalignment arises from dimension-six operators that are
generated by tree-level KK quark exchanges. They give
\begin{equation}\label{eq:misalignment}
\big[y_{u(d)}\big]_{ij}-\frac{\sqrt2}{v_W}\big[M_{u,d}\big]_{ij}\sim -
\frac{2}{3}F_{q_i} \bar Y^3 F_{u_j(d_j)}\frac{v_W^2}{m_{KK}^2}\,, 
\end{equation}
where $\bar Y$ is a typical value of the dimensionless 5D Yukawa
coupling and is in numerical analyses typically taken to be below
$\bar Y \lesssim 4$ (see, e.g., \cite{Archer:2014jca}). The Higgs
mediated FCNCs are thus suppressed by the same zero-mode wave-function
overlaps that also suppress the quark masses, giving rise to the RS
GIM mechanism~\cite{Cacciapaglia:2007fw, Agashe:2004cp,
  Agashe:2004ay}.

Using that the CKM matrix elements are given by $V_{ij}\sim
F_{q_i}/F_{q_j}$ for $i<j$, Eq.~\eqref{eq:misalignment} can be
rewritten as
\begin{equation}
\big[y_{u(d)}\big]_{ij}-\frac{\sqrt2}{v_W}\big[M_{u,d}\big]_{ij}\sim -
\frac{2}{3} \bar Y^2 \frac{v_W^2}{m_{KK}^2}  
\left\{
\begin{matrix}
\frac{m_{u_j(d_j)}}{v_W} V_{ij}\,, &\, \qquad i<j\,, \\
1\,, &\, \qquad i=j\,, \\
\frac{m_{u_i(d_i)}}{v_W} V_{ij}^{-1}\,,  &\, \qquad j<i \,.
\end{matrix}
\right.
\end{equation}
This yields the $\kappa_i$ collected in
Tables~\ref{tab:upyukawa}-\ref{tab:downFVyukawa}. In the numerical
analysis of ref.~\cite{Azatov:2009na} the diagonal values $\kappa_i$
were typically found to be smaller than one, with deviations in
$\kappa_t$ up to $30\%$, $\kappa_b$ up to $15\%$, in $\kappa_{s,c}$ up
to $\sim 5\%$, and in $\kappa_{u,d}$ of $~1\%$ (these estimates were
obtained fixing the mass of the first KK gluon excitation to
$3.7$~TeV, above the present ATLAS bound
\cite{ATLAS-CONF-2015-009}). The effective Higgs coupling to nucleons,
$f_{\cal S}^{(N)}$, thus only gets reduced by ${\mathcal O}(10\%)$,
giving a ${\mathcal O}(20\%)$ smaller DM scattering cross section on
nuclei, compared to the case of SM Yukawa couplings. The largest
effect arises in DM annihilations to top quarks, where the cross
section can be reduced by a factor of two, while the annihilation
cross section to $b\bar b$ pairs can be $\sim 30\%$ smaller than for
SM Yukawa couplings.

\subsection{Composite pseudo-Goldstone Higgs}
Finally, we investigate the possibility that the Higgs is a
pseudo-Goldstone boson arising from the spontaneous breaking of a
global symmetry in a strongly coupled sector \cite{Dugan:1984hq,
  Georgi:1984ef, Kaplan:1983sm, Kaplan:1983fs}. We assume that the SM
fermions couple linearly to composite operators
$O_{L,R}$~\cite{Kaplan:1991dc},
\begin{equation}
\lambda_{L,i}^q \bar Q_{L,i} O_R^i+\lambda_{R,j}^u \bar u_{R,j}
O_L^j+h.c. \,,
\end{equation}
where $i,j$ are flavor indices. This is the 4D dual of the fermion
mass generation in 5D RS models. The Higgs couples to the composite
sector with a typical coupling $y_*$. The SM masses and Yukawa
couplings then arise from expanding the two-point functions of the
$O_{L,R}$ operators in powers of the Higgs field~\cite{Agashe:2009di},
giving rise to four- and higher-dimensional Higgs operators, such as
in \eqref{eq:EFT:MFV}.

The new ingredient, related to the pNGB nature of the Higgs, is that
the shift symmetry dictates the form of the higher-dimensional
operators. The flavor structure and the composite Higgs coset
structure completely factorize if the SM fields couple to only one
composite operator. The general decomposition of Higgs couplings then
becomes \cite{Agashe:2009di} (see also
\cite{Gillioz:2012se,Delaunay:2013iia})
\begin{equation}\label{eq:comp}
	Y_u \bar{Q}_L H u_R + Y_u^\prime\bar{Q}_L H u_R
        \frac{(H^\dagger H)}{\Lambda^2}+\ldots \quad \to \quad
        c_{ij}^u \, P(h/f) \, \bar{Q}_L^i H u_R^j \,,
\end{equation}
and similarly for the down quarks. Here $P(h/f)=a_0+a_2 (h/f)^2+\ldots
$ is an analytic function whose form is fixed by the structure of the
spontaneous breaking and the embedding of the SM fields in the global
symmetry of the strongly coupled sector, while $f$ is the equivalent
of the pion decay constant and is of order $v_W$. Since the flavor
structure of the coefficients of the dimension-four and dimension-six
operators is the same, they can be diagonalized simultaneously. All
corrections to the quark Yukawa couplings from this effect are
therefore strictly diagonal, and we have
\begin{equation}\label{eq:kappaq:estimate}
\kappa_q\sim 1+{\mathcal O}\Big(\frac{v_W^2}{f^2}\Big)\,.
\end{equation}
For example, for the models based on the breaking of $SO(5)$ to
$SO(4)$, the diagonal Yukawa couplings can be written
as~\cite{Pomarol:2012qf}
\begin{equation}
\kappa_q=\frac{1+2m-(1+2m+n)(v_W/f)^2}{\sqrt{1-(v_W/f)^2}}\,,
\end{equation}
where $n,m$ are positive integers. The MCHM4 model corresponds to
$m=n=0$, while MCHM5 is given by $m=0,n=1$.

The flavor-violating contributions to the quark Yukawa couplings then
arise only from corrections to the quark kinetic terms. That is, they
are related to dimension-six operators of the
form~\cite{Agashe:2009di}
\begin{equation}
\bar q_L i \SlashD q_L \frac{H^\dagger H}{\Lambda^2}, \,\,
\bar u_R i \SlashD u_R \frac{H^\dagger H}{\Lambda^2},
\dots\,. 
\end{equation}
These operators arise from the exchange of composite vector resonances
with typical mass $M_* \sim \Lambda$. After using the equations of
motion they contribute to the misalignment between the fermion masses
and the corresponding Yukawa couplings. The NDA estimates for these
corrections are, neglecting relative ${\mathcal O}(1)$ contributions
in the sum~\cite{Agashe:2009di, Azatov:2014lha, Delaunay:2013pja},
\begin{equation}
\kappa_{ij}^u\sim 2 y_*^2 \frac{v_W^2}{M_*^2}
\Big(\lambda_{L,i}^q\lambda_{L,j}^q \frac{m_{u_j}}{v_W}
+\lambda_{R,i}^u\lambda_{R,j}^u \frac{m_{u_i}}{v_W}\Big)\,,
\end{equation}
and similarly for the down quarks. If the strong sector is $CP$
violating, then $\tilde \kappa_{ij}^{u,d}\sim \kappa_{ij}^{u,d}$.

The exchange of composite vector resonances contributes also to the
flavor diagonal Yukawa couplings, shifting the
estimate~\eqref{eq:kappaq:estimate} by (note the different
normalizations of $\kappa_q$ and $\kappa_{qq'}$ in \eqref{eq:Lh:CPV2})
\begin{equation}
\Delta \kappa_{q_i}\sim 2 y_*^2 \frac{v_W^2}{M_*^2}
\Big[\big(\lambda_{L,i}^q\big)^2+ \big(\lambda_{R,i}^u\big)^2\Big] \,.
\end{equation}
This shift can be large for the quarks with a large composite
component if the Higgs is strongly coupled to the vector resonances,
$y_*\sim 4\pi$, and these resonances are relatively light, $M_*\sim
4\pi v_W\sim 3$ TeV. The left-handed top and bottom, as well as the
right-handed top, are expected to be composite, explaining the large
top mass (i.e., $\lambda_{L,3}^q\sim \lambda_{R,3}^u\sim 1$). In the
anarchic flavor scenario, one expects the remaining quarks to be
mostly elementary (so the remaining $\lambda_i\ll 1$). However, if
there is some underlying flavor alignment, it is also possible that
the light quarks are composite. This is most easily achieved in the
right-handed sector~\cite{Redi:2011zi, Redi:2012uj,
  Delaunay:2013iia}. Taking all right-handed up-type quarks fully
composite, and assuming that this results in a shift $\Delta
\kappa_{u}\sim \Delta \kappa_{c}\sim \Delta \kappa_{t}\sim 1$, this
would lead to an increase in the effective Higgs coupling to nucleons,
$f_{\cal S}^{(N)}$, of about 50\%, and an increase in the DM-nucleon
scattering rate of about 100\%.

\section{Constraining the light-quark Yukawa couplings}\label{sec:light_quark_yukawas}
If DM is a thermal relic interacting with ordinary matter
predominantly via SM Higgs exchange, direct detection scattering rates
immediately give information about the light-quark Yukawa couplings
once the coupling of the DM particle to the Higgs particle is fixed.

In fact, DM scattering in direct detection searches would be one of
the very few possible probes of the light-quark Yukawa couplings. The
interactions of the Higgs boson with $u$, $d$, or $s$ quarks give rise
to flavor-conserving neutral currents. Off-shell Higgs contributions
in processes with only SM external particles always compete with
other, much larger flavor-conserving neutral currents induced by
gluon, photon, or $Z$ exchange. This leaves us with two options:
either to consider on-shell Higgs decays~\cite{Kagan:2014ila,
  Perez:2015aoa, Delaunay:2013pja}, or to use new probes, such as DM
scattering in direct detection experiments.

In principle, there is enough information to make a closed
argument. Suppose that indirect DM searches yield a positive DM
annihilation signal for $m_\chi>m_h/2$. At the end of the
high-luminosity LHC run, the Higgs couplings to $W$, $Z$, $t$, and $b$
will be precisely determined. Assuming that DM is a thermal relic
interacting only through the Higgs portal, this fixes the value of
$g_\chi$ since the annihilation cross section for $m_\chi>m_h/2$ is
otherwise almost completely controlled by the Higgs couplings to $W$,
$Z$, and $t$. In principle, a consistency check that the DM is really
interacting through a Higgs portal could be provided, for a very
limited range of DM masses $m_\chi\gtrsim m_h/2$, by a 100-TeV hadron
collider~\cite{Craig:2014lda}.

\begin{figure}[t]
\begin{center}
\includegraphics[width=0.49\textwidth]{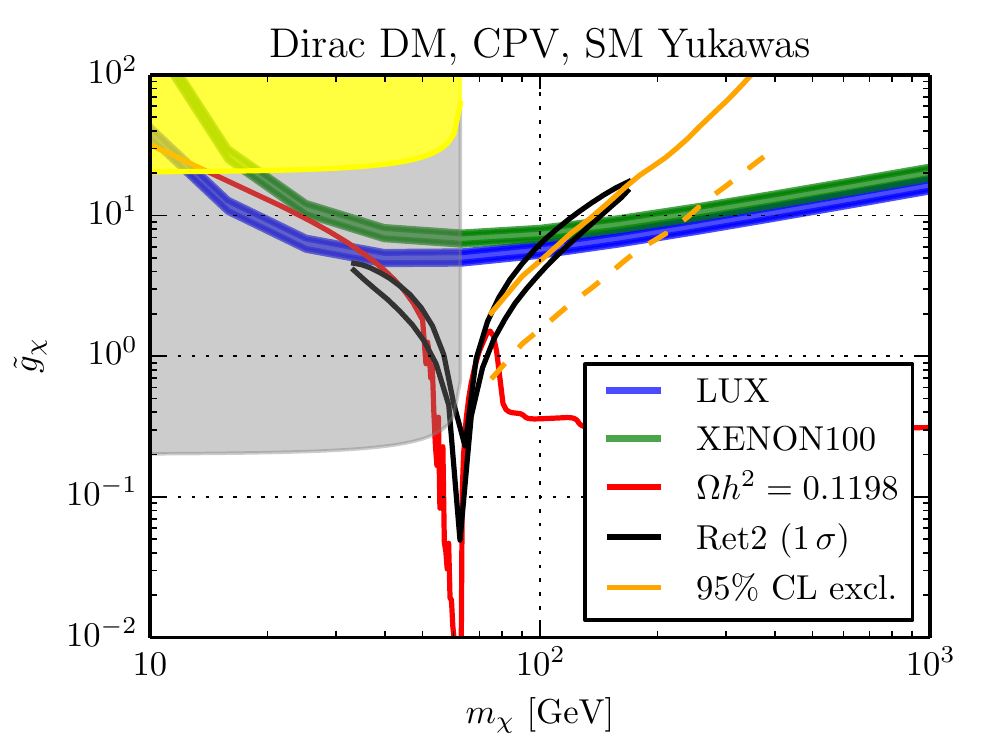}
\end{center}
\caption{The $\gamma$-ray excess in the recently discovered dwarf
  spheroidal galaxy Reticulum~2, interpreted as a signal of DM
  annihilating into $b \bar b$ pairs, is shown as the black $1 \sigma$
  contour (see Ref.~\cite{Geringer-Sameth:2015lua} for details). The
  orange lines show the 95\% CL exclusion limits at the 14-TeV LHC
  (solid line) and a prospective 100-TeV hadron collider (dashed
  line), obtained by rescaling the bounds given in
  Ref.~\cite{Craig:2014lda}. The remaining color coding is the same as
  in Fig.~\ref{fig:gchiscalar}. See text for more details.
  \label{fig:DM_example}
  }
\end{figure}

After the discovery of DM, the direct detection searches would
immediately imply an upper bound on the light-quark Yukawa
couplings. As an illustration, consider the excess in $\gamma$-ray
emmission in the recently discovered dwarf spheroidal galaxy
Reticulum~2~\cite{Drlica-Wagner:2015xua}. Let's take the bold step of
interpreting this signal as originating from DM annihilating into $b
\bar b$ pairs (see Ref.~\cite{Geringer-Sameth:2015lua} for details).
Assuming the Dirac-fermion DM scenario with purely CP-violating
couplings, we obtain a $1\sigma$ region in the $m_\chi$ -- $\tilde
g_\chi$ plane that is not yet excluded by direct detection
constraints, denoted by the orange lines in
Fig.~\ref{fig:DM_example}. Note that part of this region is consistent
with DM furnishing the dominant component of the observed relic
density while at the same time not being excluded by the invisible
Higgs decay width. Concentrating on the overlap region, $m_\chi \sim
75\,$GeV, a comparison with the ratios shown in Fig.~\ref{fig:ratio}
would immediately imply an upper bound of $\kappa_u \lesssim 10$,
$\kappa_d\lesssim 10$, $\kappa_s\lesssim 12$ from the LUX direct
detection search (allowing only one of the Yukawa couplings to float
at a time).

These estimates could potentially be loosened by uncertainties in the
DM velocity profile and the local DM density. On the other hand, if DM
is discovered in direct detection the relative size of the light-quark
Yukawas could be probed by comparing scattering rates on different
target materials.

An additional cross check of our scenario could be provided by
searches for DM production at hadron colliders. In
Fig.~\ref{fig:DM_example} we denote the 95\% CL exclusion limits,
assuming 3000/fb of data, at the 14-TeV LHC by a solid orange line and
at a prospective 100-TeV hadron collider by a dashed orange
line. These curves have been obtained by converting the bounds in
Ref.~\cite{Craig:2014lda} to the case of Dirac DM using
\texttt{FeynRules}~\cite{Alloul:2013bka} and
\texttt{MadGraph5}~\cite{Alwall:2014hca}. We see that, while the LHC
will be sensitive to a part of the interesting region in parameter
space, the scenario of $m_\chi = 75\,$GeV DM can be excluded only at a
100-TeV collider.

\section{Conclusions}\label{sec:conclusions}
Not much is known experimentally about the couplings of the Higgs to
light quarks. It is entirely possible that the Higgs couples only to
the third generation of fermions. Experimentally equally viable is the
possibility that the light-quark Yukawas are significantly enhanced,
up to ${\mathcal O}(50)$ for $\kappa_s$ and up to ${\mathcal O}(10^3)$
for $\kappa_u$ and $\kappa_d$. Such extremely large enhancements are
not natural from a model-building point of view as they require a
large fine tuning of the light-quark masses, but at present cannot be
excluded experimentally.

Modified Yukawa couplings to light quarks could have implications for
DM searches. In this paper we focused on Higgs-portal DM. We
considered constraints on scalar, vector, and fermionic Higgs-portal
models of DM from relic density, direct and indirect detection, and
the invisible Higgs width. A central result of our analysis is that,
for phenomenologically viable Higgs-portal DM, there is a relatively
small change in the predictions between the case where the Higgs is SM
like and the case where the Higgs couples only to the third generation
of fermions. For direct detection this is a consequence of the fact
that, for very small light-quark Yukawas, the scattering cross section
is dominated by the effective Higgs-gluon coupling, which is obtained
by integrating out the heavy quarks. For instance, setting all the
light quark Yukawa couplings to zero reduces the direct detection
scattering cross section by $\sim 50\%$ compared to the case where the
light quark Yukawa couplings are assumed to have SM values. Similarly,
the relic abundance and indirect detection signals are dominated by
the heaviest kinematically open annihilation channels, diminishing the
importance of Higgs couplings to light quarks.

On the other hand, saturating the experimentally allowed values for
the light-quark Yukawas, the DM direct detection rates can increase by
four orders of magnitude compared to the case where the light-quark
Yukawa couplings are kept at their SM values. Moreover, negative
values of the light-quark Yukawa couplings can result in a strong
reduction of the nucleon scattering rates.
The changes in DM annihilation rates are much smaller. The
annihilation of DM into light quarks is a subleading effect, unless
$m_\chi<m_W$. Even in this case, the dominant annihilation channel is
into $b\bar b$ pairs, while the annihilation to light quarks can
constitute at most an ${\mathcal O}(1)$ fraction if the current
experimental upper bounds on the light-quark Yukawa couplings are
saturated. A Higgs-portal for DM in this mass range is excluded either
by bounds on the invisible Higgs decay width or by indirect DM
searches.

We also investigated the expected sizes of corrections to DM
phenomenology due to changes in Yukawa couplings in a number of new
physics models. The largest deviation in expected DM scattering rate
on nucleons was found for a modified Giudice-Lebedev model of
light-quark masses where up to an order of magnitude enhancement due
to corrections to light-quark Yukawa couplings are
possible. Similarly, an ${\mathcal O}(1)$ change of the scattering
rate is anticipated in a pseudo-Goldstone Higgs scenario with
composite right-handed light quarks while in RS models with anarchic
flavor a reduction of about $20\%$ can be expected. The effects in MFV
models, multi-Higgs models with natural flavor conservation, and the
type-II two-Higgs-doublet model (i.e., the tree-level Higgs sector of
the MSSM), on the other hand, are expected to be much smaller.

Finally, we point out that a discovery of Higgs-portal DM in
indirect searches would immediately imply an upper bound on the
light-quark Yukawa couplings due to the upper bounds in direct DM
searches. 
 
\mysection{Acknowledgements}
We thank Nathaniel Craig and Matthew McCullough for the discussion
about the exclusions of the Higgs portal above the Higgs resonance. We
also thank Paddy Fox for the discussion about the UV completion of the
Higgs portal vector DM, and Andreas Crivellin for reminding us of the
significance of negative Yukawa couplings for direct DM
detection. J.~Z. is supported in part by the U.S. National Science
Foundation under CAREER Grant PHY-1151392. The research of J.~B.  is
supported by the ERC Advanced Grant EFT4LHC of the European Research
Council and the Cluster of Excellence Precision Physics, Fundamental
Interactions and Structure of Matter (PRISMA-EXC 1098).  F.~B. is
grateful to the Fermilab theory group for their hospitality.

\bibliography{paper}

\begin{thebibliography}{116}
\expandafter\ifx\csname natexlab\endcsname\relax\def\natexlab#1{#1}\fi
\expandafter\ifx\csname bibnamefont\endcsname\relax
  \def\bibnamefont#1{#1}\fi
\expandafter\ifx\csname bibfnamefont\endcsname\relax
  \def\bibfnamefont#1{#1}\fi
\expandafter\ifx\csname citenamefont\endcsname\relax
  \def\citenamefont#1{#1}\fi
\expandafter\ifx\csname url\endcsname\relax
  \def\url#1{\texttt{#1}}\fi
\expandafter\ifx\csname urlprefix\endcsname\relax\def\urlprefix{URL }\fi
\providecommand{\bibinfo}[2]{#2}
\providecommand{\eprint}[2][]{\url{#2}}

\bibitem[{\citenamefont{Patt and Wilczek}(2006)}]{Patt:2006fw}
\bibinfo{author}{\bibfnamefont{B.}~\bibnamefont{Patt}} \bibnamefont{and}
  \bibinfo{author}{\bibfnamefont{F.}~\bibnamefont{Wilczek}}
  (\bibinfo{year}{2006}), \eprint{hep-ph/0605188}.

\bibitem[{\citenamefont{March-Russell et~al.}(2008)\citenamefont{March-Russell,
  West, Cumberbatch, and Hooper}}]{MarchRussell:2008yu}
\bibinfo{author}{\bibfnamefont{J.}~\bibnamefont{March-Russell}},
  \bibinfo{author}{\bibfnamefont{S.~M.} \bibnamefont{West}},
  \bibinfo{author}{\bibfnamefont{D.}~\bibnamefont{Cumberbatch}},
  \bibnamefont{and} \bibinfo{author}{\bibfnamefont{D.}~\bibnamefont{Hooper}},
  \bibinfo{journal}{JHEP} \textbf{\bibinfo{volume}{0807}}, \bibinfo{pages}{058}
  (\bibinfo{year}{2008}), \eprint{0801.3440}.

\bibitem[{\citenamefont{Andreas et~al.}(2008)\citenamefont{Andreas, Hambye, and
  Tytgat}}]{Andreas:2008xy}
\bibinfo{author}{\bibfnamefont{S.}~\bibnamefont{Andreas}},
  \bibinfo{author}{\bibfnamefont{T.}~\bibnamefont{Hambye}}, \bibnamefont{and}
  \bibinfo{author}{\bibfnamefont{M.~H.} \bibnamefont{Tytgat}},
  \bibinfo{journal}{JCAP} \textbf{\bibinfo{volume}{0810}}, \bibinfo{pages}{034}
  (\bibinfo{year}{2008}), \eprint{0808.0255}.

\bibitem[{\citenamefont{Englert et~al.}(2011)\citenamefont{Englert, Plehn,
  Zerwas, and Zerwas}}]{Englert:2011yb}
\bibinfo{author}{\bibfnamefont{C.}~\bibnamefont{Englert}},
  \bibinfo{author}{\bibfnamefont{T.}~\bibnamefont{Plehn}},
  \bibinfo{author}{\bibfnamefont{D.}~\bibnamefont{Zerwas}}, \bibnamefont{and}
  \bibinfo{author}{\bibfnamefont{P.~M.} \bibnamefont{Zerwas}},
  \bibinfo{journal}{Phys.Lett.} \textbf{\bibinfo{volume}{B703}},
  \bibinfo{pages}{298} (\bibinfo{year}{2011}), \eprint{1106.3097}.

\bibitem[{\citenamefont{Lebedev et~al.}(2012)\citenamefont{Lebedev, Lee, and
  Mambrini}}]{Lebedev:2011iq}
\bibinfo{author}{\bibfnamefont{O.}~\bibnamefont{Lebedev}},
  \bibinfo{author}{\bibfnamefont{H.~M.} \bibnamefont{Lee}}, \bibnamefont{and}
  \bibinfo{author}{\bibfnamefont{Y.}~\bibnamefont{Mambrini}},
  \bibinfo{journal}{Phys.Lett.} \textbf{\bibinfo{volume}{B707}},
  \bibinfo{pages}{570} (\bibinfo{year}{2012}), \eprint{1111.4482}.

\bibitem[{\citenamefont{Lopez-Honorez et~al.}(2012)\citenamefont{Lopez-Honorez,
  Schwetz, and Zupan}}]{LopezHonorez:2012kv}
\bibinfo{author}{\bibfnamefont{L.}~\bibnamefont{Lopez-Honorez}},
  \bibinfo{author}{\bibfnamefont{T.}~\bibnamefont{Schwetz}}, \bibnamefont{and}
  \bibinfo{author}{\bibfnamefont{J.}~\bibnamefont{Zupan}},
  \bibinfo{journal}{Phys.Lett.} \textbf{\bibinfo{volume}{B716}},
  \bibinfo{pages}{179} (\bibinfo{year}{2012}), \eprint{1203.2064}.

\bibitem[{\citenamefont{Djouadi et~al.}(2013)\citenamefont{Djouadi, Falkowski,
  Mambrini, and Quevillon}}]{Djouadi:2012zc}
\bibinfo{author}{\bibfnamefont{A.}~\bibnamefont{Djouadi}},
  \bibinfo{author}{\bibfnamefont{A.}~\bibnamefont{Falkowski}},
  \bibinfo{author}{\bibfnamefont{Y.}~\bibnamefont{Mambrini}}, \bibnamefont{and}
  \bibinfo{author}{\bibfnamefont{J.}~\bibnamefont{Quevillon}},
  \bibinfo{journal}{Eur.Phys.J.} \textbf{\bibinfo{volume}{C73}},
  \bibinfo{pages}{2455} (\bibinfo{year}{2013}), \eprint{1205.3169}.

\bibitem[{\citenamefont{Greljo et~al.}(2013)\citenamefont{Greljo, Julio,
  Kamenik, Smith, and Zupan}}]{Greljo:2013wja}
\bibinfo{author}{\bibfnamefont{A.}~\bibnamefont{Greljo}},
  \bibinfo{author}{\bibfnamefont{J.}~\bibnamefont{Julio}},
  \bibinfo{author}{\bibfnamefont{J.~F.} \bibnamefont{Kamenik}},
  \bibinfo{author}{\bibfnamefont{C.}~\bibnamefont{Smith}}, \bibnamefont{and}
  \bibinfo{author}{\bibfnamefont{J.}~\bibnamefont{Zupan}},
  \bibinfo{journal}{JHEP} \textbf{\bibinfo{volume}{1311}}, \bibinfo{pages}{190}
  (\bibinfo{year}{2013}), \eprint{1309.3561}.

\bibitem[{\citenamefont{Fedderke et~al.}(2014)\citenamefont{Fedderke, Chen,
  Kolb, and Wang}}]{Fedderke:2014wda}
\bibinfo{author}{\bibfnamefont{M.~A.} \bibnamefont{Fedderke}},
  \bibinfo{author}{\bibfnamefont{J.-Y.} \bibnamefont{Chen}},
  \bibinfo{author}{\bibfnamefont{E.~W.} \bibnamefont{Kolb}}, \bibnamefont{and}
  \bibinfo{author}{\bibfnamefont{L.-T.} \bibnamefont{Wang}},
  \bibinfo{journal}{JHEP} \textbf{\bibinfo{volume}{1408}}, \bibinfo{pages}{122}
  (\bibinfo{year}{2014}), \eprint{1404.2283}.

\bibitem[{\citenamefont{Craig et~al.}(2014)\citenamefont{Craig, Lou,
  McCullough, and Thalapillil}}]{Craig:2014lda}
\bibinfo{author}{\bibfnamefont{N.}~\bibnamefont{Craig}},
  \bibinfo{author}{\bibfnamefont{H.~K.} \bibnamefont{Lou}},
  \bibinfo{author}{\bibfnamefont{M.}~\bibnamefont{McCullough}},
  \bibnamefont{and}
  \bibinfo{author}{\bibfnamefont{A.}~\bibnamefont{Thalapillil}}
  (\bibinfo{year}{2014}), \eprint{1412.0258}.

\bibitem[{ATL(2013)}]{ATLAS-CONF-2013-034}
\bibinfo{type}{Tech. Rep.} \bibinfo{number}{ATLAS-CONF-2013-034},
  \bibinfo{institution}{CERN}, \bibinfo{address}{Geneva}
  (\bibinfo{year}{2013}).

\bibitem[{CMS(2013)}]{CMS-PAS-HIG-13-005}
\bibinfo{type}{Tech. Rep.} \bibinfo{number}{CMS-PAS-HIG-13-005},
  \bibinfo{institution}{CERN}, \bibinfo{address}{Geneva}
  (\bibinfo{year}{2013}).

\bibitem[{\citenamefont{Kagan et~al.}(2014)\citenamefont{Kagan, Perez,
  Petriello, Soreq, Stoynev et~al.}}]{Kagan:2014ila}
\bibinfo{author}{\bibfnamefont{A.~L.} \bibnamefont{Kagan}},
  \bibinfo{author}{\bibfnamefont{G.}~\bibnamefont{Perez}},
  \bibinfo{author}{\bibfnamefont{F.}~\bibnamefont{Petriello}},
  \bibinfo{author}{\bibfnamefont{Y.}~\bibnamefont{Soreq}},
  \bibinfo{author}{\bibfnamefont{S.}~\bibnamefont{Stoynev}},
  \bibnamefont{et~al.} (\bibinfo{year}{2014}), \eprint{1406.1722}.

\bibitem[{\citenamefont{Perez et~al.}(2015)\citenamefont{Perez, Soreq, Stamou,
  and Tobioka}}]{Perez:2015aoa}
\bibinfo{author}{\bibfnamefont{G.}~\bibnamefont{Perez}},
  \bibinfo{author}{\bibfnamefont{Y.}~\bibnamefont{Soreq}},
  \bibinfo{author}{\bibfnamefont{E.}~\bibnamefont{Stamou}}, \bibnamefont{and}
  \bibinfo{author}{\bibfnamefont{K.}~\bibnamefont{Tobioka}}
  (\bibinfo{year}{2015}), \eprint{1503.00290}.

\bibitem[{\citenamefont{Delaunay et~al.}(2014)\citenamefont{Delaunay, Golling,
  Perez, and Soreq}}]{Delaunay:2013pja}
\bibinfo{author}{\bibfnamefont{C.}~\bibnamefont{Delaunay}},
  \bibinfo{author}{\bibfnamefont{T.}~\bibnamefont{Golling}},
  \bibinfo{author}{\bibfnamefont{G.}~\bibnamefont{Perez}}, \bibnamefont{and}
  \bibinfo{author}{\bibfnamefont{Y.}~\bibnamefont{Soreq}},
  \bibinfo{journal}{Phys.Rev.} \textbf{\bibinfo{volume}{D89}},
  \bibinfo{pages}{033014} (\bibinfo{year}{2014}), \eprint{1310.7029}.

\bibitem[{\citenamefont{Efrati et~al.}(2015)\citenamefont{Efrati, Falkowski,
  and Soreq}}]{Efrati:2015eaa}
\bibinfo{author}{\bibfnamefont{A.}~\bibnamefont{Efrati}},
  \bibinfo{author}{\bibfnamefont{A.}~\bibnamefont{Falkowski}},
  \bibnamefont{and} \bibinfo{author}{\bibfnamefont{Y.}~\bibnamefont{Soreq}}
  (\bibinfo{year}{2015}), \eprint{1503.07872}.

\bibitem[{\citenamefont{Porto and Zee}(2008)}]{Porto:2007ed}
\bibinfo{author}{\bibfnamefont{R.~A.} \bibnamefont{Porto}} \bibnamefont{and}
  \bibinfo{author}{\bibfnamefont{A.}~\bibnamefont{Zee}},
  \bibinfo{journal}{Phys.Lett.} \textbf{\bibinfo{volume}{B666}},
  \bibinfo{pages}{491} (\bibinfo{year}{2008}), \eprint{0712.0448}.

\bibitem[{\citenamefont{Kile and Soni}(2011)}]{Kile:2011mn}
\bibinfo{author}{\bibfnamefont{J.}~\bibnamefont{Kile}} \bibnamefont{and}
  \bibinfo{author}{\bibfnamefont{A.}~\bibnamefont{Soni}},
  \bibinfo{journal}{Phys.Rev.} \textbf{\bibinfo{volume}{D84}},
  \bibinfo{pages}{035016} (\bibinfo{year}{2011}), \eprint{1104.5239}.

\bibitem[{\citenamefont{Agrawal et~al.}(2012)\citenamefont{Agrawal, Blanchet,
  Chacko, and Kilic}}]{Agrawal:2011ze}
\bibinfo{author}{\bibfnamefont{P.}~\bibnamefont{Agrawal}},
  \bibinfo{author}{\bibfnamefont{S.}~\bibnamefont{Blanchet}},
  \bibinfo{author}{\bibfnamefont{Z.}~\bibnamefont{Chacko}}, \bibnamefont{and}
  \bibinfo{author}{\bibfnamefont{C.}~\bibnamefont{Kilic}},
  \bibinfo{journal}{Phys.Rev.} \textbf{\bibinfo{volume}{D86}},
  \bibinfo{pages}{055002} (\bibinfo{year}{2012}), \eprint{1109.3516}.

\bibitem[{\citenamefont{Masina et~al.}(2012)\citenamefont{Masina, Panci, and
  Sannino}}]{Masina:2012hg}
\bibinfo{author}{\bibfnamefont{I.}~\bibnamefont{Masina}},
  \bibinfo{author}{\bibfnamefont{P.}~\bibnamefont{Panci}}, \bibnamefont{and}
  \bibinfo{author}{\bibfnamefont{F.}~\bibnamefont{Sannino}},
  \bibinfo{journal}{JCAP} \textbf{\bibinfo{volume}{1212}}, \bibinfo{pages}{002}
  (\bibinfo{year}{2012}), \eprint{1205.5918}.

\bibitem[{\citenamefont{Lopez-Honorez and Merlo}(2013)}]{Lopez-Honorez:2013wla}
\bibinfo{author}{\bibfnamefont{L.}~\bibnamefont{Lopez-Honorez}}
  \bibnamefont{and} \bibinfo{author}{\bibfnamefont{L.}~\bibnamefont{Merlo}},
  \bibinfo{journal}{Phys.Lett.} \textbf{\bibinfo{volume}{B722}},
  \bibinfo{pages}{135} (\bibinfo{year}{2013}), \eprint{1303.1087}.

\bibitem[{\citenamefont{Batell et~al.}(2014)\citenamefont{Batell, Lin, and
  Wang}}]{Batell:2013zwa}
\bibinfo{author}{\bibfnamefont{B.}~\bibnamefont{Batell}},
  \bibinfo{author}{\bibfnamefont{T.}~\bibnamefont{Lin}}, \bibnamefont{and}
  \bibinfo{author}{\bibfnamefont{L.-T.} \bibnamefont{Wang}},
  \bibinfo{journal}{JHEP} \textbf{\bibinfo{volume}{1401}}, \bibinfo{pages}{075}
  (\bibinfo{year}{2014}), \eprint{1309.4462}.

\bibitem[{\citenamefont{Agrawal
  et~al.}(2014{\natexlab{a}})\citenamefont{Agrawal, Batell, Hooper, and
  Lin}}]{Agrawal:2014una}
\bibinfo{author}{\bibfnamefont{P.}~\bibnamefont{Agrawal}},
  \bibinfo{author}{\bibfnamefont{B.}~\bibnamefont{Batell}},
  \bibinfo{author}{\bibfnamefont{D.}~\bibnamefont{Hooper}}, \bibnamefont{and}
  \bibinfo{author}{\bibfnamefont{T.}~\bibnamefont{Lin}},
  \bibinfo{journal}{Phys.Rev.} \textbf{\bibinfo{volume}{D90}},
  \bibinfo{pages}{063512} (\bibinfo{year}{2014}{\natexlab{a}}),
  \eprint{1404.1373}.

\bibitem[{\citenamefont{Agrawal
  et~al.}(2014{\natexlab{b}})\citenamefont{Agrawal, Blanke, and
  Gemmler}}]{Agrawal:2014aoa}
\bibinfo{author}{\bibfnamefont{P.}~\bibnamefont{Agrawal}},
  \bibinfo{author}{\bibfnamefont{M.}~\bibnamefont{Blanke}}, \bibnamefont{and}
  \bibinfo{author}{\bibfnamefont{K.}~\bibnamefont{Gemmler}},
  \bibinfo{journal}{JHEP} \textbf{\bibinfo{volume}{1410}}, \bibinfo{pages}{72}
  (\bibinfo{year}{2014}{\natexlab{b}}), \eprint{1405.6709}.

\bibitem[{\citenamefont{Hamze et~al.}(2015)\citenamefont{Hamze, Kilic, Koeller,
  Trendafilova, and Yu}}]{Hamze:2014wca}
\bibinfo{author}{\bibfnamefont{A.}~\bibnamefont{Hamze}},
  \bibinfo{author}{\bibfnamefont{C.}~\bibnamefont{Kilic}},
  \bibinfo{author}{\bibfnamefont{J.}~\bibnamefont{Koeller}},
  \bibinfo{author}{\bibfnamefont{C.}~\bibnamefont{Trendafilova}},
  \bibnamefont{and} \bibinfo{author}{\bibfnamefont{J.-H.} \bibnamefont{Yu}},
  \bibinfo{journal}{Phys.Rev.} \textbf{\bibinfo{volume}{D91}},
  \bibinfo{pages}{035009} (\bibinfo{year}{2015}), \eprint{1410.3030}.

\bibitem[{\citenamefont{Kile et~al.}(2014)\citenamefont{Kile, Kobach, and
  Soni}}]{Kile:2014jea}
\bibinfo{author}{\bibfnamefont{J.}~\bibnamefont{Kile}},
  \bibinfo{author}{\bibfnamefont{A.}~\bibnamefont{Kobach}}, \bibnamefont{and}
  \bibinfo{author}{\bibfnamefont{A.}~\bibnamefont{Soni}}
  (\bibinfo{year}{2014}), \eprint{1411.1407}.

\bibitem[{\citenamefont{Kilic et~al.}(2015)\citenamefont{Kilic, Klimek, and
  Yu}}]{Kilic:2015vka}
\bibinfo{author}{\bibfnamefont{C.}~\bibnamefont{Kilic}},
  \bibinfo{author}{\bibfnamefont{M.~D.} \bibnamefont{Klimek}},
  \bibnamefont{and} \bibinfo{author}{\bibfnamefont{J.-H.} \bibnamefont{Yu}}
  (\bibinfo{year}{2015}), \eprint{1501.02202}.

\bibitem[{\citenamefont{Calibbi et~al.}(2015)\citenamefont{Calibbi, Crivellin,
  and Zaldivar}}]{Calibbi:2015sfa}
\bibinfo{author}{\bibfnamefont{L.}~\bibnamefont{Calibbi}},
  \bibinfo{author}{\bibfnamefont{A.}~\bibnamefont{Crivellin}},
  \bibnamefont{and} \bibinfo{author}{\bibfnamefont{B.}~\bibnamefont{Zaldivar}}
  (\bibinfo{year}{2015}), \eprint{1501.07268}.

\bibitem[{\citenamefont{Agrawal et~al.}(2015)\citenamefont{Agrawal, Chacko,
  Kilic, and Verhaaren}}]{Agrawal:2015tfa}
\bibinfo{author}{\bibfnamefont{P.}~\bibnamefont{Agrawal}},
  \bibinfo{author}{\bibfnamefont{Z.}~\bibnamefont{Chacko}},
  \bibinfo{author}{\bibfnamefont{C.}~\bibnamefont{Kilic}}, \bibnamefont{and}
  \bibinfo{author}{\bibfnamefont{C.~B.} \bibnamefont{Verhaaren}}
  (\bibinfo{year}{2015}), \eprint{1503.03057}.

\bibitem[{\citenamefont{Kamenik and Zupan}(2011)}]{Kamenik:2011nb}
\bibinfo{author}{\bibfnamefont{J.~F.} \bibnamefont{Kamenik}} \bibnamefont{and}
  \bibinfo{author}{\bibfnamefont{J.}~\bibnamefont{Zupan}},
  \bibinfo{journal}{Phys.Rev.} \textbf{\bibinfo{volume}{D84}},
  \bibinfo{pages}{111502} (\bibinfo{year}{2011}), \eprint{1107.0623}.

\bibitem[{\citenamefont{Bishara and Zupan}(2015)}]{Bishara:2014gwa}
\bibinfo{author}{\bibfnamefont{F.}~\bibnamefont{Bishara}} \bibnamefont{and}
  \bibinfo{author}{\bibfnamefont{J.}~\bibnamefont{Zupan}},
  \bibinfo{journal}{JHEP} \textbf{\bibinfo{volume}{1501}}, \bibinfo{pages}{089}
  (\bibinfo{year}{2015}), \eprint{1408.3852}.

\bibitem[{\citenamefont{Kim and Zurek}(2014)}]{Kim:2013ivd}
\bibinfo{author}{\bibfnamefont{I.-W.} \bibnamefont{Kim}} \bibnamefont{and}
  \bibinfo{author}{\bibfnamefont{K.~M.} \bibnamefont{Zurek}},
  \bibinfo{journal}{Phys.Rev.} \textbf{\bibinfo{volume}{D89}},
  \bibinfo{pages}{035008} (\bibinfo{year}{2014}), \eprint{1310.2617}.

\bibitem[{\citenamefont{Harnik et~al.}(2013)\citenamefont{Harnik, Kopp, and
  Zupan}}]{Harnik:2012pb}
\bibinfo{author}{\bibfnamefont{R.}~\bibnamefont{Harnik}},
  \bibinfo{author}{\bibfnamefont{J.}~\bibnamefont{Kopp}}, \bibnamefont{and}
  \bibinfo{author}{\bibfnamefont{J.}~\bibnamefont{Zupan}},
  \bibinfo{journal}{JHEP} \textbf{\bibinfo{volume}{1303}}, \bibinfo{pages}{026}
  (\bibinfo{year}{2013}), \eprint{1209.1397}.

\bibitem[{\citenamefont{Khachatryan et~al.}(2015)}]{Khachatryan:2015kon}
\bibinfo{author}{\bibfnamefont{V.}~\bibnamefont{Khachatryan}}
  \bibnamefont{et~al.} (\bibinfo{collaboration}{CMS}) (\bibinfo{year}{2015}),
  \eprint{1502.07400}.

\bibitem[{\citenamefont{Blankenburg et~al.}(2012)\citenamefont{Blankenburg,
  Ellis, and Isidori}}]{Blankenburg:2012ex}
\bibinfo{author}{\bibfnamefont{G.}~\bibnamefont{Blankenburg}},
  \bibinfo{author}{\bibfnamefont{J.}~\bibnamefont{Ellis}}, \bibnamefont{and}
  \bibinfo{author}{\bibfnamefont{G.}~\bibnamefont{Isidori}},
  \bibinfo{journal}{Phys.Lett.} \textbf{\bibinfo{volume}{B712}},
  \bibinfo{pages}{386} (\bibinfo{year}{2012}), \eprint{1202.5704}.

\bibitem[{\citenamefont{Goudelis et~al.}(2012)\citenamefont{Goudelis, Lebedev,
  and Park}}]{Goudelis:2011un}
\bibinfo{author}{\bibfnamefont{A.}~\bibnamefont{Goudelis}},
  \bibinfo{author}{\bibfnamefont{O.}~\bibnamefont{Lebedev}}, \bibnamefont{and}
  \bibinfo{author}{\bibfnamefont{J.-h.} \bibnamefont{Park}},
  \bibinfo{journal}{Phys.Lett.} \textbf{\bibinfo{volume}{B707}},
  \bibinfo{pages}{369} (\bibinfo{year}{2012}), \eprint{1111.1715}.

\bibitem[{\citenamefont{Carmi et~al.}(2012)\citenamefont{Carmi, Falkowski,
  Kuflik, Volansky, and Zupan}}]{Carmi:2012in}
\bibinfo{author}{\bibfnamefont{D.}~\bibnamefont{Carmi}},
  \bibinfo{author}{\bibfnamefont{A.}~\bibnamefont{Falkowski}},
  \bibinfo{author}{\bibfnamefont{E.}~\bibnamefont{Kuflik}},
  \bibinfo{author}{\bibfnamefont{T.}~\bibnamefont{Volansky}}, \bibnamefont{and}
  \bibinfo{author}{\bibfnamefont{J.}~\bibnamefont{Zupan}},
  \bibinfo{journal}{JHEP} \textbf{\bibinfo{volume}{1210}}, \bibinfo{pages}{196}
  (\bibinfo{year}{2012}), \eprint{1207.1718}.

\bibitem[{\citenamefont{Brod et~al.}(2013)\citenamefont{Brod, Haisch, and
  Zupan}}]{Brod:2013cka}
\bibinfo{author}{\bibfnamefont{J.}~\bibnamefont{Brod}},
  \bibinfo{author}{\bibfnamefont{U.}~\bibnamefont{Haisch}}, \bibnamefont{and}
  \bibinfo{author}{\bibfnamefont{J.}~\bibnamefont{Zupan}},
  \bibinfo{journal}{JHEP} \textbf{\bibinfo{volume}{1311}}, \bibinfo{pages}{180}
  (\bibinfo{year}{2013}), \eprint{1310.1385}.

\bibitem[{\citenamefont{Goertz et~al.}(2013)\citenamefont{Goertz,
  Papaefstathiou, Yang, and Zurita}}]{Goertz:2013kp}
\bibinfo{author}{\bibfnamefont{F.}~\bibnamefont{Goertz}},
  \bibinfo{author}{\bibfnamefont{A.}~\bibnamefont{Papaefstathiou}},
  \bibinfo{author}{\bibfnamefont{L.~L.} \bibnamefont{Yang}}, \bibnamefont{and}
  \bibinfo{author}{\bibfnamefont{J.}~\bibnamefont{Zurita}},
  \bibinfo{journal}{JHEP} \textbf{\bibinfo{volume}{1306}}, \bibinfo{pages}{016}
  (\bibinfo{year}{2013}), \eprint{1301.3492}.

\bibitem[{\citenamefont{Ferreira~de Lima et~al.}(2014)\citenamefont{Ferreira~de
  Lima, Papaefstathiou, and Spannowsky}}]{deLima:2014dta}
\bibinfo{author}{\bibfnamefont{D.~E.} \bibnamefont{Ferreira~de Lima}},
  \bibinfo{author}{\bibfnamefont{A.}~\bibnamefont{Papaefstathiou}},
  \bibnamefont{and}
  \bibinfo{author}{\bibfnamefont{M.}~\bibnamefont{Spannowsky}},
  \bibinfo{journal}{JHEP} \textbf{\bibinfo{volume}{1408}}, \bibinfo{pages}{030}
  (\bibinfo{year}{2014}), \eprint{1404.7139}.

\bibitem[{\citenamefont{Olive et~al.}(2014)}]{Agashe:2014kda}
\bibinfo{author}{\bibfnamefont{K.}~\bibnamefont{Olive}} \bibnamefont{et~al.}
  (\bibinfo{collaboration}{Particle Data Group}), \bibinfo{journal}{Chin.Phys.}
  \textbf{\bibinfo{volume}{C38}}, \bibinfo{pages}{090001}
  (\bibinfo{year}{2014}).

\bibitem[{\citenamefont{Chatrchyan et~al.}(2014)}]{Chatrchyan:2014tja}
\bibinfo{author}{\bibfnamefont{S.}~\bibnamefont{Chatrchyan}}
  \bibnamefont{et~al.} (\bibinfo{collaboration}{CMS Collaboration}),
  \bibinfo{journal}{Eur.Phys.J.} \textbf{\bibinfo{volume}{C74}},
  \bibinfo{pages}{2980} (\bibinfo{year}{2014}), \eprint{1404.1344}.

\bibitem[{\citenamefont{Aad et~al.}(2014)}]{Aad:2014iia}
\bibinfo{author}{\bibfnamefont{G.}~\bibnamefont{Aad}} \bibnamefont{et~al.}
  (\bibinfo{collaboration}{ATLAS Collaboration}),
  \bibinfo{journal}{Phys.Rev.Lett.} \textbf{\bibinfo{volume}{112}},
  \bibinfo{pages}{201802} (\bibinfo{year}{2014}), \eprint{1402.3244}.

\bibitem[{\citenamefont{Endo and Takaesu}(2015)}]{Endo:2014cca}
\bibinfo{author}{\bibfnamefont{M.}~\bibnamefont{Endo}} \bibnamefont{and}
  \bibinfo{author}{\bibfnamefont{Y.}~\bibnamefont{Takaesu}},
  \bibinfo{journal}{Phys.Lett.} \textbf{\bibinfo{volume}{B743}},
  \bibinfo{pages}{228} (\bibinfo{year}{2015}), \eprint{1407.6882}.

\bibitem[{\citenamefont{Ackermann
  et~al.}(2015{\natexlab{a}})}]{Ackermann:2015zua}
\bibinfo{author}{\bibfnamefont{M.}~\bibnamefont{Ackermann}}
  \bibnamefont{et~al.} (\bibinfo{collaboration}{Fermi-LAT})
  (\bibinfo{year}{2015}{\natexlab{a}}), \eprint{1503.02641}.

\bibitem[{\citenamefont{Drlica-Wagner et~al.}(2015)}]{Drlica-Wagner:2015xua}
\bibinfo{author}{\bibfnamefont{A.}~\bibnamefont{Drlica-Wagner}}
  \bibnamefont{et~al.} (\bibinfo{collaboration}{Fermi-LAT, DES}),
  \bibinfo{journal}{Astrophys.J.}  (\bibinfo{year}{2015}), \eprint{1503.02632}.

\bibitem[{\citenamefont{Ackermann
  et~al.}(2015{\natexlab{b}})}]{Ackermann:2015tah}
\bibinfo{author}{\bibfnamefont{M.}~\bibnamefont{Ackermann}}
  \bibnamefont{et~al.} (\bibinfo{collaboration}{Fermi-LAT})
  (\bibinfo{year}{2015}{\natexlab{b}}), \eprint{1501.05464}.

\bibitem[{\citenamefont{Cirelli et~al.}(2011)\citenamefont{Cirelli, Corcella,
  Hektor, Hutsi, Kadastik et~al.}}]{cirelli:2010xx}
\bibinfo{author}{\bibfnamefont{M.}~\bibnamefont{Cirelli}},
  \bibinfo{author}{\bibfnamefont{G.}~\bibnamefont{Corcella}},
  \bibinfo{author}{\bibfnamefont{A.}~\bibnamefont{Hektor}},
  \bibinfo{author}{\bibfnamefont{G.}~\bibnamefont{Hutsi}},
  \bibinfo{author}{\bibfnamefont{M.}~\bibnamefont{Kadastik}},
  \bibnamefont{et~al.}, \bibinfo{journal}{JCAP}
  \textbf{\bibinfo{volume}{1103}}, \bibinfo{pages}{051} (\bibinfo{year}{2011}),
  \eprint{1012.4515}.

\bibitem[{\citenamefont{Jungman et~al.}(1996)\citenamefont{Jungman,
  Kamionkowski, and Griest}}]{Jungman:1995df}
\bibinfo{author}{\bibfnamefont{G.}~\bibnamefont{Jungman}},
  \bibinfo{author}{\bibfnamefont{M.}~\bibnamefont{Kamionkowski}},
  \bibnamefont{and} \bibinfo{author}{\bibfnamefont{K.}~\bibnamefont{Griest}},
  \bibinfo{journal}{Phys.Rept.} \textbf{\bibinfo{volume}{267}},
  \bibinfo{pages}{195} (\bibinfo{year}{1996}), \eprint{hep-ph/9506380}.

\bibitem[{\citenamefont{Shifman et~al.}(1978)\citenamefont{Shifman, Vainshtein,
  and Zakharov}}]{Shifman:1978zn}
\bibinfo{author}{\bibfnamefont{M.~A.} \bibnamefont{Shifman}},
  \bibinfo{author}{\bibfnamefont{A.}~\bibnamefont{Vainshtein}},
  \bibnamefont{and} \bibinfo{author}{\bibfnamefont{V.~I.}
  \bibnamefont{Zakharov}}, \bibinfo{journal}{Phys.Lett.}
  \textbf{\bibinfo{volume}{B78}}, \bibinfo{pages}{443} (\bibinfo{year}{1978}).

\bibitem[{\citenamefont{Akerib et~al.}(2014)}]{Akerib:2013tjd}
\bibinfo{author}{\bibfnamefont{D.}~\bibnamefont{Akerib}} \bibnamefont{et~al.}
  (\bibinfo{collaboration}{LUX Collaboration}),
  \bibinfo{journal}{Phys.Rev.Lett.} \textbf{\bibinfo{volume}{112}},
  \bibinfo{pages}{091303} (\bibinfo{year}{2014}), \eprint{1310.8214}.

\bibitem[{\citenamefont{Aprile et~al.}(2012)}]{Aprile:2012nq}
\bibinfo{author}{\bibfnamefont{E.}~\bibnamefont{Aprile}} \bibnamefont{et~al.}
  (\bibinfo{collaboration}{XENON100 Collaboration}),
  \bibinfo{journal}{Phys.Rev.Lett.} \textbf{\bibinfo{volume}{109}},
  \bibinfo{pages}{181301} (\bibinfo{year}{2012}), \eprint{1207.5988}.

\bibitem[{\citenamefont{Junnarkar and Walker-Loud}(2013)}]{Junnarkar:2013ac}
\bibinfo{author}{\bibfnamefont{P.}~\bibnamefont{Junnarkar}} \bibnamefont{and}
  \bibinfo{author}{\bibfnamefont{A.}~\bibnamefont{Walker-Loud}},
  \bibinfo{journal}{Phys.Rev.} \textbf{\bibinfo{volume}{D87}},
  \bibinfo{pages}{114510} (\bibinfo{year}{2013}), \eprint{1301.1114}.

\bibitem[{\citenamefont{Alarcon et~al.}(2012)\citenamefont{Alarcon,
  Martin~Camalich, and Oller}}]{Alarcon:2011zs}
\bibinfo{author}{\bibfnamefont{J.}~\bibnamefont{Alarcon}},
  \bibinfo{author}{\bibfnamefont{J.}~\bibnamefont{Martin~Camalich}},
  \bibnamefont{and} \bibinfo{author}{\bibfnamefont{J.}~\bibnamefont{Oller}},
  \bibinfo{journal}{Phys.Rev.} \textbf{\bibinfo{volume}{D85}},
  \bibinfo{pages}{051503} (\bibinfo{year}{2012}), \eprint{1110.3797}.

\bibitem[{\citenamefont{Alvarez-Ruso et~al.}(2014)\citenamefont{Alvarez-Ruso,
  Ledwig, Martin~Camalich, and Vicente~Vacas}}]{Alvarez-Ruso:2014sma}
\bibinfo{author}{\bibfnamefont{L.}~\bibnamefont{Alvarez-Ruso}},
  \bibinfo{author}{\bibfnamefont{T.}~\bibnamefont{Ledwig}},
  \bibinfo{author}{\bibfnamefont{J.}~\bibnamefont{Martin~Camalich}},
  \bibnamefont{and}
  \bibinfo{author}{\bibfnamefont{M.}~\bibnamefont{Vicente~Vacas}},
  \bibinfo{journal}{EPJ Web Conf.} \textbf{\bibinfo{volume}{73}},
  \bibinfo{pages}{04015} (\bibinfo{year}{2014}).

\bibitem[{\citenamefont{Crivellin
  et~al.}(2014{\natexlab{a}})\citenamefont{Crivellin, Hoferichter, and
  Procura}}]{Crivellin:2013ipa}
\bibinfo{author}{\bibfnamefont{A.}~\bibnamefont{Crivellin}},
  \bibinfo{author}{\bibfnamefont{M.}~\bibnamefont{Hoferichter}},
  \bibnamefont{and} \bibinfo{author}{\bibfnamefont{M.}~\bibnamefont{Procura}},
  \bibinfo{journal}{Phys.Rev.} \textbf{\bibinfo{volume}{D89}},
  \bibinfo{pages}{054021} (\bibinfo{year}{2014}{\natexlab{a}}),
  \eprint{1312.4951}.

\bibitem[{\citenamefont{Crivellin
  et~al.}(2014{\natexlab{b}})\citenamefont{Crivellin, D'Eramo, and
  Procura}}]{Crivellin:2014qxa}
\bibinfo{author}{\bibfnamefont{A.}~\bibnamefont{Crivellin}},
  \bibinfo{author}{\bibfnamefont{F.}~\bibnamefont{D'Eramo}}, \bibnamefont{and}
  \bibinfo{author}{\bibfnamefont{M.}~\bibnamefont{Procura}},
  \bibinfo{journal}{Phys.Rev.Lett.} \textbf{\bibinfo{volume}{112}},
  \bibinfo{pages}{191304} (\bibinfo{year}{2014}{\natexlab{b}}),
  \eprint{1402.1173}.

\bibitem[{\citenamefont{Fitzpatrick et~al.}(2013)\citenamefont{Fitzpatrick,
  Haxton, Katz, Lubbers, and Xu}}]{Fitzpatrick:2012ix}
\bibinfo{author}{\bibfnamefont{A.~L.} \bibnamefont{Fitzpatrick}},
  \bibinfo{author}{\bibfnamefont{W.}~\bibnamefont{Haxton}},
  \bibinfo{author}{\bibfnamefont{E.}~\bibnamefont{Katz}},
  \bibinfo{author}{\bibfnamefont{N.}~\bibnamefont{Lubbers}}, \bibnamefont{and}
  \bibinfo{author}{\bibfnamefont{Y.}~\bibnamefont{Xu}}, \bibinfo{journal}{JCAP}
  \textbf{\bibinfo{volume}{1302}}, \bibinfo{pages}{004} (\bibinfo{year}{2013}),
  \eprint{1203.3542}.

\bibitem[{\citenamefont{Glashow and Weinberg}(1977)}]{Glashow:1976nt}
\bibinfo{author}{\bibfnamefont{S.~L.} \bibnamefont{Glashow}} \bibnamefont{and}
  \bibinfo{author}{\bibfnamefont{S.}~\bibnamefont{Weinberg}},
  \bibinfo{journal}{Phys.Rev.} \textbf{\bibinfo{volume}{D15}},
  \bibinfo{pages}{1958} (\bibinfo{year}{1977}).

\bibitem[{\citenamefont{Paschos}(1977)}]{Paschos:1976ay}
\bibinfo{author}{\bibfnamefont{E.}~\bibnamefont{Paschos}},
  \bibinfo{journal}{Phys.Rev.} \textbf{\bibinfo{volume}{D15}},
  \bibinfo{pages}{1966} (\bibinfo{year}{1977}).

\bibitem[{\citenamefont{Giudice and Lebedev}(2008)}]{Giudice:2008uua}
\bibinfo{author}{\bibfnamefont{G.~F.} \bibnamefont{Giudice}} \bibnamefont{and}
  \bibinfo{author}{\bibfnamefont{O.}~\bibnamefont{Lebedev}},
  \bibinfo{journal}{Phys.Lett.} \textbf{\bibinfo{volume}{B665}},
  \bibinfo{pages}{79} (\bibinfo{year}{2008}), \eprint{0804.1753}.

\bibitem[{\citenamefont{D'Ambrosio et~al.}(2002)\citenamefont{D'Ambrosio,
  Giudice, Isidori, and Strumia}}]{D'Ambrosio:2002ex}
\bibinfo{author}{\bibfnamefont{G.}~\bibnamefont{D'Ambrosio}},
  \bibinfo{author}{\bibfnamefont{G.}~\bibnamefont{Giudice}},
  \bibinfo{author}{\bibfnamefont{G.}~\bibnamefont{Isidori}}, \bibnamefont{and}
  \bibinfo{author}{\bibfnamefont{A.}~\bibnamefont{Strumia}},
  \bibinfo{journal}{Nucl.Phys.} \textbf{\bibinfo{volume}{B645}},
  \bibinfo{pages}{155} (\bibinfo{year}{2002}), \eprint{hep-ph/0207036}.

\bibitem[{\citenamefont{Randall and Sundrum}(1999)}]{Randall:1999ee}
\bibinfo{author}{\bibfnamefont{L.}~\bibnamefont{Randall}} \bibnamefont{and}
  \bibinfo{author}{\bibfnamefont{R.}~\bibnamefont{Sundrum}},
  \bibinfo{journal}{Phys.Rev.Lett.} \textbf{\bibinfo{volume}{83}},
  \bibinfo{pages}{3370} (\bibinfo{year}{1999}), \eprint{hep-ph/9905221}.

\bibitem[{\citenamefont{Dugan et~al.}(1985)\citenamefont{Dugan, Georgi, and
  Kaplan}}]{Dugan:1984hq}
\bibinfo{author}{\bibfnamefont{M.~J.} \bibnamefont{Dugan}},
  \bibinfo{author}{\bibfnamefont{H.}~\bibnamefont{Georgi}}, \bibnamefont{and}
  \bibinfo{author}{\bibfnamefont{D.~B.} \bibnamefont{Kaplan}},
  \bibinfo{journal}{Nucl.Phys.} \textbf{\bibinfo{volume}{B254}},
  \bibinfo{pages}{299} (\bibinfo{year}{1985}).

\bibitem[{\citenamefont{Georgi et~al.}(1984)\citenamefont{Georgi, Kaplan, and
  Galison}}]{Georgi:1984ef}
\bibinfo{author}{\bibfnamefont{H.}~\bibnamefont{Georgi}},
  \bibinfo{author}{\bibfnamefont{D.~B.} \bibnamefont{Kaplan}},
  \bibnamefont{and} \bibinfo{author}{\bibfnamefont{P.}~\bibnamefont{Galison}},
  \bibinfo{journal}{Phys.Lett.} \textbf{\bibinfo{volume}{B143}},
  \bibinfo{pages}{152} (\bibinfo{year}{1984}).

\bibitem[{\citenamefont{Kaplan et~al.}(1984)\citenamefont{Kaplan, Georgi, and
  Dimopoulos}}]{Kaplan:1983sm}
\bibinfo{author}{\bibfnamefont{D.~B.} \bibnamefont{Kaplan}},
  \bibinfo{author}{\bibfnamefont{H.}~\bibnamefont{Georgi}}, \bibnamefont{and}
  \bibinfo{author}{\bibfnamefont{S.}~\bibnamefont{Dimopoulos}},
  \bibinfo{journal}{Phys.Lett.} \textbf{\bibinfo{volume}{B136}},
  \bibinfo{pages}{187} (\bibinfo{year}{1984}).

\bibitem[{\citenamefont{Kaplan and Georgi}(1984)}]{Kaplan:1983fs}
\bibinfo{author}{\bibfnamefont{D.~B.} \bibnamefont{Kaplan}} \bibnamefont{and}
  \bibinfo{author}{\bibfnamefont{H.}~\bibnamefont{Georgi}},
  \bibinfo{journal}{Phys.Lett.} \textbf{\bibinfo{volume}{B136}},
  \bibinfo{pages}{183} (\bibinfo{year}{1984}).

\bibitem[{\citenamefont{Dery et~al.}(2014)\citenamefont{Dery, Efrati, Nir,
  Soreq, and Susi}}]{Dery:2014kxa}
\bibinfo{author}{\bibfnamefont{A.}~\bibnamefont{Dery}},
  \bibinfo{author}{\bibfnamefont{A.}~\bibnamefont{Efrati}},
  \bibinfo{author}{\bibfnamefont{Y.}~\bibnamefont{Nir}},
  \bibinfo{author}{\bibfnamefont{Y.}~\bibnamefont{Soreq}}, \bibnamefont{and}
  \bibinfo{author}{\bibfnamefont{V.}~\bibnamefont{Susi}},
  \bibinfo{journal}{Phys.Rev.} \textbf{\bibinfo{volume}{D90}},
  \bibinfo{pages}{115022} (\bibinfo{year}{2014}), \eprint{1408.1371}.

\bibitem[{\citenamefont{Dery et~al.}(2013{\natexlab{a}})\citenamefont{Dery,
  Efrati, Hiller, Hochberg, and Nir}}]{Dery:2013aba}
\bibinfo{author}{\bibfnamefont{A.}~\bibnamefont{Dery}},
  \bibinfo{author}{\bibfnamefont{A.}~\bibnamefont{Efrati}},
  \bibinfo{author}{\bibfnamefont{G.}~\bibnamefont{Hiller}},
  \bibinfo{author}{\bibfnamefont{Y.}~\bibnamefont{Hochberg}}, \bibnamefont{and}
  \bibinfo{author}{\bibfnamefont{Y.}~\bibnamefont{Nir}},
  \bibinfo{journal}{JHEP} \textbf{\bibinfo{volume}{1308}}, \bibinfo{pages}{006}
  (\bibinfo{year}{2013}{\natexlab{a}}), \eprint{1304.6727}.

\bibitem[{\citenamefont{Dery et~al.}(2013{\natexlab{b}})\citenamefont{Dery,
  Efrati, Hochberg, and Nir}}]{Dery:2013rta}
\bibinfo{author}{\bibfnamefont{A.}~\bibnamefont{Dery}},
  \bibinfo{author}{\bibfnamefont{A.}~\bibnamefont{Efrati}},
  \bibinfo{author}{\bibfnamefont{Y.}~\bibnamefont{Hochberg}}, \bibnamefont{and}
  \bibinfo{author}{\bibfnamefont{Y.}~\bibnamefont{Nir}},
  \bibinfo{journal}{JHEP} \textbf{\bibinfo{volume}{1305}}, \bibinfo{pages}{039}
  (\bibinfo{year}{2013}{\natexlab{b}}), \eprint{1302.3229}.

\bibitem[{\citenamefont{El~Hedri et~al.}(2013)\citenamefont{El~Hedri, Fox, and
  Wacker}}]{Hedri:2013wea}
\bibinfo{author}{\bibfnamefont{S.}~\bibnamefont{El~Hedri}},
  \bibinfo{author}{\bibfnamefont{P.~J.} \bibnamefont{Fox}}, \bibnamefont{and}
  \bibinfo{author}{\bibfnamefont{J.~G.} \bibnamefont{Wacker}}
  (\bibinfo{year}{2013}), \eprint{1311.6488}.

\bibitem[{\citenamefont{Goertz}(2014)}]{Goertz:2014qia}
\bibinfo{author}{\bibfnamefont{F.}~\bibnamefont{Goertz}},
  \bibinfo{journal}{Phys.Rev.Lett.} \textbf{\bibinfo{volume}{113}},
  \bibinfo{pages}{261803} (\bibinfo{year}{2014}), \eprint{1406.0102}.

\bibitem[{\citenamefont{Chivukula and Georgi}(1987)}]{Chivukula:1987py}
\bibinfo{author}{\bibfnamefont{R.~S.} \bibnamefont{Chivukula}}
  \bibnamefont{and} \bibinfo{author}{\bibfnamefont{H.}~\bibnamefont{Georgi}},
  \bibinfo{journal}{Phys.Lett.} \textbf{\bibinfo{volume}{B188}},
  \bibinfo{pages}{99} (\bibinfo{year}{1987}).

\bibitem[{\citenamefont{Gabrielli and Giudice}(1995)}]{Gabrielli:1994ff}
\bibinfo{author}{\bibfnamefont{E.}~\bibnamefont{Gabrielli}} \bibnamefont{and}
  \bibinfo{author}{\bibfnamefont{G.}~\bibnamefont{Giudice}},
  \bibinfo{journal}{Nucl.Phys.} \textbf{\bibinfo{volume}{B433}},
  \bibinfo{pages}{3} (\bibinfo{year}{1995}), \eprint{hep-lat/9407029}.

\bibitem[{\citenamefont{Ali and London}(1999)}]{Ali:1999we}
\bibinfo{author}{\bibfnamefont{A.}~\bibnamefont{Ali}} \bibnamefont{and}
  \bibinfo{author}{\bibfnamefont{D.}~\bibnamefont{London}},
  \bibinfo{journal}{Eur.Phys.J.} \textbf{\bibinfo{volume}{C9}},
  \bibinfo{pages}{687} (\bibinfo{year}{1999}), \eprint{hep-ph/9903535}.

\bibitem[{\citenamefont{Buras et~al.}(2001)\citenamefont{Buras, Gambino,
  Gorbahn, Jager, and Silvestrini}}]{Buras:2000dm}
\bibinfo{author}{\bibfnamefont{A.}~\bibnamefont{Buras}},
  \bibinfo{author}{\bibfnamefont{P.}~\bibnamefont{Gambino}},
  \bibinfo{author}{\bibfnamefont{M.}~\bibnamefont{Gorbahn}},
  \bibinfo{author}{\bibfnamefont{S.}~\bibnamefont{Jager}}, \bibnamefont{and}
  \bibinfo{author}{\bibfnamefont{L.}~\bibnamefont{Silvestrini}},
  \bibinfo{journal}{Phys.Lett.} \textbf{\bibinfo{volume}{B500}},
  \bibinfo{pages}{161} (\bibinfo{year}{2001}), \eprint{hep-ph/0007085}.

\bibitem[{\citenamefont{Buras}(2003)}]{Buras:2003jf}
\bibinfo{author}{\bibfnamefont{A.~J.} \bibnamefont{Buras}},
  \bibinfo{journal}{Acta Phys.Polon.} \textbf{\bibinfo{volume}{B34}},
  \bibinfo{pages}{5615} (\bibinfo{year}{2003}), \eprint{hep-ph/0310208}.

\bibitem[{\citenamefont{Kagan et~al.}(2009)\citenamefont{Kagan, Perez,
  Volansky, and Zupan}}]{Kagan:2009bn}
\bibinfo{author}{\bibfnamefont{A.~L.} \bibnamefont{Kagan}},
  \bibinfo{author}{\bibfnamefont{G.}~\bibnamefont{Perez}},
  \bibinfo{author}{\bibfnamefont{T.}~\bibnamefont{Volansky}}, \bibnamefont{and}
  \bibinfo{author}{\bibfnamefont{J.}~\bibnamefont{Zupan}},
  \bibinfo{journal}{Phys.Rev.} \textbf{\bibinfo{volume}{D80}},
  \bibinfo{pages}{076002} (\bibinfo{year}{2009}), \eprint{0903.1794}.

\bibitem[{\citenamefont{Aguilar-Saavedra}(2009)}]{AguilarSaavedra:2009mx}
\bibinfo{author}{\bibfnamefont{J.}~\bibnamefont{Aguilar-Saavedra}},
  \bibinfo{journal}{Nucl.Phys.} \textbf{\bibinfo{volume}{B821}},
  \bibinfo{pages}{215} (\bibinfo{year}{2009}), \eprint{0904.2387}.

\bibitem[{1244669()}]{CMS:2013xfa}
1244669 (\bibinfo{year}{2013}), \eprint{1307.7135}.

\bibitem[{1245017()}]{ATLAS:2013hta}
1245017 (\bibinfo{year}{2013}), \eprint{1307.7292}.

\bibitem[{\citenamefont{Falkowski et~al.}(2015)\citenamefont{Falkowski, Gross,
  and Lebedev}}]{Falkowski:2015iwa}
\bibinfo{author}{\bibfnamefont{A.}~\bibnamefont{Falkowski}},
  \bibinfo{author}{\bibfnamefont{C.}~\bibnamefont{Gross}}, \bibnamefont{and}
  \bibinfo{author}{\bibfnamefont{O.}~\bibnamefont{Lebedev}}
  (\bibinfo{year}{2015}), \eprint{1502.01361}.

\bibitem[{\citenamefont{Martin-Lozano et~al.}(2015)\citenamefont{Martin-Lozano,
  Moreno, and Park}}]{Martin-Lozano:2015dja}
\bibinfo{author}{\bibfnamefont{V.}~\bibnamefont{Martin-Lozano}},
  \bibinfo{author}{\bibfnamefont{J.~M.} \bibnamefont{Moreno}},
  \bibnamefont{and} \bibinfo{author}{\bibfnamefont{C.~B.} \bibnamefont{Park}}
  (\bibinfo{year}{2015}), \eprint{1501.03799}.

\bibitem[{\citenamefont{Haber and Kane}(1985)}]{Haber:1984rc}
\bibinfo{author}{\bibfnamefont{H.~E.} \bibnamefont{Haber}} \bibnamefont{and}
  \bibinfo{author}{\bibfnamefont{G.~L.} \bibnamefont{Kane}},
  \bibinfo{journal}{Phys.Rept.} \textbf{\bibinfo{volume}{117}},
  \bibinfo{pages}{75} (\bibinfo{year}{1985}).

\bibitem[{\citenamefont{Falkowski et~al.}(2013)\citenamefont{Falkowski, Riva,
  and Urbano}}]{Falkowski:2013dza}
\bibinfo{author}{\bibfnamefont{A.}~\bibnamefont{Falkowski}},
  \bibinfo{author}{\bibfnamefont{F.}~\bibnamefont{Riva}}, \bibnamefont{and}
  \bibinfo{author}{\bibfnamefont{A.}~\bibnamefont{Urbano}},
  \bibinfo{journal}{JHEP} \textbf{\bibinfo{volume}{1311}}, \bibinfo{pages}{111}
  (\bibinfo{year}{2013}), \eprint{1303.1812}.

\bibitem[{\citenamefont{Grinstein and Uttayarat}(2013)}]{Grinstein:2013npa}
\bibinfo{author}{\bibfnamefont{B.}~\bibnamefont{Grinstein}} \bibnamefont{and}
  \bibinfo{author}{\bibfnamefont{P.}~\bibnamefont{Uttayarat}},
  \bibinfo{journal}{JHEP} \textbf{\bibinfo{volume}{1306}}, \bibinfo{pages}{094}
  (\bibinfo{year}{2013}), \eprint{1304.0028}.

\bibitem[{\citenamefont{Gherghetta and Pomarol}(2000)}]{Gherghetta:2000qt}
\bibinfo{author}{\bibfnamefont{T.}~\bibnamefont{Gherghetta}} \bibnamefont{and}
  \bibinfo{author}{\bibfnamefont{A.}~\bibnamefont{Pomarol}},
  \bibinfo{journal}{Nucl.Phys.} \textbf{\bibinfo{volume}{B586}},
  \bibinfo{pages}{141} (\bibinfo{year}{2000}), \eprint{hep-ph/0003129}.

\bibitem[{\citenamefont{Grossman and Neubert}(2000)}]{Grossman:1999ra}
\bibinfo{author}{\bibfnamefont{Y.}~\bibnamefont{Grossman}} \bibnamefont{and}
  \bibinfo{author}{\bibfnamefont{M.}~\bibnamefont{Neubert}},
  \bibinfo{journal}{Phys.Lett.} \textbf{\bibinfo{volume}{B474}},
  \bibinfo{pages}{361} (\bibinfo{year}{2000}), \eprint{hep-ph/9912408}.

\bibitem[{\citenamefont{Huber and Shafi}(2001)}]{Huber:2000ie}
\bibinfo{author}{\bibfnamefont{S.~J.} \bibnamefont{Huber}} \bibnamefont{and}
  \bibinfo{author}{\bibfnamefont{Q.}~\bibnamefont{Shafi}},
  \bibinfo{journal}{Phys.Lett.} \textbf{\bibinfo{volume}{B498}},
  \bibinfo{pages}{256} (\bibinfo{year}{2001}), \eprint{hep-ph/0010195}.

\bibitem[{\citenamefont{Huber}(2003)}]{Huber:2003tu}
\bibinfo{author}{\bibfnamefont{S.~J.} \bibnamefont{Huber}},
  \bibinfo{journal}{Nucl.Phys.} \textbf{\bibinfo{volume}{B666}},
  \bibinfo{pages}{269} (\bibinfo{year}{2003}), \eprint{hep-ph/0303183}.

\bibitem[{\citenamefont{Azatov et~al.}(2009)\citenamefont{Azatov, Toharia, and
  Zhu}}]{Azatov:2009na}
\bibinfo{author}{\bibfnamefont{A.}~\bibnamefont{Azatov}},
  \bibinfo{author}{\bibfnamefont{M.}~\bibnamefont{Toharia}}, \bibnamefont{and}
  \bibinfo{author}{\bibfnamefont{L.}~\bibnamefont{Zhu}},
  \bibinfo{journal}{Phys.Rev.} \textbf{\bibinfo{volume}{D80}},
  \bibinfo{pages}{035016} (\bibinfo{year}{2009}), \eprint{0906.1990}.

\bibitem[{\citenamefont{Casagrande et~al.}(2008)\citenamefont{Casagrande,
  Goertz, Haisch, Neubert, and Pfoh}}]{Casagrande:2008hr}
\bibinfo{author}{\bibfnamefont{S.}~\bibnamefont{Casagrande}},
  \bibinfo{author}{\bibfnamefont{F.}~\bibnamefont{Goertz}},
  \bibinfo{author}{\bibfnamefont{U.}~\bibnamefont{Haisch}},
  \bibinfo{author}{\bibfnamefont{M.}~\bibnamefont{Neubert}}, \bibnamefont{and}
  \bibinfo{author}{\bibfnamefont{T.}~\bibnamefont{Pfoh}},
  \bibinfo{journal}{JHEP} \textbf{\bibinfo{volume}{0810}}, \bibinfo{pages}{094}
  (\bibinfo{year}{2008}), \eprint{0807.4937}.

\bibitem[{\citenamefont{Bauer et~al.}(2010)\citenamefont{Bauer, Casagrande,
  Haisch, and Neubert}}]{Bauer:2009cf}
\bibinfo{author}{\bibfnamefont{M.}~\bibnamefont{Bauer}},
  \bibinfo{author}{\bibfnamefont{S.}~\bibnamefont{Casagrande}},
  \bibinfo{author}{\bibfnamefont{U.}~\bibnamefont{Haisch}}, \bibnamefont{and}
  \bibinfo{author}{\bibfnamefont{M.}~\bibnamefont{Neubert}},
  \bibinfo{journal}{JHEP} \textbf{\bibinfo{volume}{1009}}, \bibinfo{pages}{017}
  (\bibinfo{year}{2010}), \eprint{0912.1625}.

\bibitem[{\citenamefont{Malm et~al.}(2014)\citenamefont{Malm, Neubert, Novotny,
  and Schmell}}]{Malm:2013jia}
\bibinfo{author}{\bibfnamefont{R.}~\bibnamefont{Malm}},
  \bibinfo{author}{\bibfnamefont{M.}~\bibnamefont{Neubert}},
  \bibinfo{author}{\bibfnamefont{K.}~\bibnamefont{Novotny}}, \bibnamefont{and}
  \bibinfo{author}{\bibfnamefont{C.}~\bibnamefont{Schmell}},
  \bibinfo{journal}{JHEP} \textbf{\bibinfo{volume}{1401}}, \bibinfo{pages}{173}
  (\bibinfo{year}{2014}), \eprint{1303.5702}.

\bibitem[{\citenamefont{Archer et~al.}(2015)\citenamefont{Archer, Carena,
  Carmona, and Neubert}}]{Archer:2014jca}
\bibinfo{author}{\bibfnamefont{P.~R.} \bibnamefont{Archer}},
  \bibinfo{author}{\bibfnamefont{M.}~\bibnamefont{Carena}},
  \bibinfo{author}{\bibfnamefont{A.}~\bibnamefont{Carmona}}, \bibnamefont{and}
  \bibinfo{author}{\bibfnamefont{M.}~\bibnamefont{Neubert}},
  \bibinfo{journal}{JHEP} \textbf{\bibinfo{volume}{1501}}, \bibinfo{pages}{060}
  (\bibinfo{year}{2015}), \eprint{1408.5406}.

\bibitem[{\citenamefont{Blanke et~al.}(2009{\natexlab{a}})\citenamefont{Blanke,
  Buras, Duling, Gori, and Weiler}}]{Blanke:2008zb}
\bibinfo{author}{\bibfnamefont{M.}~\bibnamefont{Blanke}},
  \bibinfo{author}{\bibfnamefont{A.~J.} \bibnamefont{Buras}},
  \bibinfo{author}{\bibfnamefont{B.}~\bibnamefont{Duling}},
  \bibinfo{author}{\bibfnamefont{S.}~\bibnamefont{Gori}}, \bibnamefont{and}
  \bibinfo{author}{\bibfnamefont{A.}~\bibnamefont{Weiler}},
  \bibinfo{journal}{JHEP} \textbf{\bibinfo{volume}{0903}}, \bibinfo{pages}{001}
  (\bibinfo{year}{2009}{\natexlab{a}}), \eprint{0809.1073}.

\bibitem[{\citenamefont{Blanke et~al.}(2009{\natexlab{b}})\citenamefont{Blanke,
  Buras, Duling, Gemmler, and Gori}}]{Blanke:2008yr}
\bibinfo{author}{\bibfnamefont{M.}~\bibnamefont{Blanke}},
  \bibinfo{author}{\bibfnamefont{A.~J.} \bibnamefont{Buras}},
  \bibinfo{author}{\bibfnamefont{B.}~\bibnamefont{Duling}},
  \bibinfo{author}{\bibfnamefont{K.}~\bibnamefont{Gemmler}}, \bibnamefont{and}
  \bibinfo{author}{\bibfnamefont{S.}~\bibnamefont{Gori}},
  \bibinfo{journal}{JHEP} \textbf{\bibinfo{volume}{0903}}, \bibinfo{pages}{108}
  (\bibinfo{year}{2009}{\natexlab{b}}), \eprint{0812.3803}.

\bibitem[{\citenamefont{Albrecht et~al.}(2009)\citenamefont{Albrecht, Blanke,
  Buras, Duling, and Gemmler}}]{Albrecht:2009xr}
\bibinfo{author}{\bibfnamefont{M.~E.} \bibnamefont{Albrecht}},
  \bibinfo{author}{\bibfnamefont{M.}~\bibnamefont{Blanke}},
  \bibinfo{author}{\bibfnamefont{A.~J.} \bibnamefont{Buras}},
  \bibinfo{author}{\bibfnamefont{B.}~\bibnamefont{Duling}}, \bibnamefont{and}
  \bibinfo{author}{\bibfnamefont{K.}~\bibnamefont{Gemmler}},
  \bibinfo{journal}{JHEP} \textbf{\bibinfo{volume}{0909}}, \bibinfo{pages}{064}
  (\bibinfo{year}{2009}), \eprint{0903.2415}.

\bibitem[{\citenamefont{Agashe et~al.}(2007)\citenamefont{Agashe, Perez, and
  Soni}}]{Agashe:2006wa}
\bibinfo{author}{\bibfnamefont{K.}~\bibnamefont{Agashe}},
  \bibinfo{author}{\bibfnamefont{G.}~\bibnamefont{Perez}}, \bibnamefont{and}
  \bibinfo{author}{\bibfnamefont{A.}~\bibnamefont{Soni}},
  \bibinfo{journal}{Phys.Rev.} \textbf{\bibinfo{volume}{D75}},
  \bibinfo{pages}{015002} (\bibinfo{year}{2007}), \eprint{hep-ph/0606293}.

\bibitem[{\citenamefont{Agashe et~al.}(2014)\citenamefont{Agashe, Azatov, Cui,
  Randall, and Son}}]{Agashe:2014jca}
\bibinfo{author}{\bibfnamefont{K.}~\bibnamefont{Agashe}},
  \bibinfo{author}{\bibfnamefont{A.}~\bibnamefont{Azatov}},
  \bibinfo{author}{\bibfnamefont{Y.}~\bibnamefont{Cui}},
  \bibinfo{author}{\bibfnamefont{L.}~\bibnamefont{Randall}}, \bibnamefont{and}
  \bibinfo{author}{\bibfnamefont{M.}~\bibnamefont{Son}} (\bibinfo{year}{2014}),
  \eprint{1412.6468}.

\bibitem[{\citenamefont{Dillon and Huber}(2014)}]{Dillon:2014zea}
\bibinfo{author}{\bibfnamefont{B.~M.} \bibnamefont{Dillon}} \bibnamefont{and}
  \bibinfo{author}{\bibfnamefont{S.~J.} \bibnamefont{Huber}}
  (\bibinfo{year}{2014}), \eprint{1410.7345}.

\bibitem[{\citenamefont{Cacciapaglia et~al.}(2008)\citenamefont{Cacciapaglia,
  Csaki, Galloway, Marandella, Terning et~al.}}]{Cacciapaglia:2007fw}
\bibinfo{author}{\bibfnamefont{G.}~\bibnamefont{Cacciapaglia}},
  \bibinfo{author}{\bibfnamefont{C.}~\bibnamefont{Csaki}},
  \bibinfo{author}{\bibfnamefont{J.}~\bibnamefont{Galloway}},
  \bibinfo{author}{\bibfnamefont{G.}~\bibnamefont{Marandella}},
  \bibinfo{author}{\bibfnamefont{J.}~\bibnamefont{Terning}},
  \bibnamefont{et~al.}, \bibinfo{journal}{JHEP}
  \textbf{\bibinfo{volume}{0804}}, \bibinfo{pages}{006} (\bibinfo{year}{2008}),
  \eprint{0709.1714}.

\bibitem[{\citenamefont{Agashe et~al.}(2005)\citenamefont{Agashe, Perez, and
  Soni}}]{Agashe:2004cp}
\bibinfo{author}{\bibfnamefont{K.}~\bibnamefont{Agashe}},
  \bibinfo{author}{\bibfnamefont{G.}~\bibnamefont{Perez}}, \bibnamefont{and}
  \bibinfo{author}{\bibfnamefont{A.}~\bibnamefont{Soni}},
  \bibinfo{journal}{Phys.Rev.} \textbf{\bibinfo{volume}{D71}},
  \bibinfo{pages}{016002} (\bibinfo{year}{2005}), \eprint{hep-ph/0408134}.

\bibitem[{\citenamefont{Agashe et~al.}(2004)\citenamefont{Agashe, Perez, and
  Soni}}]{Agashe:2004ay}
\bibinfo{author}{\bibfnamefont{K.}~\bibnamefont{Agashe}},
  \bibinfo{author}{\bibfnamefont{G.}~\bibnamefont{Perez}}, \bibnamefont{and}
  \bibinfo{author}{\bibfnamefont{A.}~\bibnamefont{Soni}},
  \bibinfo{journal}{Phys.Rev.Lett.} \textbf{\bibinfo{volume}{93}},
  \bibinfo{pages}{201804} (\bibinfo{year}{2004}), \eprint{hep-ph/0406101}.

\bibitem[{ATL(2015)}]{ATLAS-CONF-2015-009}
\bibinfo{type}{Tech. Rep.} \bibinfo{number}{ATLAS-CONF-2015-009},
  \bibinfo{institution}{CERN}, \bibinfo{address}{Geneva}
  (\bibinfo{year}{2015}).

\bibitem[{\citenamefont{Kaplan}(1991)}]{Kaplan:1991dc}
\bibinfo{author}{\bibfnamefont{D.~B.} \bibnamefont{Kaplan}},
  \bibinfo{journal}{Nucl.Phys.} \textbf{\bibinfo{volume}{B365}},
  \bibinfo{pages}{259} (\bibinfo{year}{1991}).

\bibitem[{\citenamefont{Agashe and Contino}(2009)}]{Agashe:2009di}
\bibinfo{author}{\bibfnamefont{K.}~\bibnamefont{Agashe}} \bibnamefont{and}
  \bibinfo{author}{\bibfnamefont{R.}~\bibnamefont{Contino}},
  \bibinfo{journal}{Phys.Rev.} \textbf{\bibinfo{volume}{D80}},
  \bibinfo{pages}{075016} (\bibinfo{year}{2009}), \eprint{0906.1542}.

\bibitem[{\citenamefont{Gillioz et~al.}(2012)\citenamefont{Gillioz, Grober,
  Grojean, Muhlleitner, and Salvioni}}]{Gillioz:2012se}
\bibinfo{author}{\bibfnamefont{M.}~\bibnamefont{Gillioz}},
  \bibinfo{author}{\bibfnamefont{R.}~\bibnamefont{Grober}},
  \bibinfo{author}{\bibfnamefont{C.}~\bibnamefont{Grojean}},
  \bibinfo{author}{\bibfnamefont{M.}~\bibnamefont{Muhlleitner}},
  \bibnamefont{and} \bibinfo{author}{\bibfnamefont{E.}~\bibnamefont{Salvioni}},
  \bibinfo{journal}{JHEP} \textbf{\bibinfo{volume}{1210}}, \bibinfo{pages}{004}
  (\bibinfo{year}{2012}), \eprint{1206.7120}.

\bibitem[{\citenamefont{Delaunay et~al.}(2013)\citenamefont{Delaunay, Grojean,
  and Perez}}]{Delaunay:2013iia}
\bibinfo{author}{\bibfnamefont{C.}~\bibnamefont{Delaunay}},
  \bibinfo{author}{\bibfnamefont{C.}~\bibnamefont{Grojean}}, \bibnamefont{and}
  \bibinfo{author}{\bibfnamefont{G.}~\bibnamefont{Perez}},
  \bibinfo{journal}{JHEP} \textbf{\bibinfo{volume}{1309}}, \bibinfo{pages}{090}
  (\bibinfo{year}{2013}), \eprint{1303.5701}.

\bibitem[{\citenamefont{Pomarol and Riva}(2012)}]{Pomarol:2012qf}
\bibinfo{author}{\bibfnamefont{A.}~\bibnamefont{Pomarol}} \bibnamefont{and}
  \bibinfo{author}{\bibfnamefont{F.}~\bibnamefont{Riva}},
  \bibinfo{journal}{JHEP} \textbf{\bibinfo{volume}{1208}}, \bibinfo{pages}{135}
  (\bibinfo{year}{2012}), \eprint{1205.6434}.

\bibitem[{\citenamefont{Azatov et~al.}(2014)\citenamefont{Azatov, Panico,
  Perez, and Soreq}}]{Azatov:2014lha}
\bibinfo{author}{\bibfnamefont{A.}~\bibnamefont{Azatov}},
  \bibinfo{author}{\bibfnamefont{G.}~\bibnamefont{Panico}},
  \bibinfo{author}{\bibfnamefont{G.}~\bibnamefont{Perez}}, \bibnamefont{and}
  \bibinfo{author}{\bibfnamefont{Y.}~\bibnamefont{Soreq}},
  \bibinfo{journal}{JHEP} \textbf{\bibinfo{volume}{1412}}, \bibinfo{pages}{082}
  (\bibinfo{year}{2014}), \eprint{1408.4525}.

\bibitem[{\citenamefont{Redi and Weiler}(2011)}]{Redi:2011zi}
\bibinfo{author}{\bibfnamefont{M.}~\bibnamefont{Redi}} \bibnamefont{and}
  \bibinfo{author}{\bibfnamefont{A.}~\bibnamefont{Weiler}},
  \bibinfo{journal}{JHEP} \textbf{\bibinfo{volume}{1111}}, \bibinfo{pages}{108}
  (\bibinfo{year}{2011}), \eprint{1106.6357}.

\bibitem[{\citenamefont{Redi}(2012)}]{Redi:2012uj}
\bibinfo{author}{\bibfnamefont{M.}~\bibnamefont{Redi}},
  \bibinfo{journal}{Eur.Phys.J.} \textbf{\bibinfo{volume}{C72}},
  \bibinfo{pages}{2030} (\bibinfo{year}{2012}), \eprint{1203.4220}.

\bibitem[{\citenamefont{Geringer-Sameth
  et~al.}(2015)\citenamefont{Geringer-Sameth, Walker, Koushiappas, Koposov,
  Belokurov et~al.}}]{Geringer-Sameth:2015lua}
\bibinfo{author}{\bibfnamefont{A.}~\bibnamefont{Geringer-Sameth}},
  \bibinfo{author}{\bibfnamefont{M.~G.} \bibnamefont{Walker}},
  \bibinfo{author}{\bibfnamefont{S.~M.} \bibnamefont{Koushiappas}},
  \bibinfo{author}{\bibfnamefont{S.~E.} \bibnamefont{Koposov}},
  \bibinfo{author}{\bibfnamefont{V.}~\bibnamefont{Belokurov}},
  \bibnamefont{et~al.} (\bibinfo{year}{2015}), \eprint{1503.02320}.

\bibitem[{\citenamefont{Alloul et~al.}(2014)\citenamefont{Alloul, Christensen,
  Degrande, Duhr, and Fuks}}]{Alloul:2013bka}
\bibinfo{author}{\bibfnamefont{A.}~\bibnamefont{Alloul}},
  \bibinfo{author}{\bibfnamefont{N.~D.} \bibnamefont{Christensen}},
  \bibinfo{author}{\bibfnamefont{C.}~\bibnamefont{Degrande}},
  \bibinfo{author}{\bibfnamefont{C.}~\bibnamefont{Duhr}}, \bibnamefont{and}
  \bibinfo{author}{\bibfnamefont{B.}~\bibnamefont{Fuks}},
  \bibinfo{journal}{Comput.Phys.Commun.} \textbf{\bibinfo{volume}{185}},
  \bibinfo{pages}{2250} (\bibinfo{year}{2014}), \eprint{1310.1921}.

\bibitem[{\citenamefont{Alwall et~al.}(2014)\citenamefont{Alwall, Frederix,
  Frixione, Hirschi, Maltoni et~al.}}]{Alwall:2014hca}
\bibinfo{author}{\bibfnamefont{J.}~\bibnamefont{Alwall}},
  \bibinfo{author}{\bibfnamefont{R.}~\bibnamefont{Frederix}},
  \bibinfo{author}{\bibfnamefont{S.}~\bibnamefont{Frixione}},
  \bibinfo{author}{\bibfnamefont{V.}~\bibnamefont{Hirschi}},
  \bibinfo{author}{\bibfnamefont{F.}~\bibnamefont{Maltoni}},
  \bibnamefont{et~al.}, \bibinfo{journal}{JHEP}
  \textbf{\bibinfo{volume}{1407}}, \bibinfo{pages}{079} (\bibinfo{year}{2014}),
  \eprint{1405.0301}.

\end{thebibliography}

\end{document}